\documentclass[a4paper,12pt]{article}
\usepackage[utf8]{inputenc}
\usepackage{jheppub}
\usepackage{tikz}
\usepackage[miktex]{gnuplottex}
\usetikzlibrary{shapes,arrows,positioning,automata,backgrounds,calc,er,patterns}
\usepackage{tikz-feynman}
\tikzfeynmanset{compat=1.0.0}
\usepackage{soul}
\usepackage{slashed} 
\usepackage{enumerate}
\usepackage{epsfig} 
\usepackage{float} 
\usepackage{subcaption} 
\usepackage{amsmath} 
\usepackage{wasysym} 
\usepackage{mathrsfs} 
\usepackage{amsfonts} 
\usepackage{amsbsy} 
\usepackage{amscd}    
\usepackage{multirow} 
\usepackage{tikz}
\usepackage{slashed}
\usepackage[miktex]{gnuplottex}
\usetikzlibrary{arrows,positioning,shapes.geometric} 
\usepackage{color}
\usepackage{verbatim}
\usepackage{xcolor}
\usepackage{aligned-overset}

\makeatletter

\def\eeq{\end{equation}}
\def\beq{\begin{equation}}

\newcommand{\Rmnum}[1]{\expandafter\@slowromancap\romannumeral #1@}

\newcommand{\cbra}[1]{\left\{#1\right\}}
\newcommand{\rbra}[1]{\left(#1\right)}

\newcommand{\bea} {\begin{eqnarray}}
\newcommand{\eea} {\end{eqnarray}}

	\newcommand{\nn}{\nonumber}

	\newcommand{\re}{\textrm{Re}}
	\newcommand{\im}{\textrm{Im}}

\makeatother

\title{Probing CP Violation in the Complex 2HDMS with a 95 GeV Scalar at Future $e^+e^-$ Colliders}

\author[a]{Jayita Lahiri,}
   \author[b]{Cheng Li,}
\author[c,d]{Gudrid Moortgat-Pick}

\affiliation[a]{Department of Physical Sciences, Indian Institute of Science Education and Research Kolkata, Mohanpur - 741246, India}
\affiliation[b]{School of Science, Shenzhen Campus of Sun Yat-Sen University, Gongchang Road 66, 518107 Shenzhen, China}
\affiliation[c]{II. Institut f{\"u}r Theoretische Physik, Universit{\"a}t Hamburg, Luruper Chaussee 149, 22761 Hamburg, Germany} 
\affiliation[d]{Deutsches Elektronen-Synchrotron DESY, Notkestr.\ 85, 22607 Hamburg, Germany}

\emailAdd{jayita.lahiri@desy.de}
\emailAdd{gudrid.moortgat-pick@desy.de}
\emailAdd{cheng.li@desy.de}

\abstract{In this study, we investigate the prospects for probing CP violation in the complex singlet extension of the two-Higgs-doublet model (2HDM) at future electron-positron colliders. In extended scalar sectors, cancellations among different contributions to electric dipole moments (EDMs) can relax the stringent constraints on CP-violating phases, potentially bringing them in the current reach of collider experiments. We consider a scenario in which both the 125 GeV Higgs boson and the additional scalar associated with the reported 95 GeV excess are CP-mixed states. We probe the CP-violating phase of the 95 GeV scalar through its production via Higgsstrahlung and subsequent decay into a tau-lepton pair, using a CP-sensitive forward-backward asymmetry in hadronic and semileptonic tau-decay channels. We estimate the sensitivity at future lepton colliders, namely the ILC, FCC-ee, and CEPC, and investigate the impact of the Standard Model Drell-Yan background. We find that the hadronic channel offers greater sensitivity than the semileptonic channel. We also compare our results with existing constraints and future projections for CP violation in the 125 GeV Higgs sector and briefly discuss the implications for electroweak baryogenesis. Our results highlight the potential of future lepton colliders to probe CP violation in extended scalar sectors while remaining consistent with current EDM constraints.   
}

\preprint{DESY-26-096}
\begin{document}

\maketitle

\newpage

\section{Introduction}

The Standard Model (SM) of particle physics has been scrutinized and validated to a great degree of precision in recent experiments, confirming that it is a successful `effective' theory.
However, there are still many unanswered questions that push us to explore beyond the Standard Model. One such pertinent question is 
the matter-antimatter asymmetry of the universe.
The dominance of matter over antimatter has been puzzling for a long time. Quantitatively, this asymmetry can be written as 

\begin{equation}
 \eta=  \frac{n_B - n_{\bar B}}{n_{\gamma}} = 6 \times 10^{-10}, 
\end{equation}

\noindent
where $n_B$ is the number of baryons, $n_{\bar B}$ the number of anti-baryons and $n_\gamma$ is the number of photons in the universe.
The process in which the observed non-zero baryon asymmetry of the universe (BAU) was produced, is called baryogenesis. For a successful baryogenesis, the three Sakharov conditions need to be satisfied~\cite{Sakharov:1967dj}:
\begin{itemize}
\item Baryon number violation,
\item C and CP-violation,
\item Departure from thermal equilibrium. 
\end{itemize}

\noindent
Interestingly, in the SM, all the elements mentioned above are present: for example, B-violation is observed in the sphaleron process, the weak interaction violates C (charge conjugation), and CP-violation is also present via the phases of Cabbibo-Kobayashi-Maskawa (CKM) matrix. But the predicted CP-violating effects within the SM are too small, and therefore, the numerical value of the observed BAU is also too low in SM. Furthermore, a strong first-order phase transition, which is an out-of-equilibrium process, is not achievable within the SM given the current observed Higgs mass. 
Therefore, in order to achieve a successful baryogenesis, we must go beyond the SM. Over time, various theories have been proposed, such as baryogenesis with Grand Unified Theories (GUT baryogenesis) \cite{Yoshimura:1978ex,Weinberg:1979bt}, baryogenesis via leptogenesis ~\cite{Fukugita:1986hr,Luty:1992un}, and electroweak baryogenesis ~\cite{Kuzmin:1985mm,Shaposhnikov:1987tw,Trodden:1998ym,Morrissey:2012db}. A large number of studies exist in the context of electroweak baryogenesis, in models with an extended scalar sector, for instance in the two Higgs Doublet Model (2HDM) and its extensions~\cite{Fromme:2006cm,Profumo:2007wc,Arhrib:2010ju,Cline:2011mm,Biekotter:2025fjx,Biekotter:2025vxl}. 

Although the observed baryon asymmetry of the Universe points to the need for additional sources of CP violation, no direct experimental evidence for CP violation beyond the Standard Model has been established to date. Therefore, considerable theoretical and experimental efforts are ongoing to search for CP violation in the BSM sector, should it be present. Low-energy electric dipole moment (EDM) experiments provide a powerful probe of CP-violating interactions. However, no statistically significant evidence for a nonzero EDM has been observed to date, leading to stringent upper bounds on the electric and neutron EDMs. On the other hand, collider experiments provide another important avenue for probing CP violation. CP-sensitive observables can be constructed and measured at colliders to search for deviations from the predictions of the SM. In particular, extensive efforts have been made to search for CP violation in the scalar sector. For instance, the CP properties of the 125 GeV Higgs boson being studied through various production and decay channels. One of the most important Higgs decay channels in this context is the $\tau\tau$ final state. The CP properties of the 125 GeV Higgs boson are being probed at the LHC and are expected to be investigated with greater precision at future lepton colliders. At the same time, CP violation in the Higgs sector is strongly constrained by electric dipole moment (EDM) experiments, as CP-violating interactions can induce nonzero EDMs. It is therefore interesting to explore the interplay between these two complementary probes and investigate the extent to which collider measurements and low-energy EDM constraints can jointly constrain CP violation in the Higgs sector.

We emphasize in this work, that the CP-violating phase if present only in the 125 GeV Higgs boson, will be strongly constrained by the EDM experiments. Consequently, such tiny phases will  be beyond the reach of future probes at $e^+e^-$ colliders. However, interesting possibilities can arise if there is an extended scalar sector and multiple CP-phases exist in the theory. In such cases, for a wide range of parameter points, cancellation among multiple contributions to EDM takes place and the EDM bounds become much relaxed~\cite{Kanemura:2020ibp,Lahiri:2026eud}. One should note that, owing to the presence of multiple phases and accidental cancellations, studying the interplay between EDM and collider probes of CP-phases, is possible only in the context of concrete models.
 In this work, we consider the complex singlet extension of two Higgs doublet model~\cite{Shah:2018elh,Heinemeyer:2021msz,Dutta:2022qeq,Dutta:2023cig,Dutta:2025nmy}. We would like to mention that this model can accommodate the recently observed 95 GeV excess at the $\gamma\gamma$~\cite{CMS:2017yta,ATLAS:2018xad,CMS:2023yay} and $\tau\tau$ modes~\cite{CMS:2022rbd} at the LHC experiment CMS  and more recently in the $\gamma\gamma$ mode at the LHC experiment ATLAS \cite{ATLAS-CONF-2023-035}. Similar excess around the same mass range has been observed at the former Large Electron-Positron Collider (LEP) in the $b\bar{b}$ mode~\cite{LEPWorkingGroupforHiggsbosonsearches:2003ing}. We would like to mention that such an excess can be explained in several BSM models with extended scalar sectors~\cite{Fox:2017uwr, Haisch:2017gql,  Benbrik:2022azi, Benbrik:2022dja, Azevedo:2023zkg, Belyaev:2023xnv,Biekotter:2019kde,Aguilar-Saavedra:2023vpd, Banik:2023ecr,Heinemeyer:2021msz,Biekotter:2021ovi, Biekotter:2023jld, Biekotter:2023oen,Cao:2016uwt, Cao:2019ofo, Choi:2019yrv, Biekotter:2021qbc, Li:2022etb,PhysRevD.100.035023, Biekotter:2017xmf, Liu:2018ryo, Liu:2018xsw, Biekotter:2019gtq, Richard:2017kot, Aguilar-Saavedra:2020wrj, Biekotter:2020cjs, Coloretti:2023wng, Bhattacharya:2023lmu, Ashanujjaman:2023etj, Escribano:2023hxj,Borah:2023hqw}. We consider the interesting scenario in our model, where along with the 125 GeV Higgs, the 95 GeV scalar is also a CP-mixed state. Here too, multiple CP-phases can lead to relaxed bounds from EDM. We probe the CP-phase in the 95 GeV scalar in its $\tau\tau$ decay mode and present the sensitivity of the same at future colliders such as ILC, FCC-ee or CEPC. We compare the results with already existing projections for the CP-phase of the 125 GeV Higgs at the future lepton colliders. We would like to mention that our model can accommodate a viable dark matter candidate, while at the same time include CP-violation and the aforementioned 95 GeV excess. The interplay between dark matter phenomenology and CP violation has been explored in earlier works~\cite{Biekotter:2025vxl,Lahiri:2026eud}.

The structure of our paper is as follows. In Section~\ref{model}, we discuss the complex singlet extension of 2HDM, the general model framework, and discuss the mass eigenstates and possible CP-mixing. In Section \ref{simplified}, we consider a simplified version of this model where the CP-phase is assumed to be in the 125 GeV and 95 GeV scalars, all other scalars are assumed to be CP-eigenstates. In Section~\ref{edm}, we discuss the constraints from electric dipole moment experiments. In Section \ref{collider}, we construct CP-sensitive angular observables and evaluate the sensitivity of future $e^+e^-$ colliders to the CP-mixing angle (or CP-violating phase) of the 95 GeV scalar. In Section~\ref{combined}, we present a combined interpretation of our results with existing CP-studies as well as future projections on the CP-mixed 125 GeV Higgs boson. We also comment on baryogenesis in this scenario. Finally, we summarize our results and conclude in Section~\ref{conclusion}.

\section{General framework}
\label{model}
The complex singlet extension of 2HDM or complex 2HDMS consists of two gauge doublet fields and one complex gauge singlet field, which are shown in the following:
\begin{equation}
    \Phi_1 = e^{i  \xi_1} \begin{pmatrix}
    \chi_1^+\\ \frac{1}{\sqrt{2}}{(v_1 + \rho_1 + i \eta_1)}
    \end{pmatrix},~\Phi_2 = \begin{pmatrix}
    \chi_2^+\\ \frac{1}{\sqrt{2}}{(v_2 + \rho_2 + i \eta_2)} 
    \end{pmatrix},~S = e^{i \xi_S}(v_S + \rho_S + i \eta_S)~.
\end{equation}
The general form of the Higgs fields can take the complex phases, which can generate CP-violation in the theory. 
Even though, both $\Phi_1$ and $\Phi_2$ have a CP-phase, it is always possible to rotate the doublet fields to a basis where only one CP-phase exists. We choose to keep the CP-phase $\xi_1$ in $\Phi_1$.

We will now discuss the scalar-fermion interaction via the Yukawa Lagrangian in our model.
The general form of the Yukawa couplings is given by the following terms of the Lagrangian
\begin{equation}
    \begin{split}
        \mathcal{L}_{yuk} = &Y_d \Phi_1^\dagger \bar{Q}_L d_R + Y_e \Phi_1^\dagger \bar{L}_L e_R -Y_u \Phi_1 \bar{Q}_L u_R \\
&+\epsilon_d\Phi_2^\dagger \bar{Q}_L d_R + \epsilon_e \Phi_2^\dagger \bar{L}_L e_R -\epsilon_u \Phi_2 \bar{Q}_L u_R
    \end{split}~,
\end{equation}
where both gauge doublets $\Phi_1$ and $\Phi_2$ can couple to the up-type quarks, down-type quarks and leptons. After electroweak symmetry breaking, the $\Phi_1$ and $\Phi_2$ can mix together and generate the masses of fermions, where the SM-like Higgs to fermions couplings can be expressed in terms of the combination of the $Y_f$ (Yukawa couplings of $\Phi_1$) and the $\epsilon_f$ (Yukawa couplings of $\Phi_2$) 
\begin{equation}
    Y^f_{SM} v_{SM}= Y_f (v_1 e^{-i\xi_1}) + \epsilon_f v_2~.
\end{equation}

\noindent
We define the coefficient $\zeta_f$ by the relation $\epsilon_f = \zeta_f Y_f$ enabling simultaneous diagonalization of the two mass matrices $Y_f$ and $\epsilon_f$~\cite{Pich:2009sp}. We can thereafter solve the $Y_f$ and $\epsilon_f$ as follows.
\begin{align}
    Y_f &= \frac{Y^f_\mathrm{SM}}{v_\mathrm{SM}}  ( v_1-\zeta_f v_2 )e^{i\xi_1},\\
    \epsilon_f &=  \frac{Y^f_\mathrm{SM}}{v_\mathrm{SM}}  (v_2  + \zeta_f v_1).
\end{align}
\begin{table}[h]
    \centering
    \begin{tabular}{c|cccc}
         & type-I& type-II& type-X& type-Y\\
         \hline
         $\zeta_u$& $\frac{1}{\tan\beta}$& $\frac{1}{\tan\beta}$& $\frac{1}{\tan\beta}$& $\frac{1}{\tan\beta}$\\
         $\zeta_d$& $\frac{1}{\tan\beta}$& $-\tan\beta$& $\frac{1}{\tan\beta}$& $-\tan\beta$\\
         $\zeta_e$& $\frac{1}{\tan\beta}$& $-\tan\beta$& $-\tan\beta$& $\frac{1}{\tan\beta}$ \\
    \end{tabular}
    \caption{The choice of $\zeta_f$ for four  types of 2HDM in the Higgs basis (discussed in \autoref{higgsbasis}). We chose to show the $\zeta_f$'s in the Higgs basis for their simple forms in the Higgs basis.}
    \label{tab:yuk_zeta}
\end{table}
\noindent
We define $\tan\beta = v_2/v_1$. In particular, we can choose the specific values for the proportionality factor $\zeta_f$ in Table~\ref{tab:yuk_zeta}, and enforce the Yukawa couplings to follow typical four-types of 2HDM. In this work, we will assume Type-II case for Yukawa couplings.

The most general scalar potential in this model is given as
\begin{equation}
\begin{split}
        V = &m_{11}^2 (\Phi_1^\dagger \Phi_1) + m_{22}^2 (\Phi_2^\dagger \Phi_2)  + \frac{\lambda_1}{2}(\Phi_1^\dagger \Phi_1)^2 + \frac{\lambda_2}{2} (\Phi_2^\dagger \Phi_2)^2 + \lambda_3 (\Phi_2^\dagger \Phi_2)(\Phi_1^\dagger \Phi_1) + \lambda_4 (\Phi_1^\dagger \Phi_2)(\Phi_2^\dagger \Phi_1)\\
        &+[-m_{12}^2(\Phi_1^\dagger \Phi_2) +\frac{\lambda_5}{2}(\Phi_1^\dagger \Phi_2)^2 + \lambda_6 (\Phi_1^\dagger \Phi_1)(\Phi_1^\dagger \Phi_2) + \lambda_7(\Phi_2^\dagger \Phi_2)(\Phi_1^\dagger \Phi_2) + \mathrm{h.c.}]\\
        & +m_S^2 (S^\dagger S) + \lambda_{1}' (S^\dagger S)(\Phi_1^\dagger \Phi_1) + \lambda_2'(S^\dagger S)(\Phi_2^\dagger \Phi_2) + \frac{\lambda_3''}{4}(S^\dagger S)^2\\
        &+[{m_S'}^2 S^2+ \mu_{S1} S^3 + \mu_{S2} S (S^\dagger S) + \mu_{11} S (\Phi_1^\dagger \Phi_1) + \mu_{22}S(\Phi_2^\dagger \Phi_2) + \mu_{12} S(\Phi_1^\dagger \Phi_2) + \mu_{21} S(\Phi_2^\dagger \Phi_1)\\
        &+\frac{\lambda_1''}{4!}S^4 + \frac{\lambda_2''}{3!}(S^\dagger S)S^2 + \lambda_3' (S^\dagger S)(\Phi_1^\dagger \Phi_2) + \lambda_4' S^2(\Phi_1^\dagger \Phi_1) + \lambda_5'S^2(\Phi_2^\dagger \Phi_2)\\&+\lambda_6' S^2(\Phi_1^\dagger \Phi_2) + \lambda_7' S^2 (\Phi_2^\dagger \Phi_1) + \mathrm{h.c.} ].
\end{split}~.
\end{equation}
In this Higgs potential, the parameters $m_{11}$, $m_{22}$,  $m_S$, $\lambda_{1,2,3,4}$, $\lambda'_{1,2}$ and $\lambda_3''$ are real. The ${m_S'}^2$, $m_{12}^2$  $\lambda_{5,6,7}$, $\lambda'_{3,4,5,6,7}$, $\lambda_{1,2}''$ and all the cubic terms are the complex parameters, which can provide possible sources of CP-violation.

After spontaneous symmetry breaking in 2HDMS, one obtains five neutral scalars and one charged scalar as physical scalars. For the neutral fields, we can do the following rotation on the interaction eigenstates. The transformation is performed in two steps. In the first step the rotation is in the pseudo-scalar sector $\eta_1, \eta_2 \rightarrow G^0, \eta$ as follows:
\begin{equation}
    \begin{pmatrix}
    \rho_1\\ \rho_2\\ \rho_S\\ G^0\\ \eta\\ \eta_S
    \end{pmatrix} = 
    \begin{pmatrix}
1 & 0 & 0 & 0 & 0 & 0 \\
0 & 1 & 0 & 0 & 0 & 0 \\
0 & 0 & 1 & 0 & 0 & 0 \\
0 & 0 & 0  & \cos\beta & \sin\beta& 0 \\
0 & 0 & 0  & -\sin\beta & \cos\beta& 0\\
0 & 0 & 0 & 0 & 0 & 1 \\
\end{pmatrix}\begin{pmatrix}\rho_1\\ \rho_2\\ \rho_S\\ \eta_1\\ \eta_2\\ \eta_S,
    \end{pmatrix}.
\end{equation}
where the Goldstone boson degree of freedom $G^0$ can be separated from the physical scalar sector.
\begin{align}
    G^0 = \eta_1\cos\beta  + \eta_2\sin\beta, \\
    \eta = \eta_2 \cos\beta -\eta_1\sin\beta.
\end{align}
In the second step, the fields $\rho_1$, $\rho_2$, $\rho_S$, $\eta$ and $\eta_S$ can be rotated by 5$\times$5 SO(5) rotation matrix $R$ and we obtain five mass eigenstates. 
\begin{equation}
\begin{pmatrix}
h_1\\ h_2\\ h_3\\ h_4\\ h_5\\ 
\end{pmatrix} = R \begin{pmatrix}
    \rho_1\\ \rho_2\\ \rho_S\\ \eta\\ \eta_S
    \end{pmatrix},~~~~
    \mathrm{diag}( m_{h_1}^2, m_{h_1}^2, m_{h_3}^2, m_{h_4}^2, m_{h_5}^2 ) = R M_H^2 R^T,
\end{equation}
\noindent
where $M_H^2$ is the mass matrix of the neutral scalars whose explicit forms are given in Eqs.~\eqref{eq:mh11}-\eqref{eq:mhc}.
The rotation matrix $R$ has 10 mixing angles 
\begin{equation}
\begin{split}
        &R = R(\alpha_{12}, \alpha_{13}, \alpha_{23}, \alpha_{14}, \alpha_{24}, \alpha_{34}, \alpha_{15}, \alpha_{25}, \alpha_{35}, \alpha_{45}), \\
        &=\begin{pmatrix}
1 & 0 & 0 & 0 & 0 \\
0 & 1 & 0 & 0 & 0 \\
0 & 0 & 1 & 0 & 0 \\
0 & 0 & 0 & c_{\alpha_{45}} & s_{\alpha_{45}} \\
0 & 0 & 0 & -s_{\alpha_{45}} & c_{\alpha_{45}}
\end{pmatrix}\begin{pmatrix}
1 & 0 & 0 & 0 & 0 \\
0 & 1 & 0 & 0 & 0 \\
0 & 0 & c_{\alpha_{35}} & 0 & s_{\alpha_{35}} \\
0 & 0 & 0 & 1 & 0 \\
0 & 0 & -s_{\alpha_{35}} & 0 & c_{\alpha_{35}}
\end{pmatrix}\begin{pmatrix}
1 & 0 & 0 & 0 & 0 \\
0 & c_{\alpha_{25}} & 0 & 0 & s_{\alpha_{25}} \\
0 & 0 & 1 & 0 & 0 \\
0 & 0 & 0 & 1 & 0 \\
0 & -s_{\alpha_{25}} & 0 & 0 & c_{\alpha_{25}}
\end{pmatrix}\begin{pmatrix}
c_{\alpha_{15}} & 0 & 0 & 0 & s_{\alpha_{15}} \\
0 & 1 & 0 & 0 & 0 \\
0 & 0 & 1 & 0 & 0 \\
0 & 0 & 0 & 1 & 0 \\
-s_{\alpha_{15}} & 0 & 0 & 0 & c_{\alpha_{15}}
\end{pmatrix}\\&~~~~\begin{pmatrix}
1 & 0 & 0 & 0 & 0 \\
0 & 1 & 0 & 0 & 0 \\
0 & 0 & c_{\alpha_{34}} & s_{\alpha_{34}} & 0 \\
0 & 0 & -s_{\alpha_{34}} & c_{\alpha_{34}} & 0 \\
0 & 0 & 0 & 0 & 1
\end{pmatrix}\begin{pmatrix}
1 & 0 & 0 & 0 & 0 \\
0 & c_{\alpha_{24}} & 0 & s_{\alpha_{24}} & 0 \\
0 & 0 & 1 & 0 & 0 \\
0 & -s_{\alpha_{24}} & 0 & c_{\alpha_{24}} & 0 \\
0 & 0 & 0 & 0 & 1
\end{pmatrix}\begin{pmatrix}
c_{\alpha_{14}} & 0 & 0 & s_{\alpha_{14}} & 0 \\
0 & 1 & 0 & 0 & 0 \\
0 & 0 & 1 & 0 & 0 \\
-s_{\alpha_{14}} & 0 & 0 & c_{\alpha_{14}} & 0 \\
0 & 0 & 0 & 0 & 1
\end{pmatrix}\\&~~~~\begin{pmatrix}
1 & 0 & 0 & 0 & 0 \\
0 & c_{\alpha_{23}} & s_{\alpha_{23}} & 0 & 0 \\
0 & -s_{\alpha_{23}} & c_{\alpha_{23}} & 0 & 0 \\
0 & 0 & 0 & 1 & 0 \\
0 & 0 & 0 & 0 & 1
\end{pmatrix}\begin{pmatrix}
c_{\alpha_{13}} & 0 & s_{\alpha_{13}} & 0 & 0 \\
0 & 1 & 0 & 0 & 0 \\
-s_{\alpha_{13}} & 0 & c_{\alpha_{13}} & 0 & 0 \\
0 & 0 & 0 & 1 & 0 \\
0 & 0 & 0 & 0 & 1
\end{pmatrix}\begin{pmatrix}
c_{\alpha_{12}} & s_{\alpha_{12}} & 0 & 0 & 0 \\
-s_{\alpha_{12}} &  c_{\alpha_{12}} & 0 & 0 & 0 \\
0 & 0 & 1 & 0 & 0 \\
0 & 0 & 0 & 1 & 0 \\
0 & 0 & 0 & 0 & 1
\end{pmatrix}.
\end{split}
\end{equation}
The explicit entries of the $R$ matrix is shown in Eqs.~\eqref{eq:r11}-\eqref{eq:r55}.
By applying this rotation in the neutral scalar sector, we can obtain the general form of the effective Higgs to fermions couplings in the mass basis
, which is given by
\begin{equation}
    c_{h_i ff}  = (c_\beta - \zeta_f s_\beta )(R_{i1}-i\gamma^5 R_{i4}c_\beta ) + (s_\beta + \zeta_f c_\beta)(R_{i2}-i\gamma^5 R_{i4}s_\beta).\label{eq:hiff}
\end{equation}

\noindent
It is clear that the mixing between the CP-even neutral states $\rho_{1,2,3}$ and the CP-odd neutral states $\eta, \eta_S$, will ensure CP-violation in the scalar sector and introduce CP-violating Yukawa couplings of all the CP-mixed scalars (see Eq.~\ref{eq:hiff}). The details of the general scenario is given in \autoref{general}. For the purpose of simplicity, we will choose a scenario, where the rotation matrix elements are chosen in such a way that the two mass eigenstates with masses 125 GeV (our observed Higgs boson) and 95 GeV (the observed excess) will be CP-mixed. All other scalars are assumed to be predominantly CP-eigenstates. We discuss the simplified scenario in the next section. 


\section{Theoretical set-up - CP-violating phases only in the 95 GeV and 125 GeV Higgs}
\label{simplified}

We will now consider a simplified scenario, where the mixing angles discussed above are chosen in a way that the CP-mixing happens in a restricted sector. Before that, we mention that when the mixing angles $\cos(\beta-\alpha_{12})$ and the rest of the mixing angles are all close to zero, 
\begin{equation}
    (c_{\beta-\alpha_{12}}, \alpha_{13}, \alpha_{23}, \alpha_{14}, \alpha_{24}, \alpha_{34}, \alpha_{15}, \alpha_{25}, \alpha_{35}, \alpha_{45}) \approx 0.
\end{equation}
the five neutral scalars would be of the following structure, which is the CP-conserving limit of the model.
\begin{itemize}
    \item $h_1 \rightarrow$ BSM CP-even doublet scalar,
    \item $h_2 \rightarrow$ the SM-like Higgs ($c_{\beta-\alpha_{12}} \approx 0$ ensures the SM-like nature of $h_2$ i.e. the alignment limit is established),
    \item $h_3 \rightarrow$ the CP-even singlet scalar,
    \item $h_4 \rightarrow$ the CP-odd doublet scalar,
    \item $h_5 \rightarrow$ can act as a dark sector.
\end{itemize}

\noindent
In the complex 2HDMS model, we have multiple possible sources of CP-violation, namely the complex parameters in the Higgs potential, the phase in scalar vev's and the complex part of the Yukawa couplings parameters $\zeta_f$.  
However, for sake of simplicity we choose the vev's to be real and choose the Type-II Yukawa structures in Table~\ref{tab:yuk_zeta}, and therefore $\zeta_f$ would be real. In this case, the complex parameters in the Higgs potential would lead to non-zero mixing angles between CP-even states and the CP-odd state.

As already mentioned, in this study, we consider type-II Yukawa couplings of the scalar doublets, and therefore the Higgs to fermions couplings are given by
\begin{align}
c_{h_itt} &= R_{i2}/\sin\beta
-i\gamma_5 R_{i4}/\tan\beta, \label{fermcoupling1}\\
c_{h_ibb} &= c_{h_i\tau\tau} = R_{i1}/\cos\beta
-i\gamma_5 R_{i4}\tan\beta.
\label{fermcoupling2}
\end{align}
The reduced fermion Yukawa couplings can be parametrized as the following:
\begin{equation}
    c_{h_iff} = a_f + i \gamma_5 b_f,
\end{equation}
\begin{eqnarray}  
 \frac{a_f}{\sqrt{a_f^2 + b_f^2}} = \cos\phi_f,~~
 \frac{b_f}{\sqrt{a_f^2 + b_f^2}} = \sin\phi_f
 \label{parameter}
\end{eqnarray}

\noindent
As we see in Eqs.~\ref{fermcoupling1} and \ref{fermcoupling2}, the complex part of the scalar to fermions couplings depend on the CP-mixing entries of $R_{i4}$, which are dominated by the mixing angles $\alpha_{14}$, $\alpha_{24}$ and $\alpha_{34}$ (see Table~\ref{tab:h123ff}). 
\begin{itemize}
    \item $\alpha_{14}\to$ mixing between BSM doublet-like CP-even $\rho_1$ and pseudo-scalar $\eta$,
    \item $\alpha_{24}\to$ mixing between CP-even $\rho_2$ and pseudo-scalar $\eta$,
    \item $\alpha_{34}\to$ mixing between singlet-like $\rho_S$ and pseudo-scalar $\eta$.
\end{itemize}

\noindent
The corresponding entries in the scalar mass matrix which lead to such mixing, depend on the imaginary part of the complex parameters, particularly the $\mathbb{Z}_2$ breaking parameters.
Here, the corresponding CP-mixing mass matrix entries leading to nonzero $R_{i4}$ are as follows:
\begin{equation}
    \begin{split}
        M_{H14}^2 &= -\frac{1}{2}v^2\mathrm{Im}(2\lambda_6c_\beta + \lambda_5 s_\beta),\\
        M_{H24}^2 &= -\frac{1}{2}v^2\mathrm{Im}(2\lambda_7s_\beta + \lambda_5 c_\beta),\\
        M_{H34}^2 &= \mathrm{Im}\Big( (\mu_{12}-\mu_{21}) v + 2(\lambda_7'-\lambda_6')v_S - 2\lambda_3' v v_S \Big).
    \end{split}
\end{equation}
Although the $\mathbb{Z}_2$ breaking parameters still exist, these parameters would only contribute to the loop-level FCNC, since we assume the type-II Yukawa coupling structure.

Furthermore, we associate $h_3$ with the 95~GeV excess at LEP and LHC, while the $h_5$ could be a dark matter in our study. The dark matter phenomenology in this model has been studied in ~\cite{Lahiri:2026eud}.
We impose a condition on the mixing angles of $h_5$ $(\alpha_{15},~\alpha_{25},~\alpha_{35},~\alpha_{45})\to0$, so that $h_5$ can act as a dark particle.
Under these circumstances, the fermionic couplings of $h_{1,2,3}$ are shown in Table~\ref{tab:h123ff}.

\begin{table}[!hptb]
    \centering
    \hspace*{-1cm}
   {\small{ \begin{tabular}{lcc}
        &    $c_{h_itt}$&  $c_{h_ibb}$ \\
        \hline\\
         $h_1$& $\frac{s_{{12}}}{s_\beta}c_{{13}}c_{{14}} -i\gamma_5\frac{s_{14}}{t_\beta} $& $\frac{c_{{12}}}{c_\beta}c_{{13}}c_{{14}}-i\gamma_5 s_{14} t_\beta$\\ \\
         $h_2$& $\bigl(c_{12}c_{23}c_{24} + s_{12}(-c_{24}s_{13}s_{23} - c_{13}s_{14}s_{24})\bigr)/{s_\beta} $& $(-c_{23}c_{24}s_{12} + c_{12}\bigl(-c_{24}s_{13}s_{23} - c_{13}s_{14}s_{24}\bigr))/c_\beta$\\
         &  $-i\gamma_5(c_{14}s_{24})/{t_\beta} $ &   $-i\gamma_5 c_{14}s_{24} t_\beta$\\ \\
         & $\bigl( c_{12}(-c_{34}s_{23} - c_{23}s_{24}s_{34})  $& $\bigl( s_{12}(c_{34}s_{23} + c_{23}s_{24}s_{34})$\\
         $h_3$&  $+ s_{12}\bigl(-s_{13}(c_{23}c_{34} - s_{23}s_{24}s_{34}) - c_{13}s_{14}c_{24}s_{34}\bigr) \bigr)/{s_\beta}$&    $+ c_{12}\bigl(-s_{13}(c_{23}c_{34} - s_{23}s_{24}s_{34}) - c_{13}s_{14}c_{24}s_{34}\bigr)\bigr)/c_\beta$\\
         &  $-i\gamma_5(c_{14}c_{24}s_{34})/{t_\beta} $ &   $-i\gamma_5 c_{14}c_{24}s_{34} t_\beta$\\ \\
         \hline
    \end{tabular}}}
    \caption{Reduced fermion couplings to $h_1$, $h_2$ and $h_3$.}
    \label{tab:h123ff}
\end{table}

In order to obtain the valid parameter points, we check the parameter points against the theoretical constraints as well as the experimental constraints. The model files  are generated using \texttt{SARAH-v4.14.3}~\cite{Staub:2013tta} and the particle spectra and decays are generated using \texttt{SPheno-v4.0.5}~\cite{Porod:2003um} We check the tree-level perturbative unitarity constraints by $\mathtt{SPheno}$ and check the boundedness from below by the method as discussed in \autoref{sec:bfb}. For the collider constraints, we use the $\mathtt{HiggsTools}$~\cite{Bahl:2022igd,Bechtle:2013xfa,Bechtle:2013wla,Bechtle:2020pkv,Bechtle:2020uwn} to check the BSM scalars against the LHC, LEP and Tevatron direct search bounds, and restrict the 125~GeV Higgs within the allowed range of Higgs measurement at 95\% C.L. Furthermore, we also impose constraints from the $STU$ observables and $B$-physics observables, both of which are calculated using $\mathtt{SPheno}$.

After performing a global scan, we obtain a benchmark point allowed by all the aforementioned constraints. This benchmark scenario, shown in Table~\ref{tab:bench}, accommodate the 95~GeV excess and the CP-violation effects are allowed by collider searches and EDM measurement (which will be discussed in detail in the next section).
\begin{table}[h]
    \centering
    {\small{\begin{tabular}{ccccccccc}
    \hline\\
        $m_{h_1}$ & $m_{h_2}$& $m_{h_3}$& $m_{h_4}$& $m_{h_5}$& $m_{H^\pm}$&  $\tilde{\mu}$& $v_S$\\
         $800$& $125$& $95$& $805$& $150$&795& $800$& 1000\\ \\
         \hline\\
         $t_\beta$&$c_{\beta-\alpha_{12}}$& $\alpha_{13}$& $\alpha_{23}$& $\alpha_{14}$& $\alpha_{24}$& $\alpha_{34}$& $\xi_{1/S}$\\
         $5$& $0.015$& $0.0$& $0.35$& $0.0$& $0.00662652$& $0.0356523$& 0\\ \\
         \hline
    \end{tabular}}}
    \caption{The parameters in the chosen benchmark scenario.}
    \label{tab:bench}
\end{table}
In this benchmark scenario, the 125~GeV Higgs $h_2$ and 95~GeV Higgs $h_3$ are both having the non-zero CP-mixing phase $\alpha_{24}$ and $\alpha_{34}$, respectively. On the other hand, this benchmark scenario can still be a valid point in the CP-conserving case, where $\alpha_{24}$ and $\alpha_{34}$ are both zero. 
We will discuss the specific values chosen for $\alpha_{24}$ and $\alpha_{34}$ in the next section, since they are guided by EDM bounds on the model. The angle $c_{\beta-\alpha_{12}} = 0.015$ ensures that the alignment is imposed on this benchmark, i.e. $h_2$ is the SM-like Higgs boson of mass 125 GeV. The value for the mixing angle $\alpha_{23}=0.35$ is chosen in order to satisfy the observed excess in the $\gamma\gamma$ final state for the 95 GeV scalar.


\section{Constraints from EDM on the model parameters}
\label{edm}

In this section, we discuss the constraints on our model parameter space from electric dipole moment (EDM) experiments. CP-violation in nature will lead to EDM of the electron, neutron and atoms. This can be understood in the following manner.

The Hamiltonian of the EDM for a non-relativistic particle with spin ${\vec{S}}$, in the electric field $\vec{E}$ can be written as follows:

\begin{equation}
H_\text{EDM}=-d_f \frac{\vec{S}}{|\vec{S}|}\cdot\vec{E},
\label{}
\end{equation}
\noindent
where $d_f$ is the electric dipole moment (EDM) of the particle.

Under the time reversal transformation: 
$\mathcal{T}(\vec{S})=-\vec{S}$ and $\mathcal{T}(\vec{E})=+\vec{E}$,
the sign of this term $H_\text{EDM}$ is flipped, which implies CP symmetry is broken by virtue of the CPT theorem.

In EFT language,
\begin{equation}
\mathcal{L}_\text{EDM}=-\frac{d_f}{2}\bar{f} \sigma^{\mu\nu}(i\gamma^5)f F_{\mu\nu},
\end{equation}
\noindent
where $F_{\mu\nu}$ is the electromagnetic field tensor.

The most recent bounds on electron EDM are

\begin{eqnarray}
|d_e| &< 1.1 \times 10^{-29}\,\mathrm{e\,cm}
\quad (\mathrm{ThO})\text{~\cite{Panda:2019dnf}}
\label{acme}\\
|d_e| &< 4.1 \times 10^{-30}\,\mathrm{e\,cm}
\quad (\mathrm{HfF}^{+})\text{~\cite{Roussy:2022cmp}}
\label{edmnew}
\end{eqnarray}


In addition to the electric dipole moment of the constituent quarks, the neutron EDM receives the contribution from the chromo-EDM (CEDM) of the quarks $d_q^C$. CEDM of a quark $q$ can be expressed as
		\begin{align}
			\mathcal{L}_\textrm{CEDM}=-\frac{d_q^C}{2}\bar{q} \sigma^{\mu\nu}(i\gamma^5)q G_{\mu\nu}
            \label{sumruleedm}
		,\end{align}
where the $G_{\mu\nu}$ is the QCD field strength tensor.
The nEDM collaboration provides the most updated constraint on the neutron EDM:

~~~~~~~~~~~~~~~~~~~~~~~~~~~~~~~~~~$|d_n|<1.8\times10^{-26}$ e~cm \cite{Abel:2020pzs}.

\noindent
The neutron EDM ($d_n$) can be estimated by using the QCD sum rule\cite{Abe:2013qla}:
		\begin{align}
			d_n=0.79d_d-0.20d_u+e(0.59d_d^C+0.30d_u^C)/g_3
		\label{eq:nEDM}
		,\end{align}

\noindent
where $g_3$ is the QCD SU(3) gauge coupling constant\footnote{The extra factor of $g_3$ in expression~\ref{eq:nEDM} as compared to \cite{Abe:2013qla} is due to an extra factor of $g_3$ in  different definition used for Eq.~\ref{sumruleedm} in \cite{Abe:2013qla}.}.

\begin{figure}[!hptb]
	\centering
        \includegraphics[width=15.0cm,height=4cm]{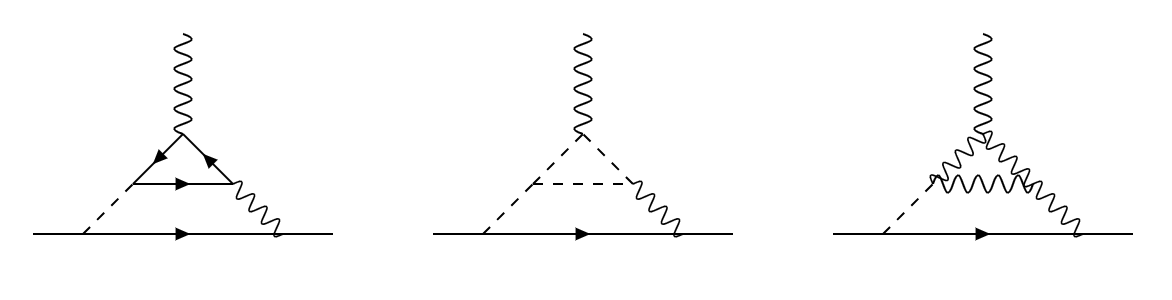}
	\caption{The two-loop Bar-Zee diagrams with fermion (left), scalar (middle) and gauge-boson (right) loop contributions. The solid lines with pointing arrows denote fermion lines (quarks and leptons), the dashed lines denote scalars ($h_{1..4}$ and $H^{\pm}$), and the wavy lines represent gauge bosons ($\gamma, Z, W$). }
	\label{barzee}
\end{figure}

\begin{figure}[!hptb]
    \centering
    \includegraphics[width=7.0cm,height=6.0cm]{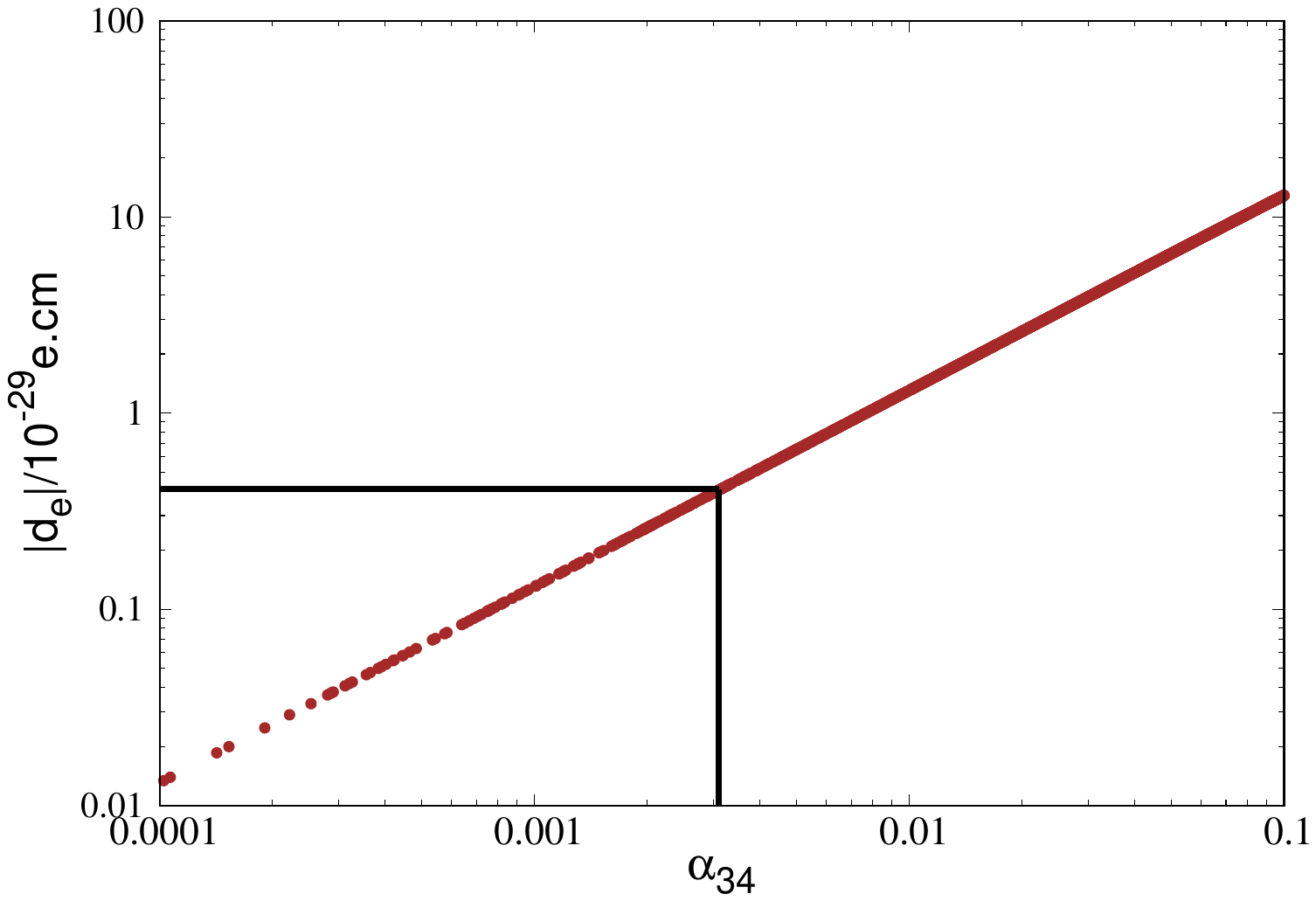}
      \includegraphics[width=7.0cm,height=6.0cm]{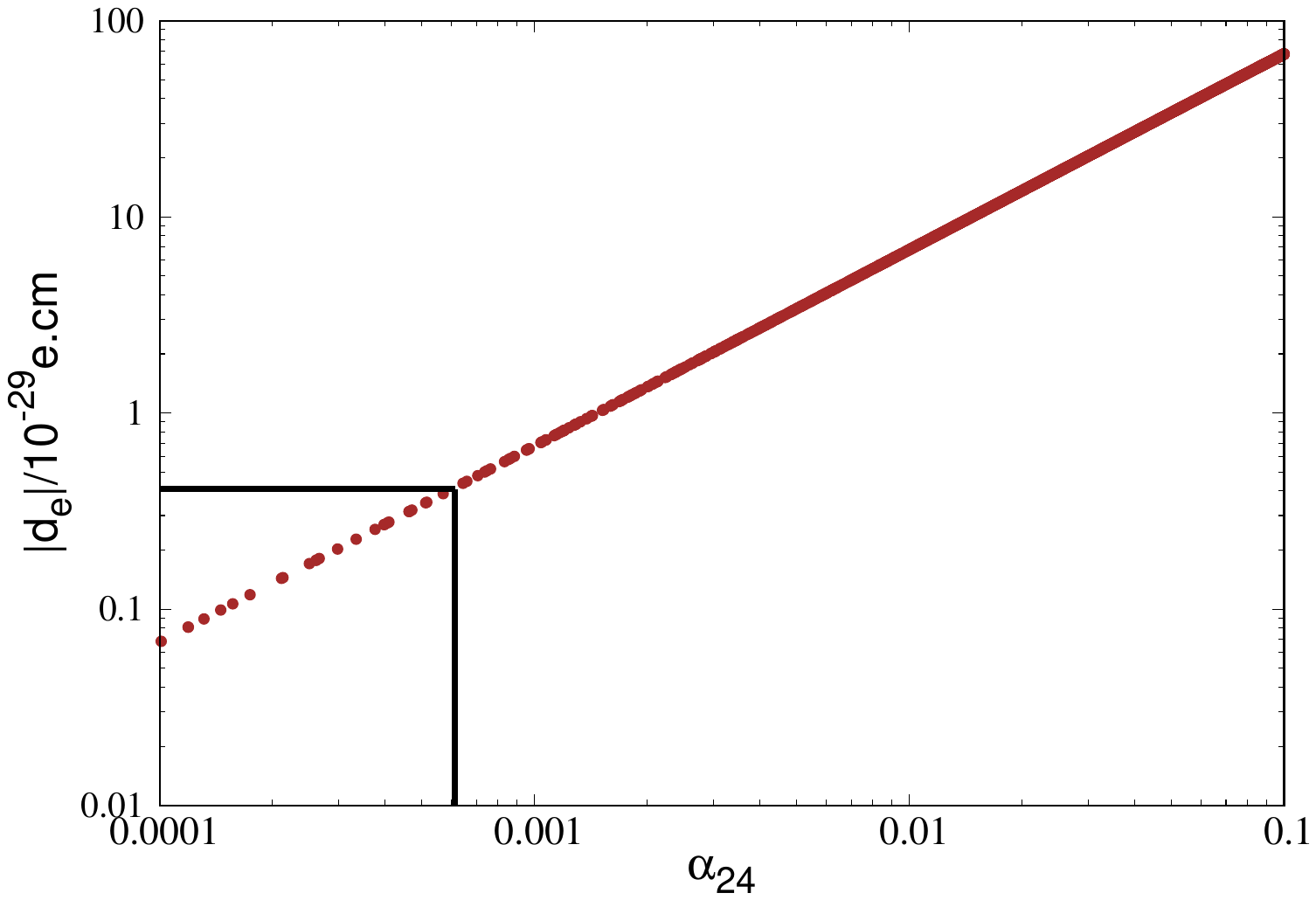}
    \caption{Left panel shows EDM as a function of the mixing angle $\alpha_{34}$ which governs the CP-odd component in the 95 GeV scalar, $\alpha_{24}=0$ is assumed for this plot. Right panel shows EDM as a function of the mixing angle $\alpha_{24}$ which governs CP-odd component in the 125 GeV scalar, $\alpha_{34}=0$ is assumed for this plot.}
    \label{fig:edm_a24_34}
\end{figure}

\begin{figure}[!hptb]
    \centering
    \begin{subfigure}{0.48\textwidth}
        \includegraphics[width=\textwidth]{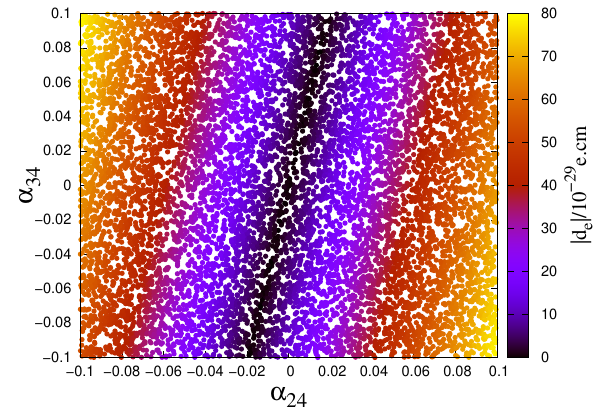}
    \end{subfigure}
    \hfill
    \begin{subfigure}{0.48\textwidth}
        \raisebox{0.8cm}{%
            \includegraphics[width=\textwidth]{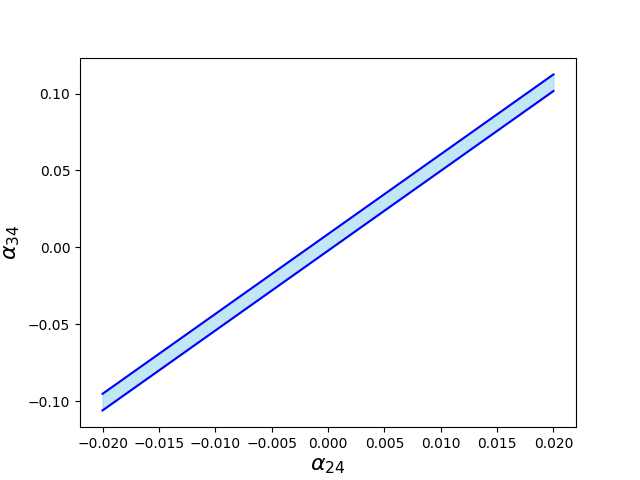}
        }
    \end{subfigure}
        \caption{Left panel: EDM in the color axis as a function of the mixing angles $\alpha_{24}$ and $\alpha_{34}$ when they are both varied simultaneously. Right panel shows the region in blue is on the $\alpha_{24}-\alpha_{34}$ plane, which is allowed by the EDM upper bound $|d_e| < 0.41 \times 10^{-29}$e.cm~\cite{Roussy:2022cmp}. }
    \label{fig:a24_34}
\end{figure}

\noindent
The major contribution to the EDM and CEDM comes from the two-loop Bar-Zee diagrams shown in Fig.~\ref{barzee}. The contribution to the Bar-Zee diagrams can be decomposed into fermion, gauge boson and scalar running in the loop: 

		\begin{equation*}
			d_f=d_f(\textrm{fermion})+d_f(\textrm{Higgs})+d_f(\textrm{gauge}).
		\end{equation*}

\noindent
Each contribution $d_f(X)$ further consists of contributions from $\gamma,Z$ and $W$ in an internal leg, 
		\begin{equation*}
			d_f(X)=d_f^\gamma(X)+d_f^Z(X)+d_f^W(X).
            \label{edmeq}
		\end{equation*}

\noindent
The detailed expressions for the aforementioned loop contributions are given in Appendix.~\ref{BarZeeformulae}. 
The gauge boson loops contribute negligibly, since the coupling strengths of the gauge bosons to the non-standard scalars are too small in the alignment limit ($R_{1j}\approx 0$ in Eqs.~\ref{gauge1} and \ref{gauge2} in the alignment limit). On the other hand, the fermion and scalar boson loop can contribute at equivalent strength. Therefore, it is crucial to take into account both contributions. One-loop contributions are suppressed by at least 4-5 orders of magnitude compared to the two-loop diagrams. 
First, we will study the impact of electron EDM on our parameter space.

In our simplified model, we have assumed that the CP-violation will lead to the CP-violating Yukawa couplings of
the two scalar mass eigenstates with 95 GeV and 125 GeV. The mixing angles $\alpha_{34}$  and $\alpha_{24}$ lead to the CP-violating components of the 95 GeV scalar and the 125 GeV Higgs boson, respectively. Consequently, CP-violating Yukawa couplings will be produced pertaining to the two scalars and these couplings will generate EDM of the electron as well as neutron. We have seen in earlier works~\cite{Kanemura:2020ibp,Lahiri:2026eud}, the neutron EDM is at least two-orders of magnitude smaller than the electron EDM in all the parameter points. Therefore, we focus on the effect of the electron EDM on the angles $\alpha_{34}$, $\alpha_{24}$. First, we show the impact of EDM on the individual angles in Fig.~\ref{fig:edm_a24_34}.
We can see that EDM bounds on individual angles are as follows:

\begin{equation}
a_{34} \lesssim 0.003,~~~~a_{24} \lesssim 0.0006.
\end{equation}

\noindent
Next, we vary both angles simultaneously and compute electron EDM for the scanned parameter space in Fig.~\ref{fig:a24_34}. One should note that there can be active cancellation between the contribution from various loops (especially fermion and scalar loops) and such cancellations lead to a narrow region that leads to negligible EDM, and is allowed by the EDM constraints (see Fig.~\ref{fig:a24_34} (left) and (right)). The fine-tuned nature of the allowed parameter space can be attributed to the presence of two independent sources of CP-violation in the simplified version of our model. However, in an earlier work~\cite{Lahiri:2026eud} it was shown that such fine-tuning can be alleviated as additional independent CP-phases are introduced. Our focus in this work is neither an exhaustive exploration of the model nor establishing validity of our model against the most general constraints. Rather, we aim to demonstrate the sensitivity of collider observables to CP-violating phases in a realistic model scenario, after imposing the existing EDM constraints. We emphasize that the interplay between collider sensitivity and EDM bounds can depend strongly on the underlying details of the model, for the reasons discussed above. We illustrate this dependence explicitly in the following section.

\section{Probing CP-violation via decay into the $\tau\tau$ channel at $e^+e^-$ colliders}
\label{collider}
Having discussed the stringent bounds from the electric dipole moment experiments on the CP-mixing angles in our model, we go on to probe the CP-violating phases in the 95 GeV scalar as well as 125 GeV Higgs boson at future lepton colliders. The CP-violation in the 125 GeV Higgs has been probed at the LHC~\cite{Berge:2008dr,Berge:2014sra,CMS:2021sdq,ATLAS:2022akr} with CP-sensitive observables and future projections on the CP-phases are also available in the context of future lepton colliders\cite{Berge:2013jra,Desch:2003rw,Rouge:2005iy}. We call the two CP-mixed scalars in our model $h_{95}$ and $h_{125}$ from now on.
We focus on the production of both $h_{95}$ and $h_{125}$ in the Higgsstrahlung process at the $e^+e^-$ collider,

\begin{equation}
    e^+ e^- \rightarrow h_{95/125} Z,~~h_{95/125}, \rightarrow \tau\tau, Z\rightarrow l^+l^-.
\end{equation}

\noindent
The $\tau$ lepton can further decay into hadronic or leptonic final states. Depending on the particular decay mode it will be 1-prong, 2-prong or 3-prong decay. Next we list all important decay modes of the $\tau$-lepton and the corresponding branching fractions in Table~\ref{tab:my_label}. 

\begin{table}[h]
    \centering
    \begin{tabular}{c|c|c}
        Notation & Decay Mode &  Branching fraction \\
        \hline
         $\ell$ & $l^{\pm}\nu\bar\nu$  & 35\% \\
         \hline
         1p0n  & $\pi^{\pm}\nu$ & 11\% \\
        \hline
        1p1n & $\rho^{\pm}\nu\rightarrow\pi^{\pm}\pi^0\nu$ & 26\% \\
        \hline
        1p2n & $a_1^{\pm}\nu\rightarrow\rho^{\pm}\pi^0\nu\rightarrow\pi^{\pm}\pi^0\pi^0\nu$ & 10\% \\
        \hline
        3p0n & $a_1^{\pm}\nu\rightarrow\rho^{0}\pi^{\pm}\nu\rightarrow\pi^{\pm}\pi^{\mp}\pi^{\pm}\nu$ & 9\% \\
        \hline
   \end{tabular}
   \caption{The decay modes of $\tau$-lepton are listed here. The notations $\ell$ for leptonic final state, `p' for charged pion and `n' for neutral pion final states are used.}
    \label{tab:my_label}
\end{table}
$\rho^{\pm,0}$ is the $\rho$ vector meson and $a_1$ is the axial vector meson. $\rho$ and $a_1$ meson further decays into $\pi^{\pm,0}$ final states. We use the notations $\ell$ for leptonic final state, `p' for charged pion and `n' for neutral pion final states.

 \begin{table}[h]
    \centering
    \begin{tabular}{c|c|c}
        Decay channel & Decay Mode combinations &  Branching fraction \\
        \hline
           &   $\ell$-1p0n & 8.1\%  \\
         
       $\tau_{lep}\tau_{had}$    &   $\ell$-1p1n & 18.3\% \\
                  &   $\ell$-1p2n & 7.6\%  \\
           &   $\ell$-3p0n & 6.9\% \\
     \hline
          &  1p0n-1p0n  &  1.3\% \\
     &  1p0n-1p1n & 6.0\% \\
  $\tau_{had}\tau_{had}$     &  1p1n-1p1n & 6.7\% \\
     &  1p0n-1p2n  & 2.5\% \\
     &  1p1n-1p2n &  5.6\% \\
     &  1p1n-3p0n &  5.1\% \\
     \hline
 \end{tabular}
  \caption{The relevant di-$\tau$ final states.}
    \label{ditau}
\end{table}

Having discussed the decay modes of $\tau$-lepton, we list the di-$\tau$ final states relevant for our analysis in Table \ref{ditau}.
One can see from Table \ref{ditau} that the semi-leptonic final states of $\tau\tau$ correspond to $\approx 41\%$ of the total events whereas the fully hadronic final states lead to $27\% $. 

\begin{figure}[h]
    \centering
    \includegraphics[width=15.3cm,height=5.5cm]{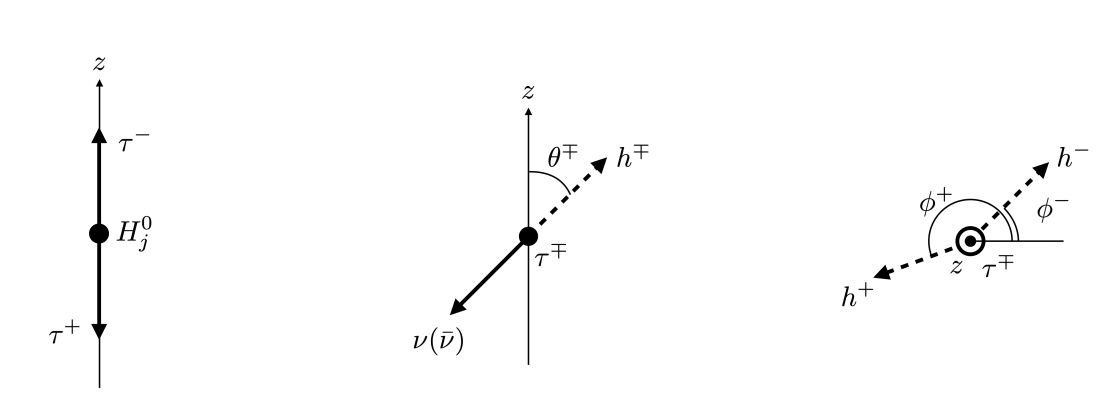}
    \caption{Construction of the angle $\Phi^{*}_{CP}$.}
    \label{fig:phicp}
\end{figure}

\noindent
Next we construct a CP-sensitive observable, involving the decay products of the $\tau$-lepton. We need to first reconstruct the rest frame of the Higgs boson. In a $e^+e^-$ collider where the initial information is fully available it will be possible to reconstruct the Higgs($h_{95}/h_{125}$) rest frame from the recoil spectrum of the associated $Z$-boson, which decays leptonically in our consideration. After reconstructing the Higgs rest frame, we boost the $\tau^+$ and $\tau^-$ 4-momenta in that frame. Next we align the $z$-axis of our co-ordinate system along the $\tau^-$ momentum. Next we boost the 4-momenta of the decay product of $\tau^+$, denoted as $h^+$($h^+$ can be $\pi^+,\rho^+,a_1^+,\ell^+$) to $\tau^+$ rest frame. Likewise, we boost the 4-momenta of the decay product of $\tau^-$, denoted as $h^-$($h^-$ can be $\pi^-,\rho^-,a_1^-,\ell^-$) to $\tau^-$ rest frame. In those frames the $\theta^{\pm},\phi^{\pm}$ are the polar and azimuthal angles of $h^{\pm}$, with the pre-defined z-axis. The Fig.~\ref{fig:phicp}, shows this construction sequentially. Now we define a CP-sensitive observable $\Phi^*_{CP}$ which is defined as follows.  

\begin{equation}
\Phi^{*}_{CP} = \Delta(\phi^+,\phi^-).
\end{equation}

\noindent
Using the $H\rightarrow \tau\tau$ spin-density matrix and the SM density matrix of polarized $\tau\rightarrow \pi\nu$, we can write the differential distribution of $\Phi^{*}_{CP}$ as follows:

\begin{equation}
\frac{1}{\Gamma}\frac{d\Gamma}{d\Phi^{*}_{CP}} = \frac{1}{2\pi}[1-\frac{\pi^2}{16}(c_1\cos\Phi^{*}_{CP} + c_2 \sin\Phi^{*}_{CP})],
\label{differential_dist}
\end{equation}

\noindent
where

\begin{eqnarray}
c_1=\frac{a_{\tau}^2 \beta_\tau - b_{\tau}^2}{a_{\tau}^2\beta_\tau + b_{\tau}^2} = \cos\phi', \label{phiprime1} \\
c_2=-\frac{2a_{\tau}b_{\tau}\beta_\tau}{a_{\tau}^2\beta_\tau + b_{\tau}^2} = \sin\phi'. 
\label{phiprime}
\end{eqnarray}

\noindent
Clearly, $c_1 = 1, c_2=0$ in the CP-conserving case and $c_1=-1, c_2=0$ in case of pure CP-odd scalar coupling to $\tau\tau$-pair.

Using the $\phi'$ parametrization (see Eqs.~\ref{phiprime1} and \ref{phiprime}), we can write Eq.~\ref{differential_dist} as follows,

\begin{equation}
\frac{1}{\Gamma}\frac{d\Gamma}{d\Phi^{*}_{CP}} = \frac{1}{2\pi}[1-\frac{\pi^2}{16}\cos(\Phi^{*}_{CP} - \phi')].
\label{eq:new}
\end{equation}

\noindent
On the other hand, we have seen (in Eq.~\ref{parameter}) earlier that

\begin{equation}
\cos\phi_{\tau}=\frac{a_{\tau}}{\sqrt{a_{\tau}^2 + b_{\tau}^2}},
~~~~~\sin\phi_{\tau}=\frac{b_{\tau}}{\sqrt{a_{\tau}^2 + b_{\tau}^2}}.
\label{phitau}
\end{equation}
\noindent

\noindent
By comparing Eq. \ref{eq:new} and \ref{phitau}, in the limit $\beta_\tau \rightarrow 1$, we obtain $\phi'=2\phi_\tau$.

Therefore, one can write
\begin{equation}
\frac{1}{\Gamma}\frac{d\Gamma}{d\Phi^{*}_{CP}} = \frac{1}{2\pi}[1-\frac{\pi^2}{16}\cos(\Phi^{*}_{CP} - 2\phi_{\tau})].
\label{eq:compare}
\end{equation}

\begin{figure}[h]
    \centering
    \includegraphics[width=7.0cm,height=6.0cm]{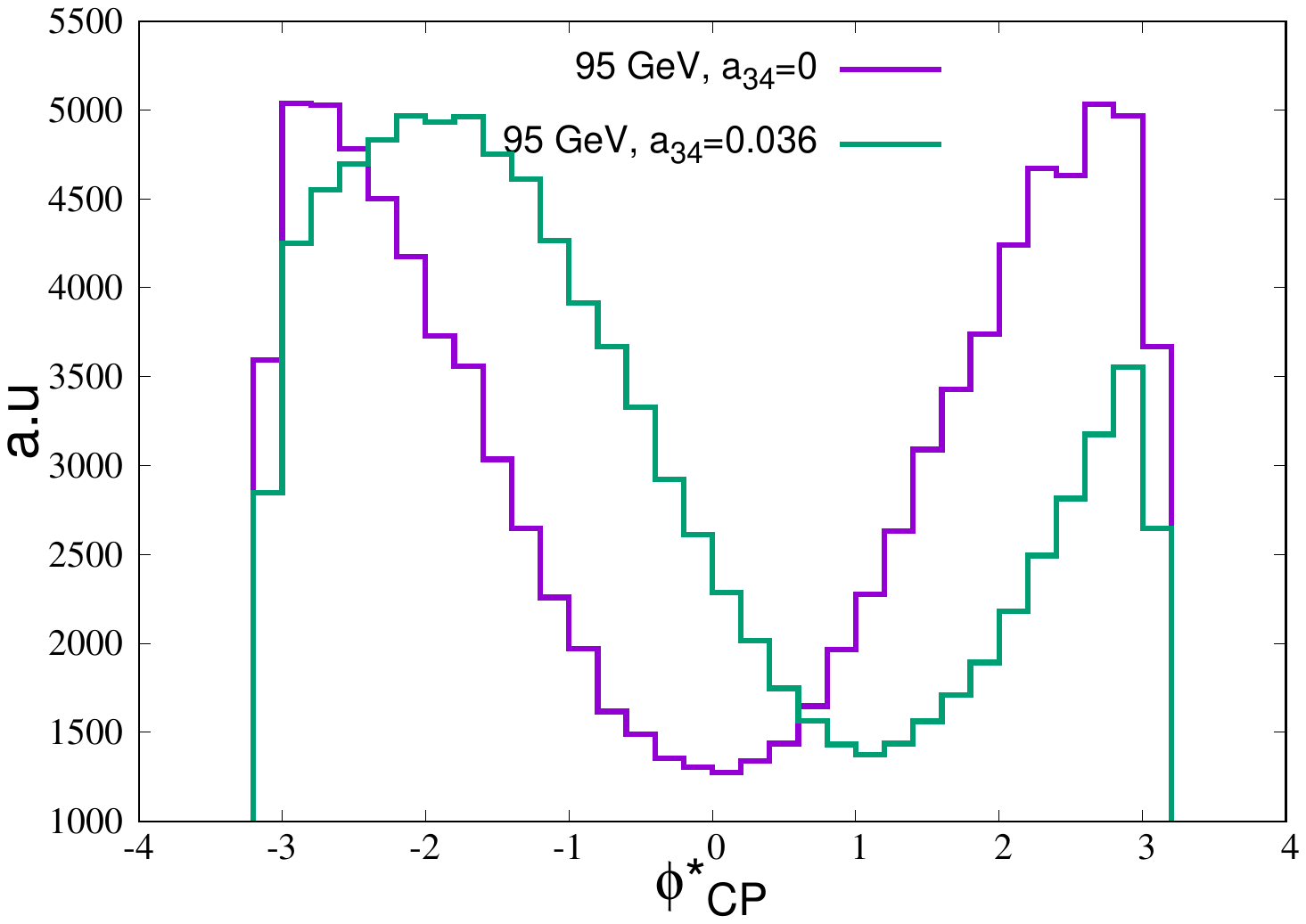}
    \includegraphics[width=7.0cm,height=6.0cm]{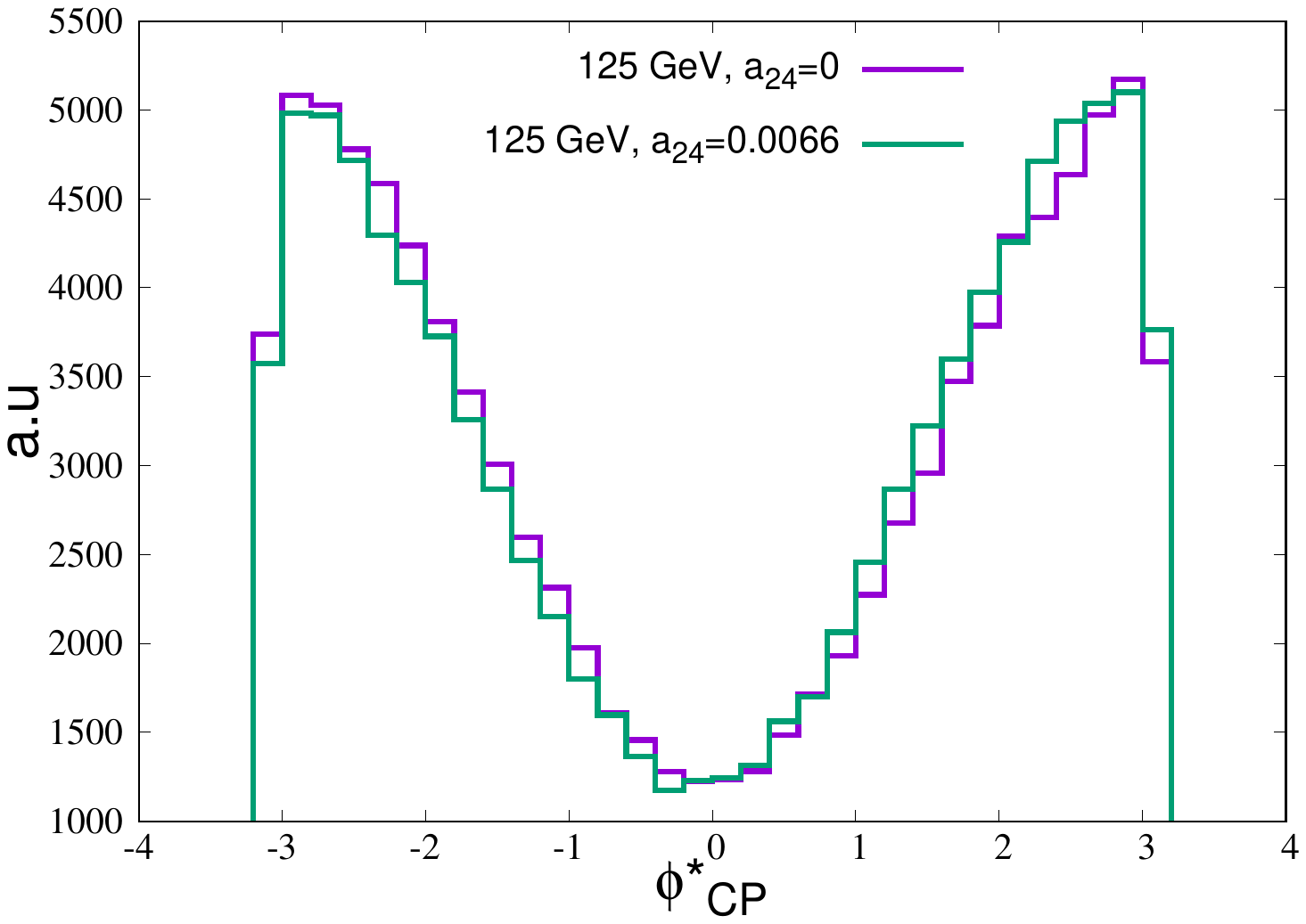}
    \caption{Normalized distribution of the angle $\Phi_{CP}^*$ for the 95 GeV (left)  and 125 GeV (right) scalars in the hadronic final state at $\sqrt{s}=250$ GeV for CP-even and CP-mixed cases.}
    \label{fig:phicphad}
\end{figure}

\begin{figure}[!hptb]
    \centering
    \includegraphics[width=7.0cm,height=6.0cm]{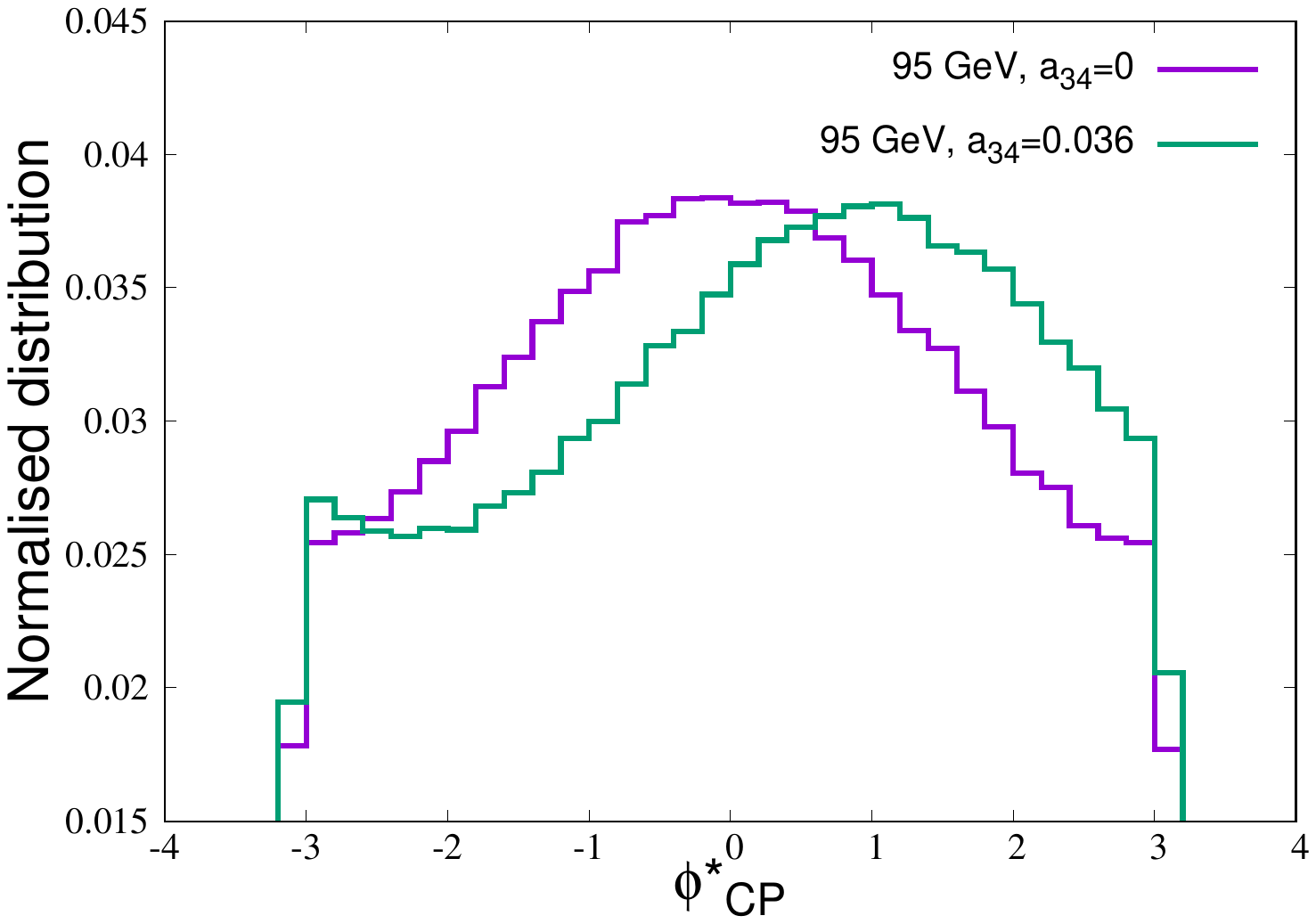}
    \includegraphics[width=7.0cm,height=6.0cm]{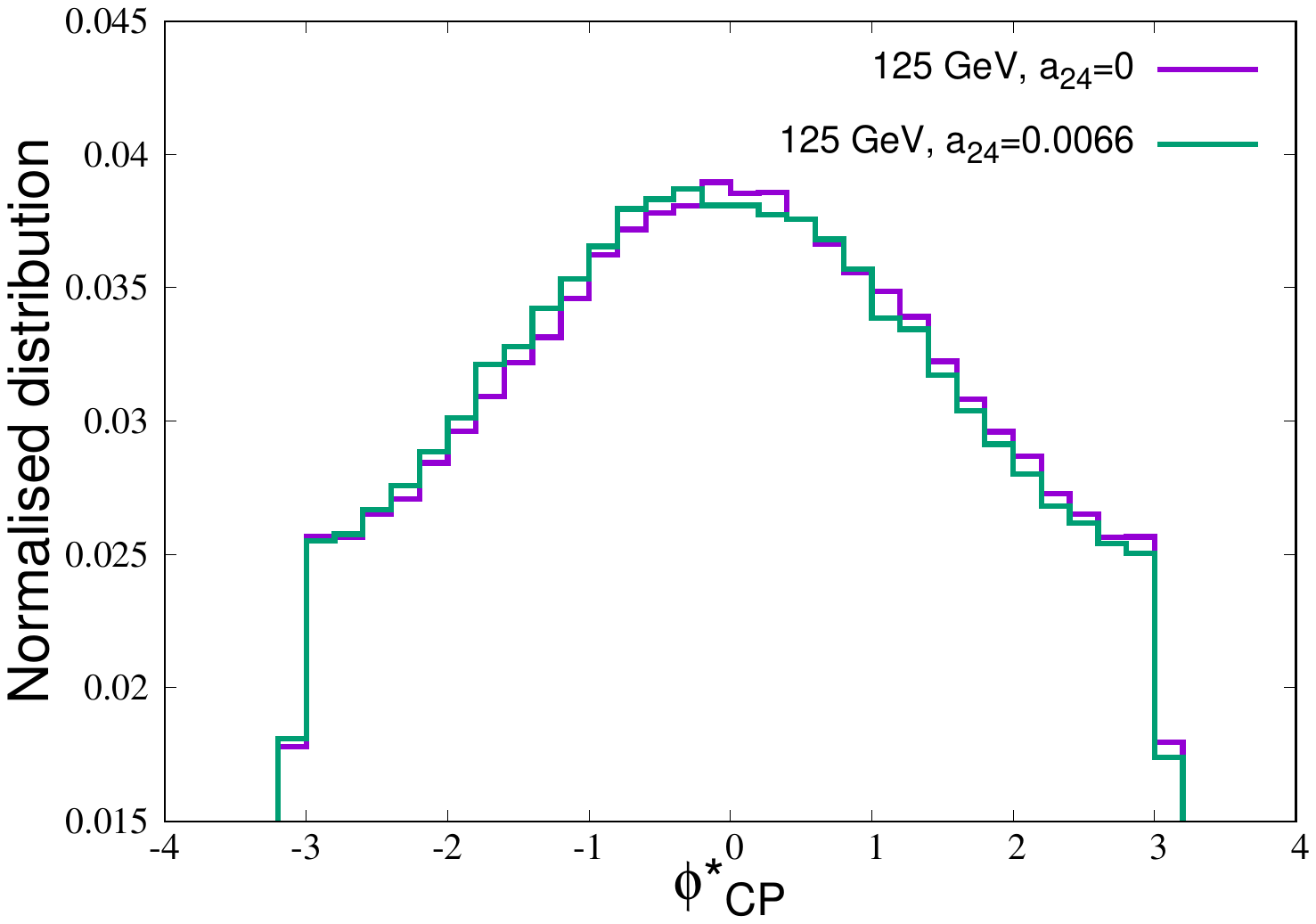}
    \caption{Normalized distribution of the angle $\Phi_{CP}^*$ for 95 GeV (left) and 125 GeV (right)  scalar in the semileptonic final state at $\sqrt{s}=250$ GeV for CP-even and CP-mixed cases.}
    \label{fig:phicpsemilep}
\end{figure}

\noindent
Next we show the distribution of the angle $\Phi_{CP}$ in both hadronic (in Fig.~\ref{fig:phicphad}) and semileptonic (in Fig.~\ref{fig:phicpsemilep}) decay modes of the $\tau$-pair. We can clearly see the angular distribution is symmetric around $\Phi^*_{CP}=0$ in case of CP-even scenario and there is non-zero forward-backward asymmetry in case of CP-mixed case. Notably, larger allowed values of the mixing angle $\alpha_{34}$ leads to visible shift ($2\phi_\tau$) in the minima of the two distributions in case 95 GeV scalar. But for the 125 GeV Higgs the more stringent limit on the mixing angle $\alpha_{24}$ leads to much smaller deviation in the CP-violating case. One can also see that the forward backward asymmetry is less in case semi-leptonic final state compared to the hadronic case.

\begin{figure}[!hptb]
    \centering
    \includegraphics[width=7.0cm,height=6.0cm]{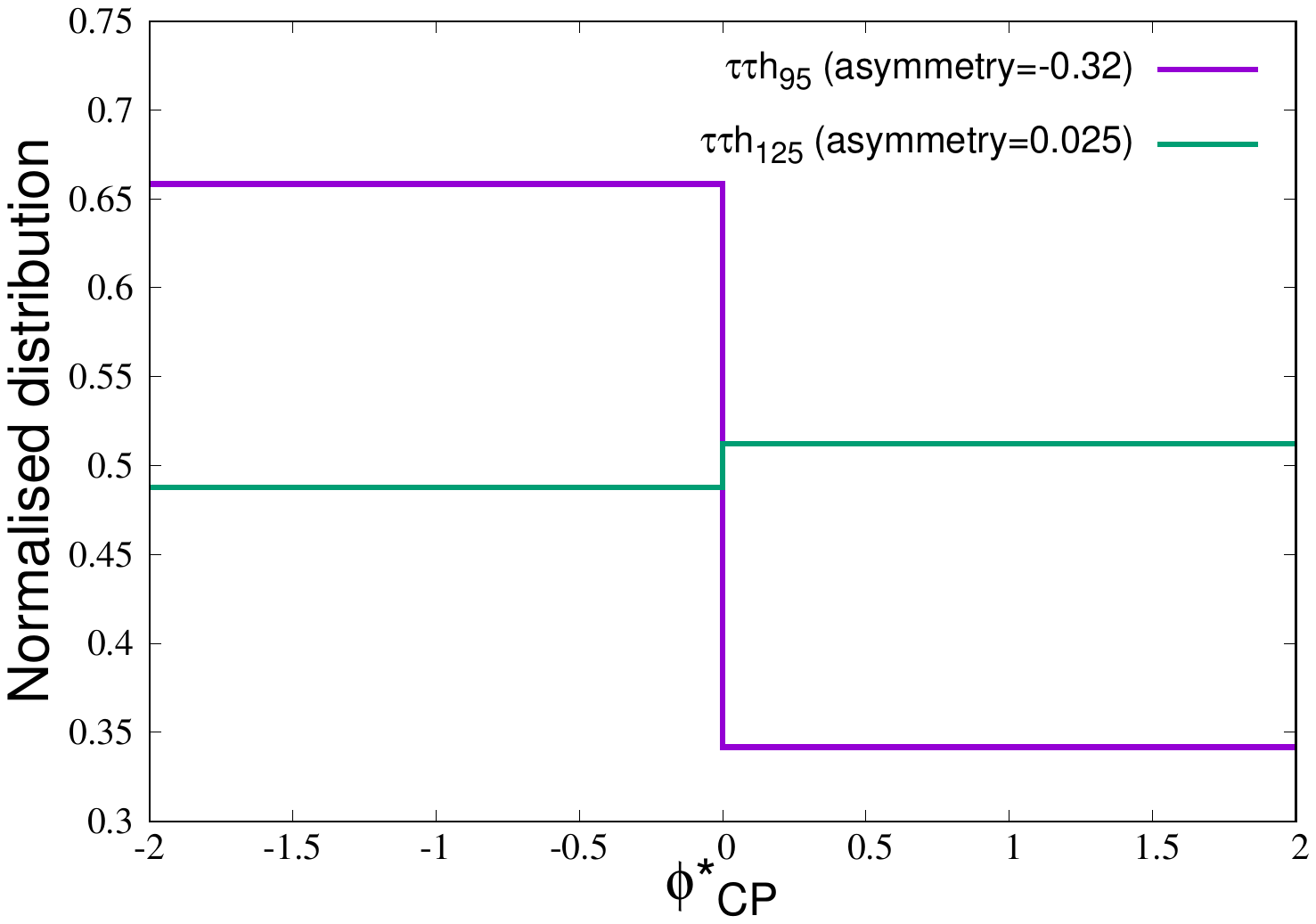}
    \includegraphics[width=7.0cm,height=6.0cm]{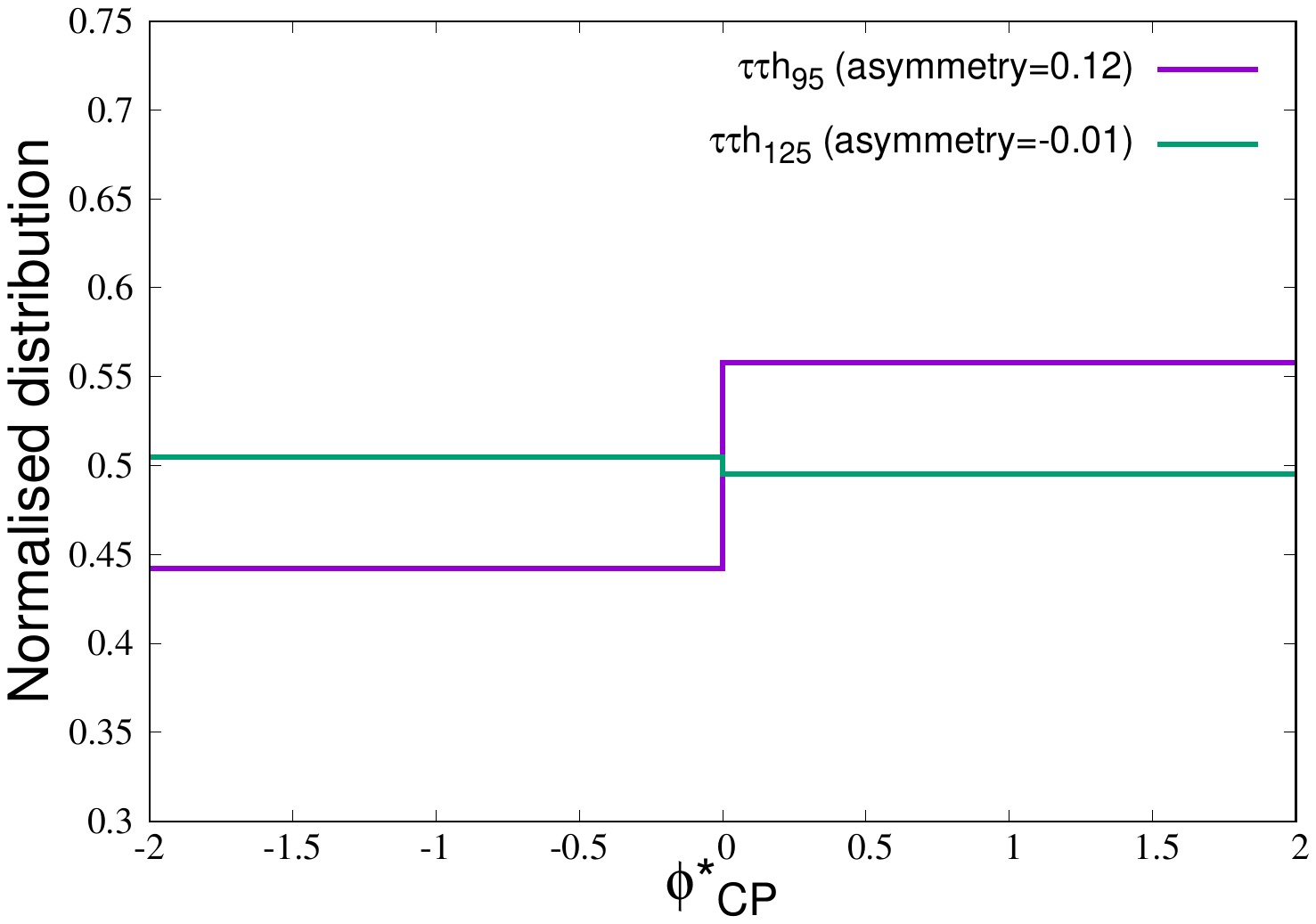}
    \caption{Distribution (two-bin) of $\Phi_{CP}^*$ for 95 GeV and 125 GeV scalars at $e^+e^-$ collider in the hadronic final state (left) and semi-leptonic final state (right) with $\sqrt{s}=250$ GeV and $\int {\cal L} dt$ = 5.6 ab$^{-1}$. We chose $\alpha_{34}$ = 0.036 and $\alpha_{24}$ = 0.0066.}
    \label{fig:asymmetry_2bin}
\end{figure}

\noindent
In Fig.~\ref{fig:asymmetry_2bin}, we show the angular distributions in the forward and backward region using two bins [$-\pi,0$], [$0,\pi$] for the chosen BP $\alpha_{34}$ = 0.036 and $\alpha_{24}$ = 0.0066. The asymmetry is much less pronounced in case of the 125 GeV scalar compared to the 95 GeV scalar because of the smaller size of EDM-allowed $\alpha_{24}$ compared to $\alpha_{34}$, as discussed above.

Our next goal is to use the asymmetries we calculated so far to probe the CP-violating phases or mixing angles in our model pertaining to the 95 GeV scalar. We will compare our results with the sensitivity
achieved in case of CP-violating phase of 125 GeV scalar in the existing literature~\cite{Gritsan:2022php}. For this purpose, we first generate the asymmetries as a function of the mixing angle pertaining to CP-violation in the 95 GeV scalar, namely $\alpha_{34}$.

 \begin{figure}[h]
    \centering
        \includegraphics[width=7.0cm,height=6.0cm]{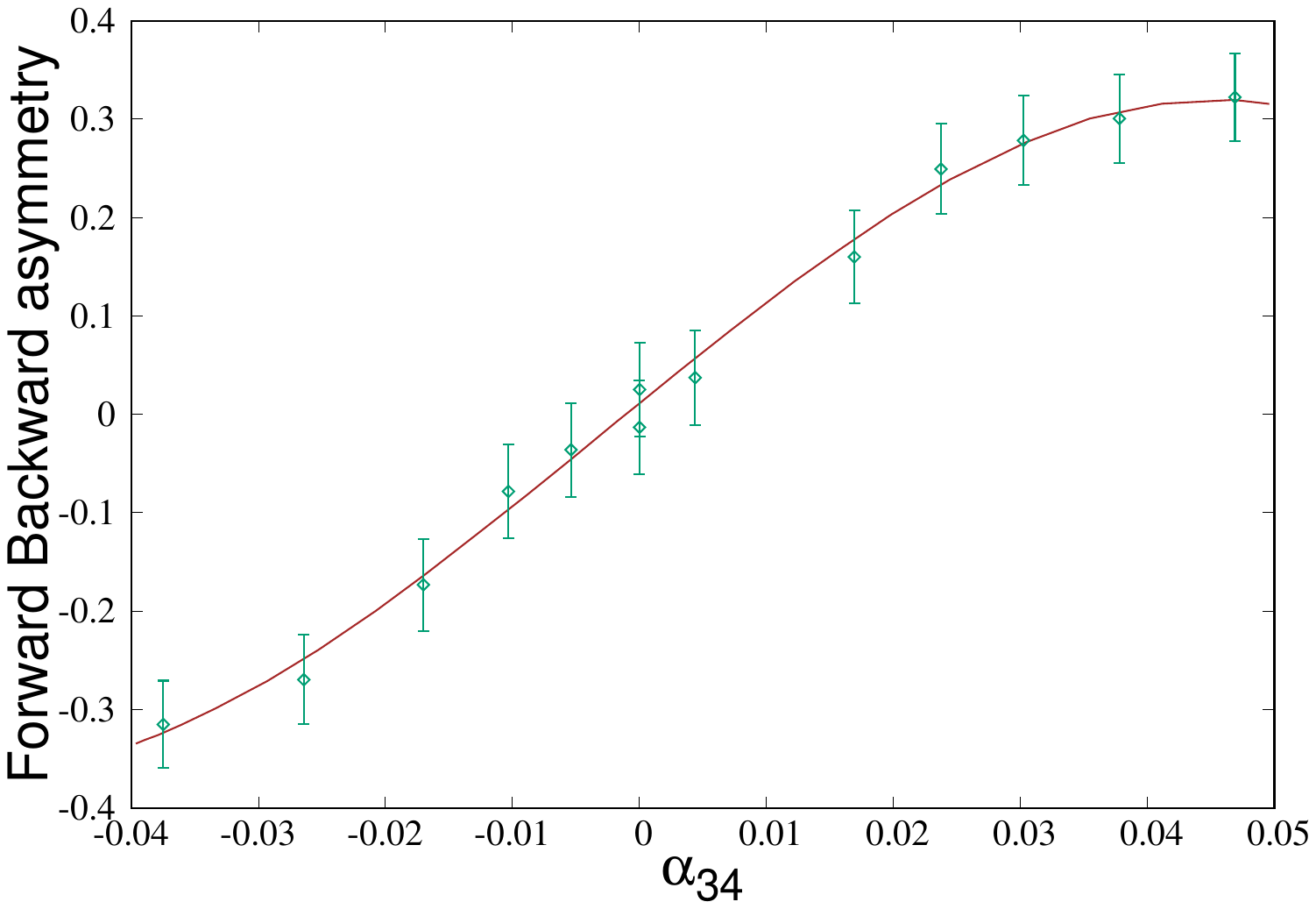}
    \includegraphics[width=7.0cm,height=6.0cm]{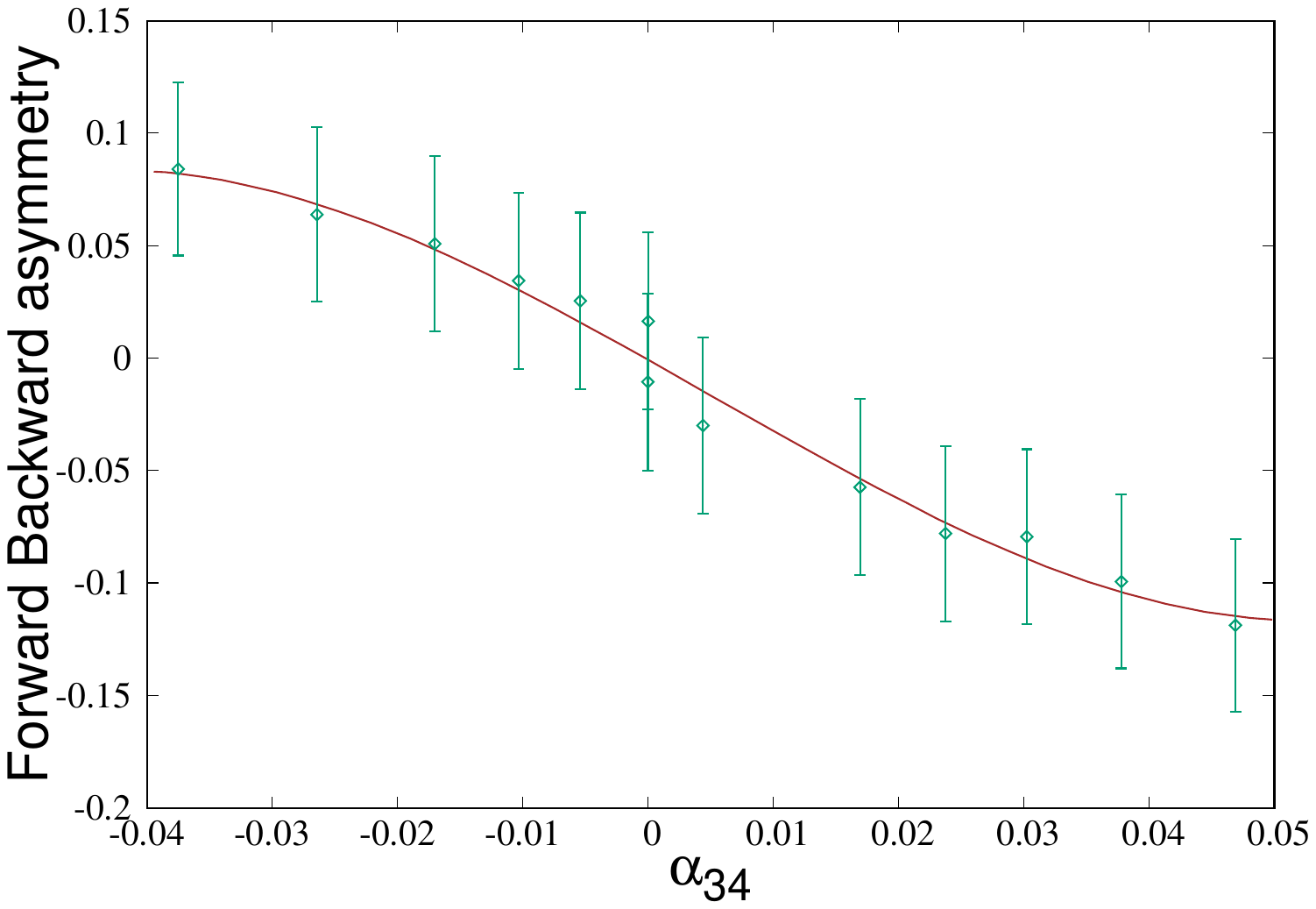}
    \caption{The forward-backward asymmetry as a function of $\alpha_{34}$ in the hadronic (left) semileptonic (right) final state.}
    \label{fig:a34_afb}
\end{figure}

\noindent
Next, we define forward-backward asymmetry $A_{FB}$ as follows:

\begin{equation}
A_{FB}=\frac{1}{\Gamma}\left(\int^\pi_0\frac{d\Gamma}{d\Phi^{*}_{CP}}d\Phi^{*}_{CP} - \int^0_{-\pi}\frac{d\Gamma}{d\Phi^{*}_{CP}}d\Phi^{*}_{CP}\right) 
\end{equation}
\noindent
We show in Fig.~\ref{fig:a34_afb}, the forward backward asymmetry as a function of the mixing angle $\alpha_{34}$ for hadronic (left) and semileptonic (right) final state of $\tau\tau$ pair coming from the decay of the 95 GeV scalar. The error in asymmetry measurement can be expressed as follows:
\begin{equation}
    \sigma_A = \sqrt{\frac{1-A^2}{N_{events}}}
    \label{error}
\end{equation}

\noindent
Eq.~\ref{error} indicates the asymmetry follows a binomial distribution.
We have shown the error bars in green for each data point, calculated with the aforesaid formula. The maroon curve is the result of fitting of the data points involving variation of $\alpha_{34}$. 

Having discussed the asymmetry distribution as a function of $\alpha_{34}$, we calculate the $\chi^2$ for a combination of independent measurements denoted by index $i$ as follows. 

\begin{equation}
\chi^2=\sum_i\frac{(A_i^{measured}-A_i(\alpha_{34}))^2}{\sigma_{A_i}^2}.
\end{equation}

\noindent
Next, using Taylor expansion around $\alpha_{34}=0$ and demanding $\chi^2 = 1$, we get 

\begin{equation}
\sigma_{34}^{-2} = \sum_i \frac{1}{\sigma_{A_i}^2}\left(\frac{dA_i}{d\alpha_{34}}\right)^2.
\label{combined}
\end{equation}

\noindent
Specifically, in the hadronic final state for $\chi^2=1$ we get,
\begin{eqnarray}
    \sigma_{\alpha_{34}}^{had} = \frac{\sigma_A^{had}}{|dA/d\alpha_{34}|^{had}}, \\
     \sigma_{\alpha_{34}}^{had} = \pm 0.0046.
\end{eqnarray}

\begin{figure}[h]
    \centering
    \includegraphics[width=9.0cm,height=7.5cm]{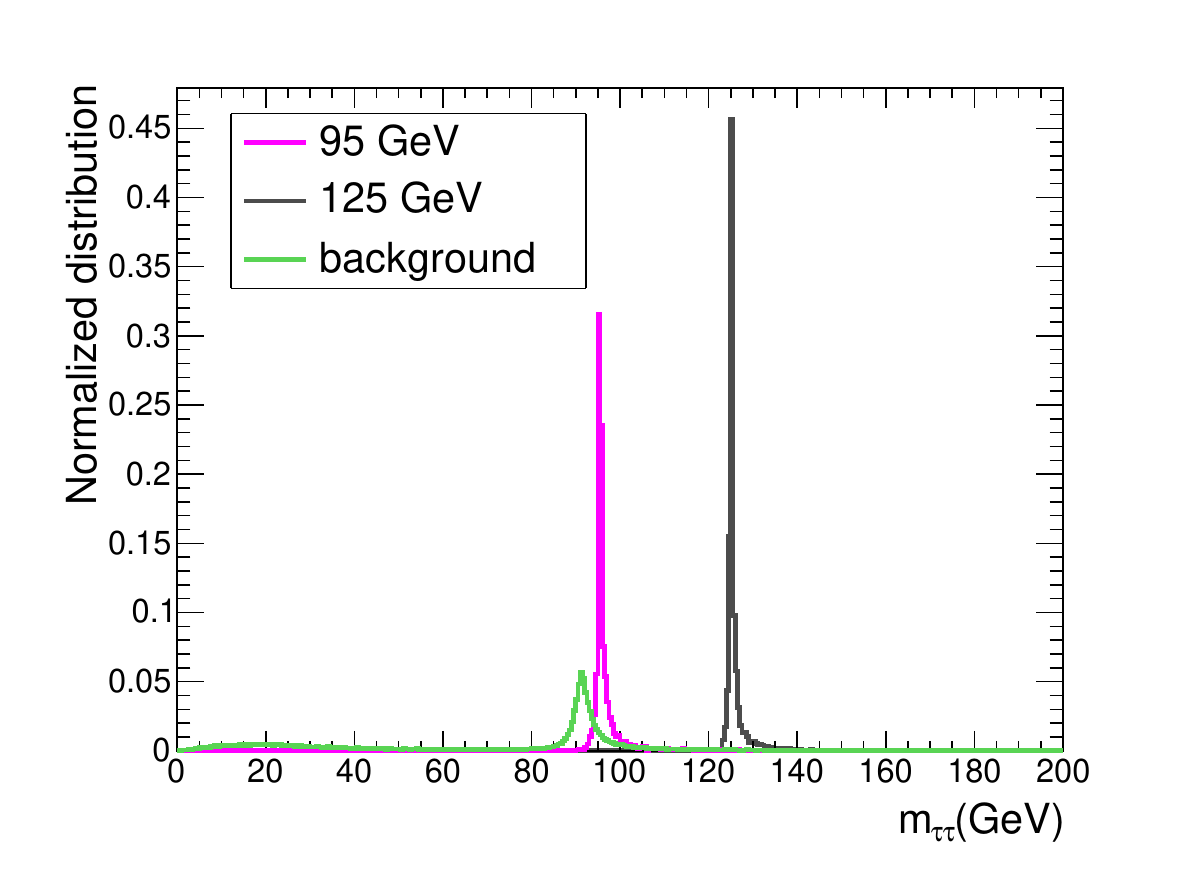}
    \caption{Normalised invariant mass ($m_{\tau\tau}$) distribution  of the $\tau$-pair, constructed out of recoil spectra of $Z(\ell^+\ell^-)$-boson, in cases of 95 GeV scalar, 125 GeV Higgs and the full SM background at $\sqrt{s} = 250$ GeV.}
    \label{fig:sig_bkg}
\end{figure}

\noindent
The uncertainty in $\alpha_{34}$ above is calculated assuming signal-only distributions. Importantly, the background events faking as signal of 95 GeV scalar is dominated by SM $Z \rightarrow \tau\tau$ process. Even after applying stringent cut $m_{\tau\tau} >$ 94.0 GeV, the total number of events contain 20\% background events.  We show the invariant mass distribution of the $\tau$-pair for signal and backgrounds in Fig.~\ref{fig:sig_bkg}. Once we take these irreducible background events into account, our asymmetry values and consequently $dA/da_{34}$ scale down as $\frac{S}{S+B}$, where S and B are the signal and background events after cuts and the error in asymmetry $\sigma_A$ scales down as $\sqrt{\frac{S}{S+B}}$. Therefore, as a combined effect of the two, $\sigma_{a_{34}}$ scales up as $\sqrt{\frac{S+B}{S}}$. 
Using this scale factor, we see that the sensitivity worsens in our case by a factor of 2.3. Including background effects we obtain

\begin{equation}
     \sigma_{\alpha_{34}}^{had} = \pm 0.01.
\end{equation}
 
 \noindent
 In the semi-leptonic final state, for $\chi^2 = 1$ we get:
\begin{eqnarray}
    \sigma_{\alpha_{34}}^{semilep} = \frac{\sigma_A^{semilep}}{|dA/d\alpha_{34}|^{semilep}}, \\
     \sigma_{\alpha_{34}}^{semilep} = \pm 0.012.
\end{eqnarray}

\noindent
 Here, including background effects discussed earlier, we obtain the following uncertainty:

\begin{equation}
     \sigma_{\alpha_{34}}^{semilep} = \pm 0.027.
\end{equation}

\noindent
Combining hadronic and semileptonic uncertainties using Eq.~\ref{combined}, with the signal-only scenario, we get :
\begin{equation}
     \sigma_{a_{34}}^{combined} = \pm 0.0044.
\end{equation}
After the inclusion of background effects, we obtain the following result: 

\begin{equation}
     \sigma_{a_{34}}^{combined} = \pm 0.01.
\end{equation}.

 \begin{figure}[h]
    \centering
    \includegraphics[width=9.0cm,height=7.5cm]{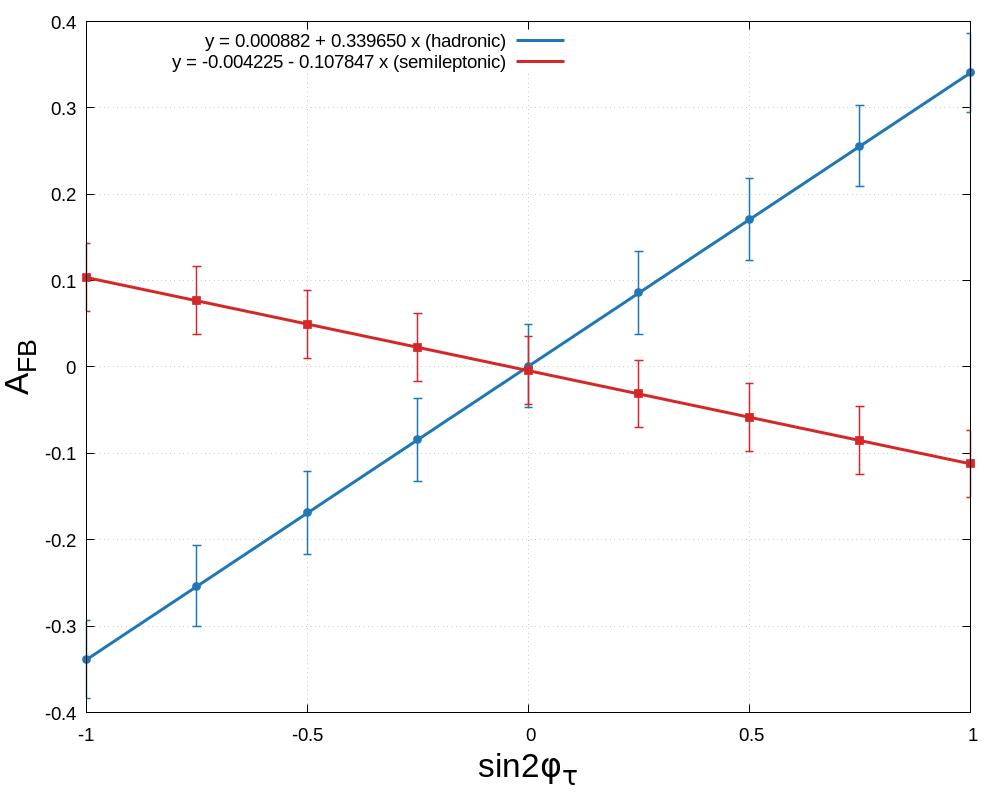}
    \caption{Forward-backward asymmetry as a function of $\sin 2\phi_{\tau}$ in the hadronic and semileptonic final state.}
    \label{asymmetry_sinphit}
\end{figure}

\noindent
Next we translate the limit on $\alpha_{34}$ to the CP-violating phase $\phi_{\tau}$ defined earlier in Eq.~\ref{phitau}. One can show that the forward backward asymmetry follows a straight-line distribution as a function of $\sin 2 \phi_{\tau}$. This can be clearly seen from the fit for $A_{FB}$ presented in Fig.~\ref{asymmetry_sinphit} as a function of $\sin 2 \phi_{\tau}$. Using the slopes of the fitted distributions and the corresponding uncertainties shown, we can calculate the limits on $\phi_{\tau}$ obtained from the hadronic, semileptonic final states as well as the combination of the two final states. We found that the limit on the CP-violating parameter $\delta f^{CP}_{h_{95}\tau\tau} = \sin^2 \phi_{\tau} \lesssim 0.00396$ in the combined case.

One should note that in our analysis, we have assumed the $\tau$-leptons can be fully reconstructed in both hadronic and semileptonic final states. In presence of neutrinos in the final state, the full reconstruction of the $\tau$'s can be challenging. Interesting methods have been proposed for $\tau$ reconstruction based on the collinear approximation, in which the visible hadronic decay products ($\pi^{\pm}$, $\rho^{\pm}$, or $a_1^{\pm}$) are assumed to be collinear with the momentum direction of the parent $\tau$ leptons. This assumption inevitably introduces an additional efficiency factor into the analysis presented above. In the collinear approximation, a $\tau$-reconstruction efficiency of up to 86\% can be achieved~\cite{ilctalk} However, the collinear approximation works best when the $\tau$'s are boosted and we lose some information by selecting only collinear events. On the other hand, at the lepton colliders, an analogue of the observable $\Phi^{*}_{CP}$ can be constructed using the impact parameter of the charged prongs ($\pi^{\pm},\rho^{\pm}$ or $a_1^{\pm}$) and the zero momentum frame (ZMF) of the charged prongs~\cite{Berge:2013jra}. Such an observable will faithfully reproduce all the results that we obtained so far.

\section{Combined interpretation of results within LHC, HL-LHC and $e^+e^-$ colliders}
\label{combined}

So far, there exist several studies on the CP properties of $h_{125}$, and the snowmass paper ~\cite{Gritsan:2022php} presented the summary for CP-studies of  $h_{125}tt$, $h_{125}\tau\tau$ and $h_{125}VV$ couplings. A recent study shows the impact of transversely polarized beams in probing the CP-violating $HZZ$ coupling~\cite{Li:2025ouv}. However, CP-violation in $HVV$ couplings will generally be loop-suppressed.   
The CP-violating $h_{125}tt$ coupling has been probed at the LHC~\cite{Gritsan:2022php} and the HL-LHC can have further improvement on this via four-top measurement~\cite{ATLAS:2025jdz}.
$tth_{125}$ associated production process would be the dominant channel to probe the top Yukawa coupling at $e^+ e^-$ collider, with center-of-mass energy $\gtrsim$ 500~GeV.  On the other hand, for the $h_{125}\tau\tau$ coupling, the $h_{125}\to\tau\tau$ decay would be the dominant process to test the CP-structure, since the 250~GeV $e^+ e^-$ collider can produce both $h_{125}$ and $h_{95}$ via Higgsstrahlung process. The snowmass paper~\cite{Gritsan:2022php} uses the following ratio to quantify the CP-violation in the Yukawa coupling:
\begin{equation}
    f_{CP}^{hff} = \frac{|\tilde{c}_{hff}|^2}{|\tilde{c}_{hff}|^2 +|{c}_{hff}|^2},
\end{equation}

\noindent
where ${c}_{hff}$ is the reduced CP-even Higgs Yukawa couplings and $\tilde{c}_{hff}$ is the reduced CP-odd Higgs Yukawa coupling.


Furthermore, the precision of the $f_{CP}$ can be extrapolated with higher luminosity at the same colliders by the following relation.
\begin{equation}
    \delta {f_{CP}}^\mathrm{extrapolate} =   \delta {f_{CP}}\sqrt{\frac{\mathcal{L}}{\mathcal{L}_\mathrm{extrapolate}}}.
\end{equation}
Therefore, we can obtain the results of $f_{CP}$ precision at $e^+ e^-$ collider with the extrapolation to $e^+ e^-$ setup \cite{Gritsan:2022php}, e.g. 5.6~ab$^{-1}$ at 250~GeV, 4~ab$^{-1}$ at 500~GeV and 8~ab$^{-1}$ at 1000~GeV, which are shown in the Table~\ref{tab:h125fcp}. Our results agree with those presented in a different work by \cite{ilctalk}.

\begin{table}[h]
    \centering
   {\scriptsize{ \begin{tabular}{l|c|cc|ccc}
    \hline
    Experiment& LHC&    \multicolumn{2}{c|}{HL-LHC}& \multicolumn{3}{c}{$e^+ e^-$}\\
    Setup& 140 fb$^{-1}$&  300 fb$^{-1}$&   3000 fb$^{-1}$&  250~GeV, 5.6 ab$^{-1}$& 500~GeV,  4 ab$^{-1}$&  1000~GeV,  8 ab$^{-1}$ \\
    \hline 
     $\delta f^{htt}_{CP}$&     $0.35$& $0.24$& $0.05$& -& $0.1$& $0.028$\\
     $\delta f^{h\tau\tau}_{CP}$ &   0.13&   0.07&   0.008&  0.0003&  0.007&   0.02\\
          \hline
    \end{tabular}}}
    \caption{The $f^{h_{125}ff}_{CP}$ precision (local 68\% C.L.) at hadron colliders and $e^+ e^-$ collider with different setups.}
    \label{tab:h125fcp}
\end{table}

Using these results, one can determine the detection limit of model parameters by combining the $tt$ and $\tau\tau$ CP-measurement via the following $\chi^2$ fit
\begin{equation}
    \chi^2 = \left( \frac{f^{htt}_{CP}}{\delta f^{htt}_{CP}}\right)^2 + \left( \frac{f^{h\tau\tau}_{CP}}{\delta f^{h\tau\tau}_{CP}}\right)^2\leq \chi^2(n=2, \mathrm{C.L.}=0.95),
    \label{eq:h125cpfit}
\end{equation}
where the experimental central $f_{CP}$ values are zero, since we did not observe any CP-violation in Higgs couplings.

Note that, we did not take the $HVV$ CP-structure into account for the combined results of $h_{125}$ CP-studies. Since the CP-violating $HVV$ coupling is a loop induced contribution, the $f^{HVV}_{CP}$ would be strongly suppressed in the model and would not bring a significant improvement when we combine the $HVV$ channel. However, the further improvement of the experimental precision can potentially enhance the sensitivity to the tiny $f^{HVV}_{CP}$, and this analysis will be left for a future study.

\begin{figure}[h]
    \centering
    \includegraphics[width=.6\linewidth]{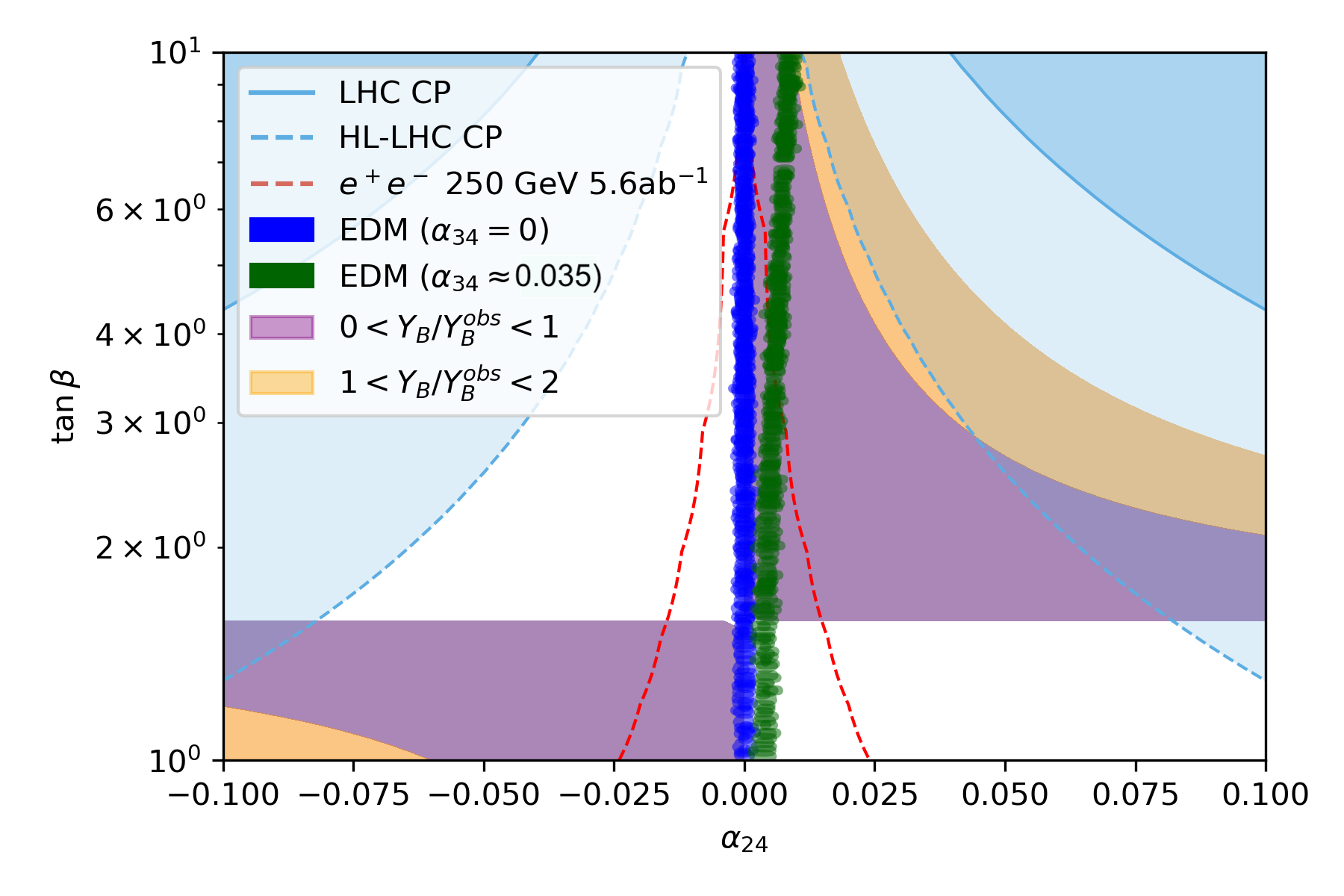}
    \caption{The parameter space spanned by $\tan\beta$ and $\alpha_{24}$. The cyan solid and dashed contours denote bounds from LHC CP-measurement of the Higgs, and future projection from the HL-LHC respectively. The red dashed line shows future projection from $e^+e^-$ collider at $\sqrt{s}$=250 GeV, and 5.6 $ab^{-1}$ integrated luminosity. The purple and beige regions indicate different values of baryon asymmetry. The blue (green) strips denote region allowed by the EDM bounds with $\alpha_{34}=0(0.035)$.}
        \label{fig:tba24}
\end{figure}

\begin{figure}[h]
    \centering
    \includegraphics[width=.45\linewidth]{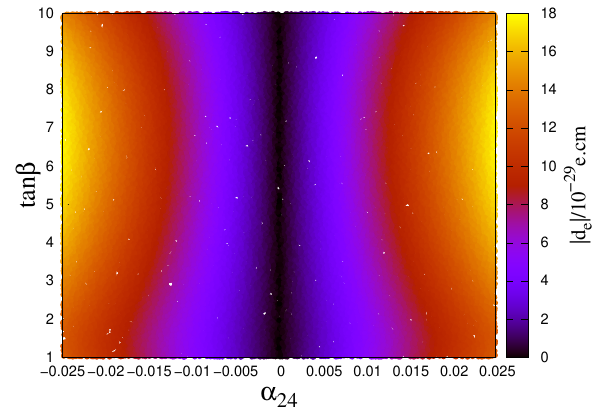}
        \includegraphics[width=.45\linewidth]{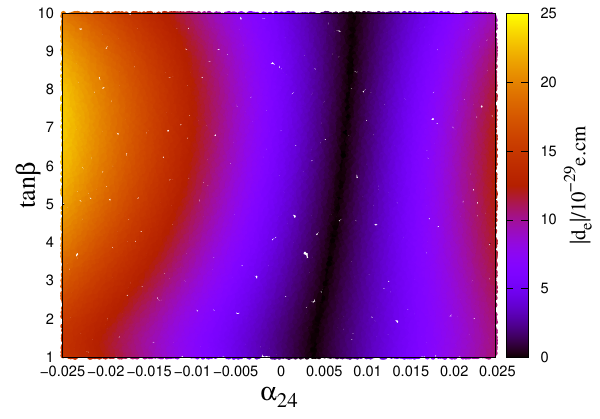}
    \caption{The plots show the parameter space of $\tan\beta$ vs $\alpha_{24}$ with electron EDM in the color axis, while $\alpha_{34}$ is kept at two benchmark values : (left) $\alpha_{34}=0$, (right) $\alpha_{34}=0.035$.}
    \label{fig:tba24_edm}
\end{figure}

\noindent
In Fig.~\ref{fig:tba24}, we present limits of CP-violation on the $\tan\beta$ vs $\alpha_{24}$ parameter space, using Eq.\ref{eq:h125cpfit}. In this figure, we show the projection of various CP-studies of the 125~GeV Higgs. We also present the EDM allowed region for both CP-conserving $h_{95}$ ($\alpha_{34}=0$) and CP-violating $h_{95}$ ($\alpha_{34}\approx 0.035$), where the EDM of $\alpha_{34}=0$ case is independent of $\tan\beta$ and excludes the $|\alpha_{24}|\gtrsim 0.002$. However, if $\alpha_{34}\neq0$, the CP-conserving $h_{125}$ ($\alpha_{24}=0$) would be excluded by EDM bounds, and in this case the larger $\tan\beta$ can shift the EDM allowed region to the larger $\alpha_{24}$ region.
We show the behavior of the EDM as a function of the two parameters $\tan\beta$ and $\alpha_{24}$ in Fig.~\ref{fig:tba24_edm}. On the left panel, $\alpha_{34}$ is assumed to be 0 and on the right panel, $\alpha_{34}$ is kept at our benchmark value i.e. $\alpha_{34}=0.035$. One can see the presence of additional angle significantly modifies EDM prediction and constraints in our model parameter space.

In addition, we perform an effective baryon asymmetry calculation by using the effective fermionic couplings \cite{AharonyShapira:2021ize},
where the calculation therein assumes that the strong first-order phase-transition takes place. In Fig.~\ref{fig:tba24}, the purple region is showing that the baryon ratio $0<Y_B/Y_B^\mathrm{obs}<1$ and the yellow region indicates $1<Y_B/Y_B^\mathrm{obs}<2$. The other blank region has either $Y_B/Y_B^\mathrm{obs}<0$ or $Y_B/Y_B^\mathrm{obs}>2$. One should note that, the EDM bound would actually exclude the region of $Y_B/Y_B^\mathrm{obs}\sim 1$, when $h_{95}$ is CP-conserving ($\alpha_{34}=0$). When $\alpha_{34}\neq0$, the larger $\tan\beta$ region can potentially have the correct baryon number ratio $Y_B$ as well as be allowed by EDM constraint. 

 As we see in Fig~\ref{fig:tba24}, the LHC bound on $\alpha_{24}$ is more stringent in large $\tan\beta$ region. The reason being, the experimental precision in the $h_{125}\tau\tau$ measurement is better than the $h_{125}tt$ measurement, and type-II Yukawa coupling structure results in enhanced $\tau$-Yukawa coupling in large $\tan\beta$ region. Consequently, the sensitivity of CP-measurement in the $\tau$-Yukawa coupling at HL-LHC and $e^+ e^-$ (250~GeV) collider also become more precise at large $\tan\beta$ region for the same reason. Comparing with the baryongenesis results, we found that the HL-LHC detection limit can probe the region of $Y_B/Y_B^\mathrm{obs}\sim 1$ for lower $\tan\beta$, while current LHC is still not able to probe the region with $Y_B/Y_B^\mathrm{obs}\sim 1$. For the $e^+ e^-$ collider at 250~GeV with 5.6~ab$^{-1}$, the detection precision would be improved compared to HL-LHC, and most of the $Y_B/Y_B^\mathrm{obs}\sim 1$ region can be probed.

\begin{figure}
    \centering
    \includegraphics[width=.475\linewidth]{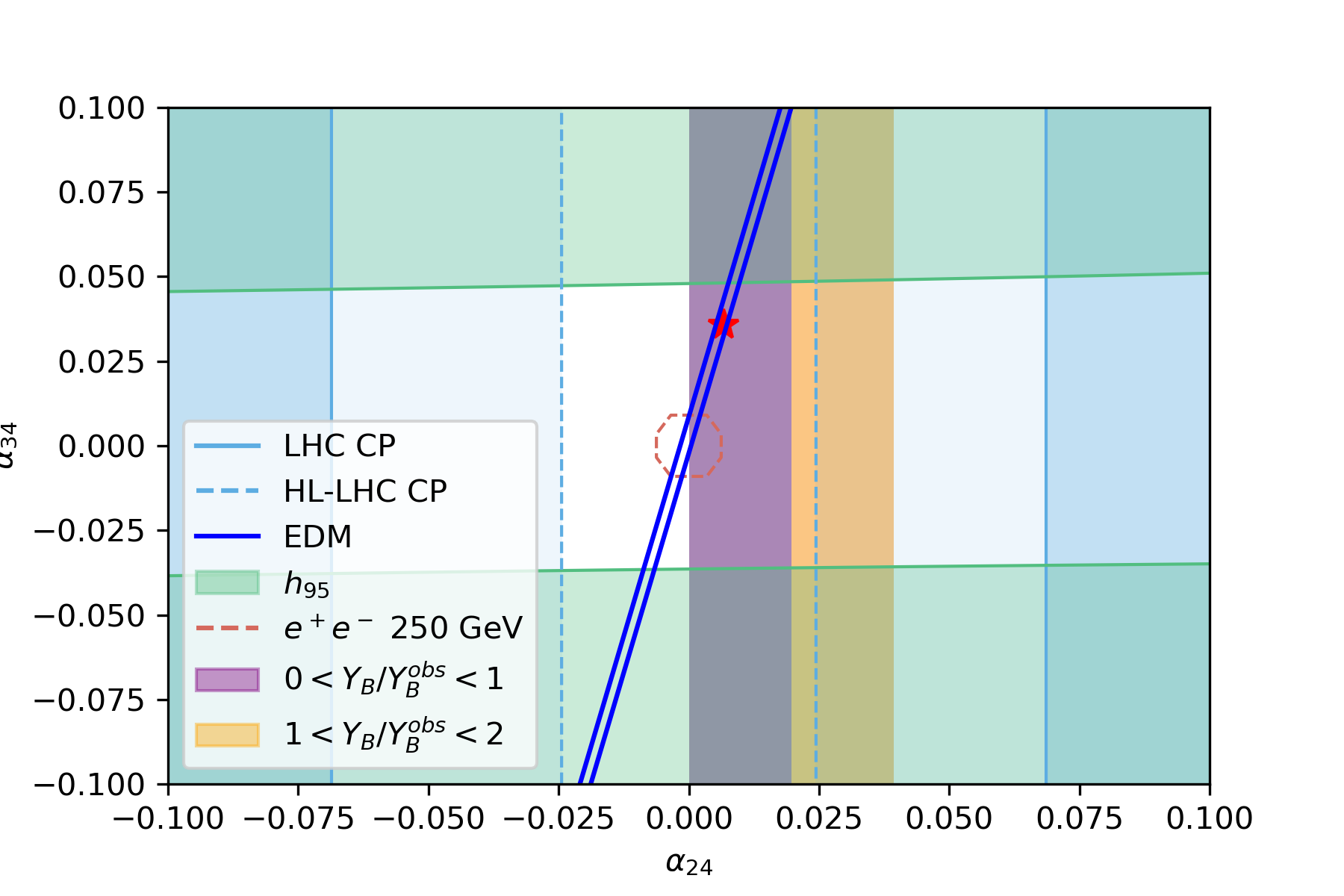}
    \includegraphics[width=.475\linewidth]{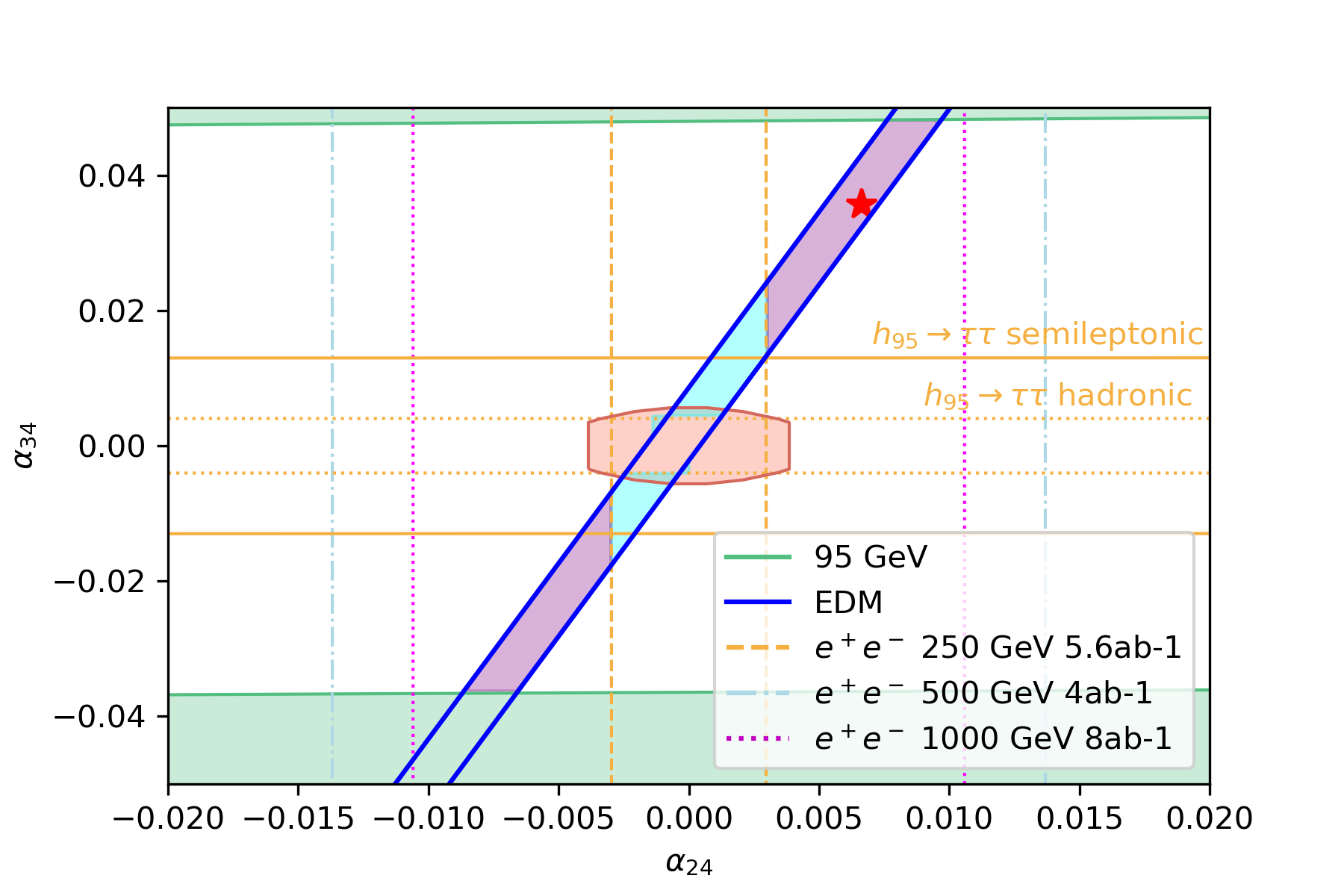}
    \caption{In the left panel, limits from LHC and HL-LHC are shown along with the EDM constraint. The purple and beige areas show regions with various baryon asymmetry values. In the right panel, we show the projected limits on the CP-phase of the 125 GeV Higgs and 95 GeV scalar at future lepton colliders. In the purple regions both $h_{95}$ CP phase and $h_{125}$ CP-phase can be observed at 95\% C.L. The cyan regions mark where $h_{125}$ would appear to be identical with the SM-like CP-even Higgs boson at 95\% C.L, while still a non-zero CP-phase in the $h_{95}$ can be observed at 95\% C.L. Our chosen benchmark is shown as a red star in both plots. }
    \label{fig:a24a34}
\end{figure}

Finally, we can combine all of these relevant CP-studies in the previous discussions, including the existing $h_{125}$ CP-studies, the EDM bound and the $h_{95}\to\tau\tau$ CP-studies, and project the results on the mixing angles $\alpha_{24}$ vs $\alpha_{34}$ plane, which are shown in Fig.~\ref{fig:a24a34}. 
In the left panel, we present the parameter space within the range of $|\alpha_{24}|\leq0.1$ and $|\alpha_{34}|\leq0.1$ based on the benchmark point in Table~\ref{tab:bench}, where the 95\% CL upper bound on $\alpha_{24}$ from LHC CP-studies, EDM bound, 68\% C.L. limit on $\alpha_{34}$ from the 95~GeV excess in the $\gamma\gamma$ channel, and the baryon asymmetry results are all shown. In addition, the 95\% C.L. projected upper bound on $\alpha_{24}$ from HL-LHC, and 250~GeV $e^+ e^-$ collider with ${\cal L} dt$ 5.6~ab$^{-1}$ are also shown.
The $e^+ e^-$ collider limits include both $h_{125}\to\tau\tau$ and $h_{95}\to\tau\tau$ channels. 

One can see that, the current LHC CP measurement would constrain the parameter space within the range $|\alpha_{24}|\lesssim 0.065$ (95\% C.L.) and the HL-LHC projects a limit $|\alpha_{24}|\lesssim 0.025$ (95\% C.L.). On the other hand, the CP-mixing angle $\alpha_{34}$ would modify the properties of the 95~GeV scalar substantially, for example, the $|\alpha_{34}|\gtrsim 0.3$ region would lead to reduced $h_{95}\gamma\gamma$ coupling, which does not match the observed experimental excess in the $\gamma\gamma$ final state. This constraint restricts the CP-mixing angle in the 95 GeV scalar $|\alpha_{34}| \lesssim 0.04$.

We see that the EDM bound constrains the parameter space within a thin band. However, it is important to remember from earlier discussions that the EDM bound would allow larger $|\alpha_{24}|$ when $|\alpha_{34}|\neq 0$, compared to the case when $|\alpha_{34}| \approx 0$. Interesting to note that the restrictive EDM constraint on $\alpha_{24} (\lesssim 6 \times 10^{-4})$ at $\alpha_{34}=0$, would directly exclude the region with sizeable ($\frac{Y_B}{Y_B^{obs}}\lesssim 1$) baryon number asymmetry.
Note that, we still present the effective baryogenesis results in this figure, where this baryon asymmetry calculation did not take into account the 95~GeV scalar ($h_{95}$) contribution. For example, we see that, the benchmark point (red star) is within the EDM-allowed region, but the baryon asymmetry is still under the observed $Y_B^\mathrm{obs}$. However, the actual baryon asymmetry corresponding to the benchmark point can still potentially match the observation, when the CP-violating $h_{95}$ contributions are taken into account. Furthermore, the strong bound from EDM on the CP-violating phase of the 125 GeV Higgs boson, in the absence of any other phase in the theory, also implies that in case the CP-violating phase is observed in the Higgs boson in future experiments, while EDM experiments yields negative results, the only explanation would be an existence of extended scalar sector with additional CP-violating phases therein.

In the right panel of Fig.~\ref{fig:a24a34}, we zoomed in on the EDM-allowed region of the left panel of Fig.~\ref{fig:a24a34} and the parameter space is within $|\alpha_{24}|\leq0.02$ and $|\alpha_{34}|\leq0.05$. Furthermore, we can project more CP-measurement $1\sigma$~precision (68\% C.L.) of local channels at the future collider, where we present the results for FCC-ee/CEPC at 250~GeV and ILC at 500~GeV and 1~TeV setups. In particular, we present the combined result at 68\% C.L. in the middle orange region, which include the CP studies of $h_{125}\to\tau\tau$ channel, $h_{95}\to\tau\tau$ hadronic channel and semi-leptonic channel. As we see, the 250~GeV setup (orange line) provides  best sensitivity to the CP-phase of $h_{125}$ in $\tau\tau$ channel. Although higher center-of-mass energies can include the $tt$-associated production channel, the overall sensitivity to $h_{125}$ CP-phase worsens compared to the 250 GeV case. Since ILC~1~TeV case can access higher integrated luminosity compared to the 500~GeV case, the detection limit in 1~TeV case ($|\alpha_{24}|\sim0.013$) is more precise than the 500~GeV case ($|\alpha_{24}|\sim0.017$).
By comparing the detection limit and EDM bound, we can determine the two parameter space regions (purple and cyan colored space) in the figure. The purple regions of the right plot are where both CP-phases (if non-zero) of $h_{125}$ and $h_{95}$ can be detected at the $e^+e^-$ collider at $\sqrt{s}=$ 250~GeV, while the $h_{95}$ matches the observed 95~GeV excess in the $\gamma\gamma$ final state, and the EDM bound is also satisfied. In the cyan region in Fig.~\ref{fig:a24a34} (right panel), the $h_{125}$ CP-violation is not observable at the $e^+e^-$ collider with $\sqrt{s}=$ 250~GeV, while the $h_{95}$ CP-violation can be detected. Therefore, this cyan region demonstrates an interesting scenario where the 125~GeV Higgs can be identified as a CP-even state, while the 95~GeV scalar can be observed as a CP-mixed state. If the parameter point belongs to the orange shaded area in the centre which is below the detection limit (68\% C.L.) of $|\alpha_{34}|\lesssim 0.004$ ($\delta f^{CP}_{h_{95}\tau\tau} \sim 0.00396$), the observations would be consistent with both $\alpha_{24}\sim 0$ and $\alpha_{34}\sim 0$, i.e. the $h_{95}$ and $h_{125}$ will both be identified as the CP-even states in the experiments.

\section{Conclusion}
\label{conclusion}

In this study, we probe CP-violating phases in 2HDM+singlet scenario at future lepton colliders. As a consequence of CP-violation, such models receive strong constraints from the non-observation of electric dipole moment. As a result, the CP-phases, allowed by the EDM bounds, become extremely challenging to probe at the colliders, owing to their small magnitude. However, earlier studies have demonstrated that extended scalar sectors can, in principle, alleviate this problem through cancellations among different contributions, provided that multiple CP-violating phases, and hence multiple sources of CP violation, are present. We consider such an extended scalar scenario, namely complex singlet extension of 2HDM, where such cancellations are possible. An interesting feature of this model is that one can accommodate the recently observed excess in the $\gamma\gamma$ final state at 95 GeV mass at the CMS experiments. Similar excess was previously reported at the LEP experiments in the $b\bar b$ final state. Alongside, this model can also provide a suitable dark matter candidate, although we do not focus on dark matter in this work. Here, we assume a simplified scenario, where the CP-violation is present only in the observed 125 GeV Higgs boson and the 95 GeV excess. All other scalars are assumed to be CP-even. We reiterate that, this is a simplification with minimal CP-violating sources, consistent with all the existing data, at the same time addressing the CP-mixing in an already observed excess (95 GeV), which can be probed in the near future. 

With this goal in mind, we choose the $\tau\tau$ decay mode of the 95 GeV scalar produced via Higgsstrahlung process, and consider a CP-sensitive angular observable to probe the CP-violating phase in the hadronic as well as semileptonic decay modes of the $\tau\tau$ pair. In this context, we consider the forward-backward asymmetry $(A_{FB})$ and using the distribution of $A_{FB}$, we compute the sensitivity of the CP-violating phase in the future lepton colliders, namely ILC, FCC-ee or CEPC. We show that the hadronic final state is more sensitive than the semileptonic one. We present our results combining the two modes. We show that in case of the 95 GeV scalar the SM Drell-Yan background is non-negligible and we also present the projected sensitivity in the presence of SM background.  We also compare our results with existing limits and future projections on the CP-violating phase of the 125 GeV Higgs boson. Furthermore, we comment on the prospect of baryogenesis, i.e. the generated baryon asymmetry in the universe, in the relevant regions of parameter space in our model. Our study provides a guideline for probing CP-violating phases at future colliders while remaining consistent with current electric dipole moment (EDM) constraints. In addition to the 125 GeV Higgs boson, we explore the intriguing possibility of probing the CP properties of the 95 GeV excess at future colliders. Our study shows two interesting results: (a) In the presence of an extended scalar sector, the EDM-allowed CP-violating phase of the 125 GeV Higgs boson can be larger, and therefore it can be probed at the future lepton colliders. (b) If CP violation is observed in the 125 GeV Higgs sector at future colliders while no corresponding signal is found in low-energy EDM experiments, this would point towards an extended scalar sector with additional CP-violating phases. In such a scenario, probing CP violation in the additional scalar states becomes particularly important. The 95 GeV excess, if confirmed, would itself provide a potential indication of an extended scalar sector and, consequently, a promising laboratory for studying its CP properties. Even in the absence of an observable CP-violating signal in the 125 GeV Higgs boson, searches for CP violation in additional scalar states remain important for addressing the long-standing question of the origin of the matter-antimatter asymmetry of the Universe. Finally, we reiterate that, in extended scalar sectors, moderate cancellations among different contributions to the EDM can occur, therefore larger CP-phases are allowed. This highlights the importance of model-specific studies when assessing the prospects for probing CP-violating phases at future colliders.

\section*{Acknowledgement}

The authors JL and GMP acknowledge support by the Deutsche Forschungsgemeinschaft (DFG, German Research Foundation) under Germany's Excellence Strategy EXC 2121 "Quantum Universe" - 390833306. 
CL is supported by the Natural Science Foundation of China (NSFC) under grant numbers 12305115, Shenzhen Science and Technology Program (Grant No. 202206193000001, 20220816094256002), Guangdong Provincial Key Laboratory of Gamma-Gamma Collider and Its Comprehensive Applications (2024KSYS001), and Guangdong Provincial Key Laboratory of Advanced Particle Detection Technology (2024B1212010005). The work of JL is supported by ANRF grant no-RJF/2025/000544.
 
\newpage
\appendix
\section{General model}
\label{general}
Complex parameters:
$$ M_{12}^2 = m_{12}^2 e^{-i \xi_1} = |m_{12}^2| e^{i(\theta_{m12}-\xi_1)}, $$
$$ \Lambda_5 = \lambda_5 e^{-i 2\xi_1} = |\lambda_5| e^{i(\theta_{\lambda 5}-2\xi_1)}, $$
$$ \Lambda_6 = \lambda_6 e^{-i \xi_1} = |\lambda_6| e^{i(\theta_{\lambda 6}-\xi_1)}, $$
$$ \Lambda_7 = \lambda_7 e^{-i \xi_1} = |\lambda_7| e^{i(\theta_{\lambda 7}-\xi_1)}, $$

$$ M^2_{SP} = {m_S'}^2 e^{i 2\xi_S} = |{m_S'}^2| e^{i(\theta_{{m_S'}}+2\xi_S)}, $$
$$ \mathcal{M}_{S1} = \mu_{S1} e^{i 3\xi_S} = |\mu_{S1}| e^{i(\theta_{\mu_{S1}}+3\xi_S)}, $$
$$ \mathcal{M}_{S2}=\mu_{S2} e^{i \xi_S} = |\mu_{S2}| e^{i(\theta_{\mu_{S2}}+\xi_S)}, $$
$$ \mathcal{M}_{11}=\mu_{11} e^{i \xi_S} = |\mu_{11}| e^{i(\theta_{\mu_{11}}+\xi_S)}, $$
$$ \mathcal{M}_{22}=\mu_{22} e^{i \xi_S} = |\mu_{22}| e^{i(\theta_{\mu_{22}}+\xi_S)}, $$
$$ \mathcal{M}_{12}=\mu_{12} e^{i (\xi_S-\xi_1)} = |\mu_{12}| e^{i(\theta_{\mu_{12}}+\xi_S-\xi_1)}, $$
$$ \mathcal{M}_{21}=\mu_{21} e^{i (\xi_S+\xi_1)} = |\mu_{21}| e^{i(\theta_{\mu_{21}}+\xi_S+\xi_1)}, $$
$$ \Lambda_1''=\lambda_1'' e^{i 4\xi_S} = |\lambda_1''| e^{i(\theta_{\lambda_1''}+4\xi_S)}, $$
$$ \Lambda_2''=\lambda_2'' e^{i 2\xi_S} = |\lambda_2''| e^{i(\theta_{\lambda_2''}+2\xi_S)}, $$
$$ \Lambda_3'=\lambda_3' e^{-i \xi_1} = |\lambda_3'| e^{i(\theta_{\lambda_3'}-\xi_1)}, $$
$$ \Lambda_4'=\lambda_4' e^{i 2\xi_S} = |\lambda_4'| e^{i(\theta_{\lambda_4'}+2\xi_S)}, $$
$$ \Lambda_5'=\lambda_5' e^{i 2\xi_S} = |\lambda_5'| e^{i(\theta_{\lambda_5'}+2\xi_S)}, $$
$$ \Lambda_6'=\lambda_6' e^{i (2\xi_S-\xi_1)} = |\lambda_6'| e^{i(\theta_{\lambda_6'}+2\xi_S-\xi_1)}, $$
$$ \Lambda_7'=\lambda_7' e^{i (2\xi_S+\xi_1)} = |\lambda_7'| e^{i(\theta_{\lambda_7'}+2\xi_S+\xi_1)}. $$

Defining  abbreviation:
$$ \lambda_{345} = \lambda_3+\lambda_4 + \operatorname{Re}(\Lambda_5), $$
$$ \lambda_{367}' = \operatorname{Re}(\Lambda_3'+\Lambda_6'+\Lambda_7'), $$
$$ \tilde{\mu}_{1221} = \mathcal{M}_{12}+\mathcal{M}_{21},$$
$$ \tilde{\mu}^2 = \frac{\operatorname{Re}(M_{12}^2) - \operatorname{Re}(\tilde{\mu}_{1221})v_S-\lambda_{367}'v_S^2}{s_\beta c_\beta},$$
$$ \tilde{\mu}_{1S} = \mathcal{M}_{11} + 2\Lambda_4'v_S, $$
$$ \tilde{\mu}_{2S} = \mathcal{M}_{22} + 2\Lambda_5'v_S,$$
$$ \tilde{\mu}_{1267}' =\mathcal{M}_{21}-\mathcal{M}_{12} + 2  (\Lambda_7'-\Lambda_6') v_S. $$
\noindent
Mass matrices are as follows:
{\small{\begin{align}
 M_{H11}^2 =& \lambda_1 v^2c^2_\beta + \tilde{\mu}^2 s^2_\beta + \frac{3}{2} \operatorname{Re}(\Lambda_6) v^2 s_\beta c_\beta -\frac{1}{2}\operatorname{Re}(\Lambda_7) v^2  s_\beta^2 t_\beta \label{eq:mh11},\\
  M_{H22}^2 =& \lambda_2 v^2 s_\beta^2 + \tilde{\mu}^2 c_\beta^2-\frac{1}{2}\operatorname{Re}(\Lambda_6)v^2 c_\beta^2 / t_\beta + \frac{3}{2}\operatorname{Re}(\Lambda_7)v^2 s_\beta c_\beta, \\
 M_{H12}^2 =&[ \lambda_{345}v^2-\tilde{\mu}^2]s_\beta c_\beta +\frac{3}{2}v^2 (\operatorname{Re}(\Lambda_6)c_\beta^2 + \operatorname{Re}(\Lambda_7)s_\beta^2),\\
 M_{H13}^2 = &v [2 (\lambda_1'v_S + \operatorname{Re}(\tilde{\mu}_{1S})) c_\beta + (\operatorname{Re}(\tilde{\mu}_{1221}) + 2\lambda_{367}'v_S )s_\beta ],\\
 M_{H23}^2 =& v [2 (\lambda_2' v_S + \operatorname{Re}(\tilde{\mu}_{2S}) )s_\beta + ( \operatorname{Re}(\tilde{\mu}_{1221}) + 2\lambda_{367}'v_S )c_\beta ],\\
M_{H33}^2 =& 2\frac{\operatorname{Re}(\Lambda_1''+4\Lambda_2'') + \lambda_3''}{3}v_S^2+ \operatorname{Re}(\mathcal{M}_{S1} + 3\mathcal{M}_{S2})v_S \nonumber\\
&- \frac{v^2}{v_S} \operatorname{Re}\Big(
\mathcal{M}_{11}c_\beta^2 + \tilde{\mu}_{1221}s_\beta c_\beta+  \mathcal{M}_{22}s_\beta^2 \Big),\\
M_{H14}^2 =& -\frac{1}{2}v^2 \operatorname{Im}\Big( 2 \Lambda_6c_\beta + \Lambda_5s_\beta\Big),\\
M_{H24}^2 =& -\frac{1}{2}v^2 \operatorname{Im}\Big( 2 \Lambda_7s_\beta +\Lambda_5c_\beta\Big),\\
M_{H34}^2 =& v\operatorname{Im}(\tilde{\mu}_{1267}'-2\Lambda_3'v_S),\\
M_{H44}^2 =& \tilde{\mu}^2 - \operatorname{Re}(\Lambda_5)v^2 - \frac{1}{2}v^2(\operatorname{Re}(\Lambda_6)/t_\beta + \operatorname{Re}(\Lambda_7)t_\beta), \\
M_{H15}^2 =& -v\operatorname{Im}\Big( 2\tilde{\mu}_{1S}c_\beta + (\tilde{\mu}_{1221}+2(\Lambda_6'+\Lambda_7')v_S)s_\beta \Big),\\
M_{H25}^2 =& -v\operatorname{Im}\Big( 2\tilde{\mu}_{2S}s_\beta + (\tilde{\mu}_{1221}+2(\Lambda_6'+\Lambda_7')v_S)c_\beta \Big),\\
M_{H35}^2 =& \frac{v^2}{v_S}\operatorname{Im}(\mathcal{M}_{11}c_\beta^2+\mathcal{M}_{22}s_\beta^2+\tilde{\mu}_{1221}s_\beta c_\beta) - \operatorname{Im}(\mathcal{M}_{S1}+\mathcal{M}_{S2})v_S-\frac{2}{3}\operatorname{Im}(\Lambda_1'' + 2\Lambda_2'')v_S^2,\\
M_{H55}^2 =& \operatorname{Re}\Big( -4 M_{SP}^2 - (3\mathcal{M}_{S1}+\mathcal{M}_{S2})v_S -\frac{4}{3}(\Lambda_1''+\Lambda_2'')v_S^2\notag\\
        &-v^2 \Big[ \frac{\tilde{\mu}_{1S}}{v_S}c_\beta^2  + \frac{\tilde{\mu}_{2S}}{v_S}s_\beta^2 +2 (\Lambda_4'c^2_\beta+\Lambda_5's^2_\beta + (\Lambda_6'+\Lambda_7')s_{2\beta})\Big]\Big), \\
M_{H45}^2 =& v\operatorname{Re}(\tilde{\mu}_{1267}'),\\
M_{H^\pm}^2 =& \tilde{\mu}^2 -\frac{1}{2}v^2(\lambda_4+\operatorname{Re}(\Lambda_5)+\operatorname{Re}(\Lambda_6)/t_\beta + \operatorname{Re}(\Lambda_7)t_\beta). \label{eq:mhc}
\end{align}}}

\medskip
\noindent
5$\times$5 Mixing matrix $R$:
\allowdisplaybreaks
{\small{\begin{align}
R_{11} &= c_{12} c_{13} c_{14} c_{15}, \label{eq:r11}\\
R_{12} &= c_{13} c_{14} c_{15} s_{12}, \\
R_{13} &= c_{14} c_{15} s_{13}, \\
R_{14} &= c_{15} s_{14}, \\
R_{15} &= s_{15}, \\
R_{21} &= -c_{23}c_{24}c_{25}s_{12} + c_{12}\bigl(-c_{24}c_{25}s_{13}s_{23} + c_{13}(-c_{25}s_{14}s_{24} - c_{14}s_{15}s_{25})\bigr), \\
R_{22} &= c_{12}c_{23}c_{24}c_{25} + s_{12}\bigl(-c_{24}c_{25}s_{13}s_{23} + c_{13}(-c_{25}s_{14}s_{24} - c_{14}s_{15}s_{25})\bigr), \\
R_{23} &= c_{13}c_{24}c_{25}s_{23} + s_{13}\bigl(-c_{25}s_{14}s_{24} - c_{14}s_{15}s_{25}\bigr), \\
R_{24} &= c_{14}c_{25}s_{24} - s_{14}s_{15}s_{25}, \\
R_{25} &= c_{15}s_{25}, \\
R_{31} &= -s_{12}\bigl(-c_{34}c_{35}s_{23} + c_{23}(-c_{35}s_{24}s_{34} - c_{24}s_{25}s_{35})\bigr) \nonumber \\ 
& \qquad + c_{12}\Bigl(-s_{13}\bigl(c_{23}c_{34}c_{35} + s_{23}(-c_{35}s_{24}s_{34} - c_{24}s_{25}s_{35})\bigr) \notag\\
      &\qquad + c_{13}\bigl(-c_{14}c_{25}s_{15}s_{35} - s_{14}(c_{24}c_{35}s_{34} - s_{24}s_{25}s_{35})\bigr)\Bigr), \\
R_{32} &= c_{12}\bigl(-c_{34}c_{35}s_{23} + c_{23}(-c_{35}s_{24}s_{34} - c_{24}s_{25}s_{35})\bigr) \nonumber\\
&\qquad + s_{12}\Bigl(-s_{13}\bigl(c_{23}c_{34}c_{35} + s_{23}(-c_{35}s_{24}s_{34} - c_{24}s_{25}s_{35})\bigr) \notag\\
      &\qquad + c_{13}\bigl(-c_{14}c_{25}s_{15}s_{35} - s_{14}(c_{24}c_{35}s_{34} - s_{24}s_{25}s_{35})\bigr)\Bigr), \\
R_{33} &= c_{13}\bigl(c_{23}c_{34}c_{35} + s_{23}(-c_{35}s_{24}s_{34} - c_{24}s_{25}s_{35})\bigr)  \nonumber \\
& \quad + s_{13}\bigl(-c_{14}c_{25}s_{15}s_{35} - s_{14}(c_{24}c_{35}s_{34} - s_{24}s_{25}s_{35})\bigr), \\
R_{34} &= -c_{25}s_{14}s_{15}s_{35} + c_{14}\bigl(c_{24}c_{35}s_{34} - s_{24}s_{25}s_{35}\bigr), \\
R_{35} &= c_{15}c_{25}s_{35}, \\
R_{41} &= -s_{12}\Bigl(-s_{23}\bigl(-c_{45}s_{34} - c_{34}s_{35}s_{45}\bigr)  + c_{23}\bigl(-c_{24}c_{35}s_{25}s_{45} - s_{24}(c_{34}c_{45} - s_{34}s_{35}s_{45})\bigr)\Bigr) \nonumber \\
      &\quad + c_{12}\Bigl(c_{13}\bigl(-c_{14}c_{25}c_{35}s_{15}s_{45} - s_{14}\bigl(-c_{35}s_{24}s_{25}s_{45} + c_{24}(c_{34}c_{45} - s_{34}s_{35}s_{45})\bigr)\bigr) \nonumber \\
      &\qquad - s_{13}\bigl(c_{23}(-c_{45}s_{34} - c_{34}s_{35}s_{45}) +
s_{23}\bigl(-c_{24}c_{35}s_{25}s_{45} - s_{24}(c_{34}c_{45} - s_{34}s_{35}s_{45})\bigr)\bigr)\Bigr), \\
R_{42} &= c_{12}\Bigl(-s_{23}\bigl(-c_{45}s_{34} - c_{34}s_{35}s_{45}\bigr)  + c_{23}\bigl(-c_{24}c_{35}s_{25}s_{45} - s_{24}(c_{34}c_{45} - s_{34}s_{35}s_{45})\bigr)\Bigr) \notag\\
      &\quad + s_{12}\Bigl(c_{13}\bigl(-c_{14}c_{25}c_{35}s_{15}s_{45} - s_{14}\bigl(-c_{35}s_{24}s_{25}s_{45} + c_{24}(c_{34}c_{45} - s_{34}s_{35}s_{45})\bigr)\bigr) \nonumber \\
      &\qquad - s_{13}\bigl(c_{23}(-c_{45}s_{34} - c_{34}s_{35}s_{45}) + s_{23}\bigl(-c_{24}c_{35}s_{25}s_{45} - s_{24}(c_{34}c_{45} - s_{34}s_{35}s_{45})\bigr)\bigr)\Bigr), \\
R_{43} &= s_{13}\bigl(-c_{14}c_{25}c_{35}s_{15}s_{45} - s_{14}\bigl(-c_{35}s_{24}s_{25}s_{45} + c_{24}(c_{34}c_{45} - s_{34}s_{35}s_{45})\bigr)\bigr) \notag\\
      &\quad + c_{13}\bigl(c_{23}(-c_{45}s_{34} - c_{34}s_{35}s_{45}) + s_{23}\bigl(-c_{24}c_{35}s_{25}s_{45} - s_{24}(c_{34}c_{45} - s_{34}s_{35}s_{45})\bigr)\bigr) ,\\
R_{44} &= -c_{25}c_{35}s_{14}s_{15}s_{45} + c_{14}\bigl(-c_{35}s_{24}s_{25}s_{45} + c_{24}(c_{34}c_{45} - s_{34}s_{35}s_{45})\bigr), \\
R_{45} &= c_{15}c_{25}c_{35}s_{45}, \\
R_{51} &= -s_{12}\Bigl(-s_{23}\bigl(-c_{34}c_{45}s_{35} + s_{34}s_{45}\bigr)  + c_{23}\bigl(-c_{24}c_{35}c_{45}s_{25} - s_{24}\bigl(-c_{45}s_{34}s_{35} - c_{34}s_{45}\bigr)\bigr)\Bigr) \notag\\
      &\quad + c_{12}\Bigl(c_{13}\bigl(-c_{14}c_{25}c_{35}c_{45}s_{15} - s_{14}\bigl(-c_{35}c_{45}s_{24}s_{25} + c_{24}(-c_{45}s_{34}s_{35} - c_{34}s_{45})\bigr)\bigr) \notag \\
      &\qquad - s_{13}\bigl(c_{23}(-c_{34}c_{45}s_{35} + s_{34}s_{45}) + s_{23}\bigl(-c_{24}c_{35}c_{45}s_{25} - s_{24}(-c_{45}s_{34}s_{35} - c_{34}s_{45})\bigr)\bigr)\Bigr), \\
R_{52} &= c_{12}\Bigl(-s_{23}\bigl(-c_{34}c_{45}s_{35} + s_{34}s_{45}\bigr)  + c_{23}\bigl(-c_{24}c_{35}c_{45}s_{25} - s_{24}\bigl(-c_{45}s_{34}s_{35} - c_{34}s_{45}\bigr)\bigr)\Bigr) \notag\\
      &\quad + s_{12}\Bigl(c_{13}\bigl(-c_{14}c_{25}c_{35}c_{45}s_{15} - s_{14}\bigl(-c_{35}c_{45}s_{24}s_{25} + c_{24}(-c_{45}s_{34}s_{35} - c_{34}s_{45})\bigr)\bigr) \notag \\
      &\qquad - s_{13}\bigl(c_{23}(-c_{34}c_{45}s_{35} + s_{34}s_{45}) + s_{23}\bigl(-c_{24}c_{35}c_{45}s_{25} - s_{24}(-c_{45}s_{34}s_{35} - c_{34}s_{45})\bigr)\bigr)\Bigr),\\
R_{53} &= s_{13}\bigl(-c_{14}c_{25}c_{35}c_{45}s_{15} - s_{14}\bigl(-c_{35}c_{45}s_{24}s_{25} + c_{24}(-c_{45}s_{34}s_{35} - c_{34}s_{45})\bigr)\bigr) \notag\\
      &\quad + c_{13}\bigl(c_{23}(-c_{34}c_{45}s_{35} + s_{34}s_{45}) + s_{23}\bigl(-c_{24}c_{35}c_{45}s_{25} - s_{24}(-c_{45}s_{34}s_{35} - c_{34}s_{45})\bigr)\bigr), \\
R_{54} &= -c_{25}c_{35}c_{45}s_{14}s_{15} + c_{14}\bigl(-c_{35}c_{45}s_{24}s_{25} + c_{24}(-c_{45}s_{34}s_{35} - c_{34}s_{45})\bigr), \\
R_{55} &= c_{15}c_{25}c_{35}c_{45}.\label{eq:r55}
\end{align}}}

\section{Higgs basis}
\label{higgsbasis}
In the Higgs basis the fields are defined as 
\begin{equation}
\Phi_1 = \begin{pmatrix}
G^+\\ \frac{1}{\sqrt{2}}(v + h + i G_0)
\end{pmatrix},~~~\Phi_2 = \begin{pmatrix}
H^+\\ \frac{1}{\sqrt{2}}( H + i A)
\end{pmatrix}.
\end{equation}

The mass matrix in the Higgs basis is 
\begin{equation}
    \begin{split}
        M^H_{11} &= \lambda_1^H v^2,\\
        M^H_{12} &= \mathrm{Re}(\lambda_6^H),\\
        M^H_{22} &= \frac{\mu_{H}^2 }{v^2}+ v_S^2 ({\lambda_2'}^H + 2 \mathrm{Re}({\lambda_5'}^H)) + v^2\frac{\lambda_3^H+\lambda_4^H+\mathrm{Re}(\lambda_5^H)}{2},\\
        M^H_{13} &= 2v(\mathrm{Re}(\mu_{11}^H+\mu_{22}^H) + v_S({\lambda_1'}^H + 2 \mathrm{Re}({\lambda_4'}^H)),\\
        M^H_{23} & = v \mathrm{Re}\Big(2v_S ({\lambda_3'}^H + {\lambda_6'}^H + {\lambda_7'}^H) + \mu_{12}^H + \mu_{21}^H\Big),\\
        M^H_{33} &= 8 v_S^2 {\lambda_{3}''}^H + \mathrm{Re} \Big( 6v_S (\mu_{S1}^H + \mu_{S2}^H) + v_S^2\frac{2{\lambda_1''}^H +8 {\lambda_2''}^H}{3} \Big),\\
        M^H_{14} & = -v^2 \mathrm{Im} (\lambda_6^H),\\
        M^H_{24} & = -\frac{v^2}{2} \mathrm{Im} (\lambda_5^H),\\
        M^H_{34} & = v\mathrm{Im} \Big( \mu_{21}^H-\mu_{12}^H -2v_S({\lambda_3'}^H+{\lambda_6'}^H-{\lambda_7'}^H) \Big),\\
        M^H_{44} & = \mu_H^2 - v^2\frac{\lambda_3^H + \lambda_4^H - \mathrm{Re}(\lambda_5^H)}{2} +
v_S^2({\lambda_2'}^H + 2\mathrm{Re}({\lambda_5'}^H)),\\
        M^H_{15} & = -2v \mathrm{Im}( 2v_S {\lambda_4'}^H + \mu_{11}^H + \mu_{22}^H ),\\
        M^H_{25} & = -v \mathrm{Im}( 2v_S({\lambda_6'}^H + {\lambda_7'}^H) + \mu_{12}^H + \mu_{21}^H ),\\
        M^H_{35} &= \mathrm{Im}\Big( \frac{v^2}{v_S}(\mu_{11}^H + \mu_{22}^H) - 2v_S (3\mu_{S1}^H+\mu_{S2}^H)-\frac{2v_S^2}{3}({\lambda_1''}^H + 2 {\lambda_2''}^H)\Big),\\
        M^H_{45}&= v\mathrm{Re}( \mu_{21}^H-\mu_{12}^H +2v_S({\lambda_7'}^H-{\lambda_6'}^H) ),\\
        M^H_{55}&=4\mu_S^2 +2v^2{\lambda_1'}^H+8v_S^2{\lambda_3''}^H  \\
        &\quad + \mathrm{Re}\Big( \frac{v^2}{v_S}(\mu_{11}^H+\mu_{22}^H)-2v_S(3\mu_{S1}^H-5\mu_{S2}^H)-\frac{2v_S^2}{3}({\lambda_1''}^H-2{\lambda_2''}^H ) \Big),
    \end{split}
\end{equation}
and 
\begin{equation}
    m_{H^\pm}^2 = \mu_H^2 + \frac{v^2\lambda_3}{2} +v_S^2 {\lambda_2'}^H + 2v_S^2 \mathrm{Re}({\lambda_5'}^H).
\end{equation}

The basis change of Higgs basis to interaction basis.
\begin{equation}
    \begin{split}
        {m^H_{11}}^2 &=m_{11}^2 c_\beta^2 + m_{22}^2 s_\beta^2 -\mathrm{Re}(m_{12}^2) s_{2\beta},\\
        \mu_H^2 &= m_{11}^2 s_\beta^2 + m_{22}^2 c_\beta^2 +\mathrm{Re}(m_{12}^2) s_{2\beta},\\
        {m_{12}^H}^2&=\mathrm{Re}(m_{12}^2)c_{2\beta}+(m_{22}^2-m_{11}^2)s_\beta c_\beta + i \mathrm{Im}(m_{12}^2),\\
        \lambda_1^H&=\lambda_1 c_\beta^4  +\lambda_2 s_\beta^4  +2  (\lambda_3+\lambda_4+\operatorname{Re}\lambda_5)c_\beta^2 s_\beta^2 + 4 (c_\beta^2  \operatorname{Re}\lambda_6 +  s_\beta^2 \operatorname{Re}\lambda_7 )s_\beta c_\beta,\\
                \lambda_2^H&=\lambda_1 c_\beta^4  +\lambda_2 s_\beta^4  +2  (\lambda_3+\lambda_4+\operatorname{Re}\lambda_5)c_\beta^2 s_\beta^2 - 4 (s_\beta^2  \operatorname{Re}\lambda_6 +  c_\beta^2 \operatorname{Re}\lambda_7 )s_\beta c_\beta,\\
        \lambda_3^H&=\lambda_3 +  \Big({\lambda_1 + \lambda_2} - 2(\lambda_3+\lambda_4+\operatorname{Re}\lambda_5)\Big)s^2_{ \beta}c^2_\beta-   \operatorname{Re} (\lambda_6-\lambda_7  )s_{ 2\beta} c_{2\beta},   \\
        \lambda_4^H&= \lambda_4 + \Big({\lambda_1 + \lambda_2} - 2(\lambda_3+\lambda_4+\operatorname{Re}\lambda_5)\Big)s^2_{ \beta}c^2_\beta-\operatorname{Re} (\lambda_6-\lambda_7  )s_{ 2\beta} c_{2\beta}, \\
        \lambda_5^H &=\operatorname{Re}\lambda_5 +\Big(\lambda_1+\lambda_2 - 2(\lambda_3+\lambda_4+\operatorname{Re}\lambda_5)\Big)s^2_{\beta}c^2_{\beta}-\operatorname{Re} (\lambda_6-\lambda_7  )s_{ 2\beta} c_{2\beta}\\
        &\quad +i\operatorname{Im}\Big(\lambda_5 c_{2\beta}-(\lambda_6-\lambda_7)s_{2\beta}\Big),\\
        \lambda_6^H&=\Big(\lambda_2 s^2_\beta- \lambda_1 c^2_\beta +( \lambda_3 + \lambda_4 + \operatorname{Re}{\lambda_5})c_{2\beta} \Big)s_\beta c_\beta + \operatorname{Re} {(\lambda_6 c^2_\beta+\lambda_7 s^2_\beta)}c_{2\beta} - \operatorname{Re}(\lambda_6-\lambda_7)2s^2_\beta c^2_\beta \\&+i \operatorname{Im} \Big( \lambda_5 s_\beta c_\beta + \lambda_6 c^2_\beta + \lambda_7 s^2_\beta \Big),\\
\lambda_7^H&=\Big( \lambda_2 c^2_\beta-\lambda_1 s^2_\beta -( \lambda_3 + \lambda_4 + \operatorname{Re}{\lambda_5})c_{2\beta} \Big)s_\beta c_\beta + \operatorname{Re} {(\lambda_6 s^2_\beta+\lambda_7 c^2_\beta)}c_{2\beta} + \operatorname{Re}(\lambda_6-\lambda_7)2s^2_\beta c^2_\beta \\&-i \operatorname{Im} \Big( \lambda_5 s_\beta c_\beta - \lambda_6 c^2_\beta - \lambda_7 s^2_\beta \Big),\\
{\lambda_1'}^H &= \lambda_1' c_\beta^2 + \lambda_2' s_\beta^2 + \operatorname{Re}\lambda_3' s_{2\beta},\\
{\lambda_2'}^H&=\lambda_1' s_\beta^2 + \lambda_2' c_\beta^2 -\operatorname{Re}\lambda_3' s_{2\beta},\\
{\lambda_3'}^H&=(\lambda_2'-\lambda_1')s_\beta c_\beta +\operatorname{Re}\lambda_3' c_{2\beta} +i\operatorname{Im}\lambda_3',\\
{\lambda_4'}^H&= c_\beta^2 \, (\mathrm{Re}\lambda_{4p} + i\, \mathrm{Im}\lambda_{4p}) 
+ s_\beta^2 \, (\mathrm{Re}\lambda_{5p} + i\, \mathrm{Im}\lambda_{5p}) \\
&
\quad + c_\beta s_\beta \big[ (\mathrm{Re}\lambda_{6p} + \mathrm{Re}\lambda_{7p}) 
+ i\, (\mathrm{Im}\lambda_{6p} + \mathrm{Im}\lambda_{7p}) \big],\\
{\lambda_5'}^H&= s_\beta^2 \, (\mathrm{Re}\lambda_{4p} + i\, \mathrm{Im}\lambda_{4p}) 
+ c_\beta^2 \, (\mathrm{Re}\lambda_{5p} + i\, \mathrm{Im}\lambda_{5p}) \\
& 
\quad - c_\beta s_\beta \big[ (\mathrm{Re}\lambda_{6p} + \mathrm{Re}\lambda_{7p}) 
+ i\, (\mathrm{Im}\lambda_{6p} + \mathrm{Im}\lambda_{7p}) \big],\\
{\lambda_6'}^H&=\big[ c_\beta^2 \mathrm{Re}\lambda_{7p} + c_\beta s_\beta (\mathrm{Re}\lambda_{5p} - \mathrm{Re}\lambda_{4p}) - s_\beta^2 \mathrm{Re}\lambda_{6p} \big] \\
&\quad + i\big[ c_\beta^2 \mathrm{Im}\lambda_{7p} + c_\beta s_\beta (\mathrm{Im}\lambda_{5p} - \mathrm{Im}\lambda_{4p}) - s_\beta^2 \mathrm{Im}\lambda_{6p} \big],\\
{\lambda_7'}^H&=\big[ c_\beta^2 \mathrm{Re}\lambda_{6p} + c_\beta s_\beta (\mathrm{Re}\lambda_{5p} - \mathrm{Re}\lambda_{4p}) - s_\beta^2 \mathrm{Re}\lambda_{7p} \big]\\
&\quad + i\big[ c_\beta^2 \mathrm{Im}\lambda_{6p} + c_\beta s_\beta (\mathrm{Im}\lambda_{5p} - \mathrm{Im}\lambda_{4p}) - s_\beta^2 \mathrm{Im}\lambda_{7p} \big],\\
\mu_{11}^H&=\big[ c_\beta^2 \mathrm{Re}\mu_{11} + s_\beta^2 \mathrm{Re}\mu_{22} + c_\beta s_\beta (\mathrm{Re}\mu_{12} + \mathrm{Re}\mu_{21}) \big]\\
& \quad + i\big[ c_\beta^2 \mathrm{Im}\mu_{11} + s_\beta^2 \mathrm{Im}\mu_{22} + c_\beta s_\beta (\mathrm{Im}\mu_{12} + \mathrm{Im}\mu_{21}) \big],\\
\mu_{22}^H&=\big[ s_\beta^2 \mathrm{Re}\mu_{11} + c_\beta^2 \mathrm{Re}\mu_{22} - c_\beta s_\beta (\mathrm{Re}\mu_{12} + \mathrm{Re}\mu_{21}) \big]\\
& \quad + i\big[ s_\beta^2 \mathrm{Im}\mu_{11} + c_\beta^2 \mathrm{Im}\mu_{22} - c_\beta s_\beta (\mathrm{Im}\mu_{12} + \mathrm{Im}\mu_{21}) \big],\\
\mu_{12}^H& =  \big[ c_\beta^2 \mathrm{Re}\mu_{12} - s_\beta^2 \mathrm{Re}\mu_{21} + c_\beta s_\beta (\mathrm{Re}\mu_{22} - \mathrm{Re}\mu_{11}) \big]\\
& \quad + i\big[ c_\beta^2 \mathrm{Im}\mu_{12} - s_\beta^2 \mathrm{Im}\mu_{21} + c_\beta s_\beta (\mathrm{Im}\mu_{22} - \mathrm{Im}\mu_{11}) \big],\\
\mu_{21}^H&=\big[ c_\beta^2 \mathrm{Re}\mu_{21} - s_\beta^2 \mathrm{Re}\mu_{12} + c_\beta s_\beta (\mathrm{Re}\mu_{22} - \mathrm{Re}\mu_{11}) \big]\\
& \quad + i\big[ c_\beta^2 \mathrm{Im}\mu_{21} - s_\beta^2 \mathrm{Im}\mu_{12} + c_\beta s_\beta (\mathrm{Im}\mu_{22} - \mathrm{Im}\mu_{11}) \big].
    \end{split}
\end{equation}

\section{Boundedness from below}
\label{sec:bfb}
The boundedness from below constraint is testing positivity of the Higgs potential at the $\Phi_i\to\infty$ in any directions, where only the quartic terms would play a role. In the 2HDMS, the Higgs potential without quadratic and cubic terms would consist of the following bilinear forms:
\begin{equation}
    \Phi_1^\dagger \Phi_1 = h_1^2,~\Phi_2^\dagger \Phi_2 = h_2^2,~\Phi_1^\dagger \Phi_2 = h_1 h_2\rho e^{i\theta}, S^\dagger S = h_S^2, S^2 = h_S^2 e^{2i\phi_S}, {S^\dagger}^2 = h_S^2 e^{-2i\phi_S},
\end{equation}
where $\Phi_1^\dagger \Phi_2$, $S^2$ and ${S^\dagger}^2$ can have the complex phase. In this case, the Higgs potential can be expressed in terms of the following basis:
\begin{equation}
    X = (h_1^2, h_1 h_2, h_2^2,h_S^2) ^T,
\end{equation}
and the following coefficient matrix $A$
%



\begin{equation}
\resizebox{\linewidth}{!}{$
\begin{bmatrix}
    \frac{\lambda_1}{2} & 
\lambda_6 \rho c_{\theta + \phi_{\lambda6}} & 
\frac{\lambda_3}{2} & 
\frac{\lambda'_{1}}{2} + \lambda'_{4} c_{2\phi_s + \phi_{\lambda'_4}} \\
\lambda_6 \rho c_{\theta + \phi_{\lambda6}} & 
\rho^2 \left(\lambda_4 + \lambda_5 c_{2\theta + \phi_{\lambda5}}\right) & 
\lambda_7 \rho c_{\theta + \phi_{\lambda7}} & 
\rho \left(
\lambda'_{3} c_{\theta + \phi_{\lambda'_3}} 
+ \lambda'_{6} c_{\theta + 2\phi_s + \phi_{\lambda'_6}} 
+ \lambda'_{7} c_{\theta - 2\phi_s - \phi_{\lambda'_7}}
\right) \\
\frac{\lambda_3}{2} & 
\lambda_7 \rho c_{\theta + \phi_{\lambda7}} & 
\frac{\lambda_2}{2} & 
\frac{\lambda'_{2}}{2} + \lambda'_{5} c_{2\phi_s + \phi_{\lambda'_5}} \\
\frac{\lambda'_{1}}{2} + \lambda'_{4} c_{2\phi_s + \phi_{\lambda'_4}} & 
\rho \left(
\lambda'_{3} c_{\theta + \phi_{\lambda'_3}} 
+ \lambda'_{6} c_{\theta + 2\phi_s + \phi_{\lambda'_6}} 
+ \lambda'_{7} c_{\theta - 2\phi_s - \phi_{\lambda'_7}}
\right) & 
\frac{\lambda'_{2}}{2} + \lambda'_{5} c_{2\phi_s + \phi_{\lambda'_5}} & 
\frac{1}{12}\left(
3 \lambda''_{3} 
+ \lambda''_{1} c_{4\phi_s + \phi_{\lambda''_1}} 
+ 2 \lambda''_{2} c_{2\phi_s + \phi_{\lambda''_2}}
\right),
\end{bmatrix}
$}
\end{equation}

where we have part of the Higgs potential with only quartic terms $V_4$
\begin{equation}
    V_4 = X^T A X.
\end{equation}
Here the parameters $\rho$ can be taken from $|\rho|\in [0,1]$, while the phases $\theta$ and $\phi_S$ can be chosen from $[0,2\pi]$.

In principle, the boundedness from below condition can be fulfilled when the matrix $A$ is co-positive~\cite{Kannike:2012pe}. However, we have three undetermined parameters in matrix $A$ and the co-positivity condition should be checked for any possible values of the $\rho$, $\theta$ and $\phi_S$. One can actually use the Gröbner basis method \cite{Ivanov:2018jmz} to do rigorous check for the most general case of this complex potential. However, we found that, this method is difficult to implement into the code and is inefficient to perform the calculation. Therefore, we make a numerical approximation for doing this check, where we scan the phases within the range of $|\rho|\in [0,1]$, $\theta\in[0,2\pi]$ and $\phi_S\in[0,2\pi]$ for discrete points, and do the co-positivity check for each point. When the whole ranges of $\rho$, $\theta$ and $\phi_S$ are co-positive, we can infer that such a point is approximately bounded from below.

\section{Exact formulae of the Barr-Zee type contributions}\label{BarZeeformulae}

Here we list the Barr-Zee contributions to the EDM (CEDM) for a fermion (quark) $d_f^\gamma$, $d_f^Z$ and $d_f^W$ $(d_q^C)$, given in Eqs.~\eqref{edmeq}.
The fermion-loop contributions to the EDM are as follows:
			\small
		\begin{align}
				d_f^V(f')	&=\frac{em_f}{(16\pi^2)^2}8g_{Vff}^vg_{Vf'f'}^v
											Q_{f'}N_C\frac{m_{f'}^2}{v^2}
									\nn\\&\quad\quad
									\times\sum_j^4\int_0^1dz\Bigg\{
												\im[\kappa_f^j]\re[\kappa_{f'}^j]
												\rbra{\frac{1}{z}-2(1-z)}
												+\re[\kappa_f^j]\im[\kappa_{f'}^j]
												\frac{1}{z}
											\Bigg\}
									C^{VH_j}_{f'f'}(z)
				,\\
				d_{f(I_f=-\frac{1}{2})}^W(tb)&=\frac{eg_2^2m_f}{(16\pi^2)^2}N_C
									\int_0^1dz\Bigg\{
											\frac{m_t^2}{v^2}\im[\zeta_f^*\zeta_u]\frac{2-z}{z}
											+\frac{m_b^2}{v^2}\im[\zeta_f^*\zeta_d]
									\Bigg\}\\
									&\qquad \phantom{frac{eg_2^2m_f}{(16\pi^2)^2}}
									\times [Q_t(1-z)+Q_bz]
									C^{WH^\pm}_{tb}(z)
				,\label{tbup}\\
				d_{f(I_f=+\frac{1}{2})}^W(tb)&=\frac{eg_2^2m_f}{(16\pi^2)^2}N_C
									\int_0^1dz\Bigg\{
											\frac{m_t^2}{v^2}\im[\zeta_f\zeta_u^*]
											+\frac{m_b^2}{v^2}\im[\zeta_f\zeta_d^*]\frac{1+z}{1-z}
									\Bigg\}\\
									&\qquad\phantom{frac{eg_2^2m_f}{(16\pi^2)^2}}\times [Q_t(1-z)+Q_bz]
									C^{WH^\pm}_{tb}(z)
		,\label{tbdown}\end{align}
		\normalsize
		\noindent
        where $d_f^V(f')$ denotes the contribution to electric dipole moment (EDM) of fermion $f$, with fermion($f'$) loop and $V=\gamma$, and $Z$ internal leg. $d_{f}^W(tb)$ indicates top-bottom loop contribution to fermion EDM, with $W$ in the internal leg. The internal $W$ leg introduces the weak coupling constant $g_2$ in the expressions~\ref{tbup} and \ref{tbdown}.

The Higgs boson-loop contributions to EDM are
		\begin{align}
				d_f^V(H^\pm)		&=\frac{em_f}{(16\pi^2)^2}4g_{Vff}^v(ig_{H^+H^-V})
									\sum_j^4\im[\kappa_f^j]\frac{g_{H^\pm H^\mp H_j}}{v}
										\int_0^1dz(1-z)
									C^{VH_j}_{H^{^\pm}H^{^\pm}}(z)
				,\\
				d_f^W(H^\pm H)		&=\frac{eg_2^2m_f}{(16\pi^2)^2}\frac{1}{2}
									\sum_j^4\im[\kappa_f^j]\frac{g_{H^{^\pm} H^{^\mp} H_j}}{v}
										\int_0^1dz(1-z)
										C^{WH^\pm}_{H^{^\pm}H_j}(z)
		,\end{align}
\noindent
    where $d_f^V(H^\pm)$ is the contribution to EDM of fermion $f$, via charged Higgs loop, where $V=
    \gamma,Z$ are in the internal leg. $d_f^W(H^\pm H)$ on the other hand implies the charged($H^\pm$) and neutral scalar ($H_i$) loop with $W$ boson in the internal leg.
The gauge-loop contributions to the EDM are
		\small
		\begin{align}
				d_f^V(W)		&=\frac{em_f}{(16\pi^2)^2}8g_{Vff}^v(ig_{WWV})\frac{m_W^2}{v^2}
									\sum_j^4\mathcal{R}_{1j}\im[\kappa_f^j]
									\nn\\&
										\times\int_0^1dz\Bigg[
											\cbra{
												\rbra{6-\frac{m_V^2}{m_W^2}}
												+\rbra{1-\frac{m_V^2}{2m_W^2}}\frac{m_{H_j}^2}{m_W^2}
											}\frac{(1-z)}{2}
											-\rbra{4-\frac{m_V^2}{m_W^2}}\frac{1}{z}
										\Bigg]
									C^{VH^0_j}_{WW}(z)
				,\label{gauge1}\\
				d_f^W(WH)	&=\frac{eg_2^2m_f}{(16\pi^2)^2}\frac{1}{2}\frac{m_W^2}{v^2}
									\sum_j^4\mathcal{R}_{1j}\im[\kappa_f^j]
										\int_0^1dz\Bigg\{
										\frac{4-z}{z}-\frac{m_{H^\pm}^2-m_{H_j}^2}{m_W^2}
										\Bigg\}(1-z)
									C^{WH^\pm}_{WH_j}(z)
		,\label{gauge2}\end{align}
		\normalsize

        \noindent
        where $d_f^V(W)$ is the EDM contribution to fermion $f$, through $W$-loop, where the internal leg is $V=\gamma,Z$. $d_f^W(WH)$ is mixed $W$ and neutral scalar loop with $W$ in the internal leg.
    Finally, the quark-loop contributions to chromo-electric dipole moment (CEDM) are
		\small
		\begin{align}
				d_q^C(q')	&=\frac{m_q}{(16\pi^2)^2}4g_3^3
											\frac{m_{q'}^2}{v^2}
									\sum_j^4\int_0^1dz\Bigg\{
												\im[\kappa_q^j]\re[\kappa_{q'}^j]
												\rbra{\frac{1}{z}-2(1-z)}
												+\re[\kappa_q^j]\im[\kappa_{q'}^j]
												\frac{1}{z}
											\Bigg\}
									C^{gH_j}_{q'q'}(z)
		,\label{cedm}\end{align}
		\normalsize
\noindent
   where $d_q^C(q')$ is the contribution to CEDM of quark (q) with quark ($q'$) loop and gluons in the internal as well as external gauge boson legs. The three interaction vertices of gluon with quarks introduce the third power of the strong coupling constant $g_3^3$ in the expression~\ref{cedm}.

 In all the EDM and CEDM expressions given above, $N_C$ is the color factor.
The coupling constants are given by
\begin{eqnarray}
g_{\gamma ff}^v&=&eQ_f, \nonumber \\
g_{Z ff}^v&=&g_z(I_f/2-Q_fs_W^2), \nonumber \\
g_{H^+ H^- \gamma}&=&-ie, \nonumber \\
g_{H^+ H^- Z}&=&-ig_Z c_{2W}/2, \nonumber \\
    g_{H^\pm H^\mp h^0_j}&=&(\lambda_3\mathcal{R}_{1j}+\text{Re}[\lambda_7]\mathcal{R}_{2j}-\text{Im}[\lambda_7]\mathcal{R}_{3j})v + ((\lambda_2'+2\lambda_5')\mathcal{R}_{4j})v_S, \nonumber \\
g_{WWA}&=&-ie, \nonumber \\
g_{WWZ}&=&-ig_Zc_W^2,
\label{definitions}
\end{eqnarray}
where $Q_f$ are the electric charges of the fermion, $g_Z=\sqrt{g_1^2+g_2^2}$, $s_W=\sin\theta_W$, $c_W=\cos\theta_W$ and $c_{2W}=\cos2\theta_W$. One can write
		\begin{align}
				C^{GH}_{XY}(z)	&=C_0\rbra{0,0;m_G^2,m_H^2,\frac{(1-z)m^2_X+zm_Y^2}{z(1-z)}},
		\end{align}
and where $C_0$ is the Passarino-Veltman function\cite{Passarino:1978jh},
		\begin{align}
			C_0(0,0;m_1,m_2,m_3)=\frac{1}{m_1^2-m_2^2}
									\cbra{
										\frac{m_1^2}{m_1^2-m_3^2}\log\rbra{\frac{m_3^2}{m_1^2}}
										-\frac{m_2^2}{m_2^2-m_3^2}\log\rbra{\frac{m_3^2}{m_2^2}}
										}
		.\end{align}
We confirmed the consistency of our results with Refs.~\cite{Leigh:1990kf,Bowser-Chao:1997kjp,Jung:2013hka,Abe:2013qla,Cheung:2014oaa,Cheung:2020ugr}.

\bibliographystyle{JHEP}
\bibliography{ref,ref1,ref2}

\providecommand{\href}[2]{#2}\begingroup\raggedright\begin{thebibliography}{10}

\bibitem{Sakharov:1967dj}
A.~D. Sakharov, \emph{{Violation of CP Invariance, C asymmetry, and baryon
  asymmetry of the universe}},
  \href{https://doi.org/10.1070/PU1991v034n05ABEH002497}{\emph{Pisma Zh. Eksp.
  Teor. Fiz.} {\bfseries 5} (1967) 32--35}.

\bibitem{Yoshimura:1978ex}
M.~Yoshimura, \emph{{Unified Gauge Theories and the Baryon Number of the
  Universe}}, \href{https://doi.org/10.1103/PhysRevLett.41.281}{\emph{Phys.
  Rev. Lett.} {\bfseries 41} (1978) 281--284}.

\bibitem{Weinberg:1979bt}
S.~Weinberg, \emph{{Cosmological Production of Baryons}},
  \href{https://doi.org/10.1103/PhysRevLett.42.850}{\emph{Phys. Rev. Lett.}
  {\bfseries 42} (1979) 850--853}.

\bibitem{Fukugita:1986hr}
M.~Fukugita and T.~Yanagida, \emph{{Baryogenesis Without Grand Unification}},
  \href{https://doi.org/10.1016/0370-2693(86)91126-3}{\emph{Phys. Lett. B}
  {\bfseries 174} (1986) 45--47}.

\bibitem{Luty:1992un}
M.~A. Luty, \emph{{Baryogenesis via leptogenesis}},
  \href{https://doi.org/10.1103/PhysRevD.45.455}{\emph{Phys. Rev. D} {\bfseries
  45} (1992) 455--465}.

\bibitem{Kuzmin:1985mm}
V.~A. Kuzmin, V.~A. Rubakov and M.~E. Shaposhnikov, \emph{{On the Anomalous
  Electroweak Baryon Number Nonconservation in the Early Universe}},
  \href{https://doi.org/10.1016/0370-2693(85)91028-7}{\emph{Phys. Lett. B}
  {\bfseries 155} (1985) 36}.

\bibitem{Shaposhnikov:1987tw}
M.~E. Shaposhnikov, \emph{{Baryon Asymmetry of the Universe in Standard
  Electroweak Theory}},
  \href{https://doi.org/10.1016/0550-3213(87)90127-1}{\emph{Nucl. Phys. B}
  {\bfseries 287} (1987) 757--775}.

\bibitem{Trodden:1998ym}
M.~Trodden, \emph{{Electroweak baryogenesis}},
  \href{https://doi.org/10.1103/RevModPhys.71.1463}{\emph{Rev. Mod. Phys.}
  {\bfseries 71} (1999) 1463--1500},
  [\href{https://arxiv.org/abs/hep-ph/9803479}{{\ttfamily hep-ph/9803479}}].

\bibitem{Morrissey:2012db}
D.~E. Morrissey and M.~J. Ramsey-Musolf, \emph{{Electroweak baryogenesis}},
  \href{https://doi.org/10.1088/1367-2630/14/12/125003}{\emph{New J. Phys.}
  {\bfseries 14} (2012) 125003},
  [\href{https://arxiv.org/abs/1206.2942}{{\ttfamily 1206.2942}}].

\bibitem{Fromme:2006cm}
L.~Fromme, S.~J. Huber and M.~Seniuch, \emph{{Baryogenesis in the two-Higgs
  doublet model}},
  \href{https://doi.org/10.1088/1126-6708/2006/11/038}{\emph{JHEP} {\bfseries
  11} (2006) 038}, [\href{https://arxiv.org/abs/hep-ph/0605242}{{\ttfamily
  hep-ph/0605242}}].

\bibitem{Profumo:2007wc}
S.~Profumo, M.~J. Ramsey-Musolf and G.~Shaughnessy, \emph{{Singlet Higgs
  phenomenology and the electroweak phase transition}},
  \href{https://doi.org/10.1088/1126-6708/2007/08/010}{\emph{JHEP} {\bfseries
  08} (2007) 010}, [\href{https://arxiv.org/abs/0705.2425}{{\ttfamily
  0705.2425}}].

\bibitem{Arhrib:2010ju}
A.~Arhrib, E.~Christova, H.~Eberl and E.~Ginina, \emph{{CP violation in charged
  Higgs production and decays in the Complex Two Higgs Doublet Model}},
  \href{https://doi.org/10.1007/JHEP04(2011)089}{\emph{JHEP} {\bfseries 04}
  (2011) 089}, [\href{https://arxiv.org/abs/1011.6560}{{\ttfamily 1011.6560}}].

\bibitem{Cline:2011mm}
J.~M. Cline, K.~Kainulainen and M.~Trott, \emph{{Electroweak Baryogenesis in
  Two Higgs Doublet Models and B meson anomalies}},
  \href{https://doi.org/10.1007/JHEP11(2011)089}{\emph{JHEP} {\bfseries 11}
  (2011) 089}, [\href{https://arxiv.org/abs/1107.3559}{{\ttfamily 1107.3559}}].

\bibitem{Biekotter:2025fjx}
T.~Biek{\"o}tter and M.~O. Olea-Romacho, \emph{{Benchmarking a fading window:
  electroweak baryogenesis in the C2HDM, LHC constraints after Run 2 and
  prospects for LISA}},
  \href{https://doi.org/10.1007/JHEP12(2025)040}{\emph{JHEP} {\bfseries 12}
  (2025) 040}, [\href{https://arxiv.org/abs/2505.09670}{{\ttfamily
  2505.09670}}].

\bibitem{Biekotter:2025vxl}
T.~Biek{\"o}tter, P.~Gabriel, M.~M. M{\"u}hlleitner and R.~Santos,
  \emph{{Towards a unified framework for pseudo-Nambu-Goldstone Dark Matter and
  electroweak baryogenesis}},
  \href{https://doi.org/10.1007/JHEP04(2026)038}{\emph{JHEP} {\bfseries 04}
  (2026) 038}, [\href{https://arxiv.org/abs/2509.01682}{{\ttfamily
  2509.01682}}].

\bibitem{Kanemura:2020ibp}
S.~Kanemura, M.~Kubota and K.~Yagyu, \emph{{Aligned CP-violating Higgs sector
  canceling the electric dipole moment}},
  \href{https://doi.org/10.1007/JHEP08(2020)026}{\emph{JHEP} {\bfseries 08}
  (2020) 026}, [\href{https://arxiv.org/abs/2004.03943}{{\ttfamily
  2004.03943}}].

\bibitem{Lahiri:2026eud}
J.~Lahiri and G.~Moortgat-Pick, \emph{{CP-violation and its implications in a
  complex singlet extension of 2HDM}},
  \href{https://arxiv.org/abs/2603.21483}{{\ttfamily 2603.21483}}.

\bibitem{Shah:2018elh}
N.~R. Shah, \emph{{The Higgs and WIMP DM Lamp Posts for New Weak Scale Physics:
  EFT Perspectives and the NMSSM}},
  \href{https://doi.org/10.1016/j.nuclphysbps.2019.03.018}{\emph{Nucl. Part.
  Phys. Proc.} {\bfseries 303-305} (2018) 92--97},
  [\href{https://arxiv.org/abs/1812.10882}{{\ttfamily 1812.10882}}].

\bibitem{Heinemeyer:2021msz}
S.~Heinemeyer, C.~Li, F.~Lika, G.~Moortgat-Pick and S.~Paasch,
  \emph{{Phenomenology of a 96~GeV Higgs boson in the 2HDM with an additional
  singlet}}, \href{https://doi.org/10.1103/PhysRevD.106.075003}{\emph{Phys.
  Rev. D} {\bfseries 106} (2022) 075003},
  [\href{https://arxiv.org/abs/2112.11958}{{\ttfamily 2112.11958}}].

\bibitem{Dutta:2022qeq}
J.~Dutta, G.~Moortgat-Pick and M.~Schreiber, \emph{{Phenomenology of the dark
  matter sector in the 2HDM extended with complex scalar singlet}},
  \href{https://doi.org/10.1140/epjp/s13360-025-06015-7}{\emph{Eur. Phys. J.
  Plus} {\bfseries 140} (2025) 87},
  [\href{https://arxiv.org/abs/2203.05509}{{\ttfamily 2203.05509}}].

\bibitem{Dutta:2023cig}
J.~Dutta, J.~Lahiri, C.~Li, G.~Moortgat-Pick, S.~F. Tabira and J.~A. Ziegler,
  \emph{{Dark matter phenomenology in 2HDMS in light of the 95 GeV excess}},
  \href{https://doi.org/10.1140/epjc/s10052-024-13176-9}{\emph{Eur. Phys. J. C}
  {\bfseries 84} (2024) 926},
  [\href{https://arxiv.org/abs/2308.05653}{{\ttfamily 2308.05653}}].

\bibitem{Dutta:2025nmy}
J.~Dutta, J.~Lahiri, C.~Li, G.~Moortgat-Pick, S.~F. Tabira and J.~A. Ziegler,
  \emph{{Search for Dark Matter in 2HDMS at LHC and future Lepton Colliders}},
  \href{https://arxiv.org/abs/2504.14529}{{\ttfamily 2504.14529}}.

\bibitem{CMS:2017yta}
{\scshape CMS} collaboration, \emph{{Search for new resonances in the diphoton
  final state in the mass range between 70 and 110 GeV in pp collisions at
  $\sqrt{s}=$ 8 and 13 TeV}}, .

\bibitem{ATLAS:2018xad}
{\scshape ATLAS} collaboration, \emph{{Search for resonances in the 65 to 110
  GeV diphoton invariant mass range using 80 fb$^{-1}$ of $pp$ collisions
  collected at $\sqrt{s}=13$ TeV with the ATLAS detector}}, .

\bibitem{CMS:2023yay}
{\scshape CMS} collaboration, \emph{{Search for a standard model-like Higgs
  boson in the mass range between 70 and 110$~\mathrm{GeV}$ in the diphoton
  final state in proton-proton collisions at $\sqrt{s}=13~\mathrm{TeV}$}},
  \href{https://arxiv.org/abs/2405.18149}{{\ttfamily 2405.18149}}.

\bibitem{CMS:2022rbd}
{\scshape CMS} collaboration, \emph{{Searches for additional Higgs bosons and
  vector leptoquarks in $\tau\tau$ final states in proton-proton collisions at
  $\sqrt{s}=13~\mathrm{TeV}$}}, .

\bibitem{ATLAS-CONF-2023-035}
{\scshape ATLAS} collaboration, \emph{{Search for diphoton resonances in the 66
  to 110 GeV mass range using 140 fb$^{-1}$ of 13 TeV pp collisions collected
  with the ATLAS detector}},  tech. rep., CERN, Geneva, 2023.

\bibitem{LEPWorkingGroupforHiggsbosonsearches:2003ing}
{\scshape LEP Working Group for Higgs boson searches, ALEPH, DELPHI, L3, OPAL}
  collaboration, R.~Barate et~al., \emph{{Search for the standard model Higgs
  boson at LEP}},
  \href{https://doi.org/10.1016/S0370-2693(03)00614-2}{\emph{Phys. Lett. B}
  {\bfseries 565} (2003) 61--75},
  [\href{https://arxiv.org/abs/hep-ex/0306033}{{\ttfamily hep-ex/0306033}}].

\bibitem{Fox:2017uwr}
P.~J. Fox and N.~Weiner, \emph{{Light Signals from a Lighter Higgs}},
  \href{https://doi.org/10.1007/JHEP08(2018)025}{\emph{JHEP} {\bfseries 08}
  (2018) 025}, [\href{https://arxiv.org/abs/1710.07649}{{\ttfamily
  1710.07649}}].

\bibitem{Haisch:2017gql}
U.~Haisch and A.~Malinauskas, \emph{{Let there be light from a second light
  Higgs doublet}}, \href{https://doi.org/10.1007/JHEP03(2018)135}{\emph{JHEP}
  {\bfseries 03} (2018) 135},
  [\href{https://arxiv.org/abs/1712.06599}{{\ttfamily 1712.06599}}].

\bibitem{Benbrik:2022azi}
R.~Benbrik, M.~Boukidi, S.~Moretti and S.~Semlali, \emph{{Explaining the 96 GeV
  Di-photon anomaly in a generic 2HDM Type-III}},
  \href{https://doi.org/10.1016/j.physletb.2022.137245}{\emph{Phys. Lett. B}
  {\bfseries 832} (2022) 137245},
  [\href{https://arxiv.org/abs/2204.07470}{{\ttfamily 2204.07470}}].

\bibitem{Benbrik:2022dja}
R.~Benbrik, M.~Boukidi and B.~Manaut, \emph{{$W$-mass and 96 GeV excess in
  type-III 2HDM}},  \href{https://arxiv.org/abs/2204.11755}{{\ttfamily
  2204.11755}}.

\bibitem{Azevedo:2023zkg}
D.~Azevedo, T.~Biek\"otter and P.~M. Ferreira, \emph{{2HDM interpretations of
  the CMS diphoton excess at 95 GeV}},
  \href{https://arxiv.org/abs/2305.19716}{{\ttfamily 2305.19716}}.

\bibitem{Belyaev:2023xnv}
A.~Belyaev, R.~Benbrik, M.~Boukidi, M.~Chakraborti, S.~Moretti and S.~Semlali,
  \emph{{Explanation of the Hints for a 95 GeV Higgs Boson within a 2-Higgs
  Doublet Model}},  \href{https://arxiv.org/abs/2306.09029}{{\ttfamily
  2306.09029}}.

\bibitem{Biekotter:2019kde}
T.~Biek\"otter, M.~Chakraborti and S.~Heinemeyer, \emph{{A 96 GeV Higgs boson
  in the N2HDM}},
  \href{https://doi.org/10.1140/epjc/s10052-019-7561-2}{\emph{Eur. Phys. J. C}
  {\bfseries 80} (2020) 2}, [\href{https://arxiv.org/abs/1903.11661}{{\ttfamily
  1903.11661}}].

\bibitem{Aguilar-Saavedra:2023vpd}
J.~A. Aguilar-Saavedra, H.~B. C\^amara, F.~R. Joaquim and J.~F. Seabra,
  \emph{{Confronting the 95 GeV excesses within the UN2HDM}},
  \href{https://arxiv.org/abs/2307.03768}{{\ttfamily 2307.03768}}.

\bibitem{Banik:2023ecr}
S.~Banik, A.~Crivellin, S.~Iguro and T.~Kitahara, \emph{{Asymmetric Di-Higgs
  Signals of the N2HDM-$U(1)$}},
  \href{https://arxiv.org/abs/2303.11351}{{\ttfamily 2303.11351}}.

\bibitem{Biekotter:2021ovi}
T.~Biek\"otter and M.~O. Olea-Romacho, \emph{{Reconciling Higgs physics and
  pseudo-Nambu-Goldstone dark matter in the S2HDM using a genetic algorithm}},
  \href{https://doi.org/10.1007/JHEP10(2021)215}{\emph{JHEP} {\bfseries 10}
  (2021) 215}, [\href{https://arxiv.org/abs/2108.10864}{{\ttfamily
  2108.10864}}].

\bibitem{Biekotter:2023jld}
T.~Biek\"otter, S.~Heinemeyer and G.~Weiglein, \emph{{The CMS di-photon excess
  at 95 GeV in view of the LHC Run 2 results}},
  \href{https://arxiv.org/abs/2303.12018}{{\ttfamily 2303.12018}}.

\bibitem{Biekotter:2023oen}
T.~Biek\"otter, S.~Heinemeyer and G.~Weiglein, \emph{{The 95.4 GeV di-photon
  excess at ATLAS and CMS}},
  \href{https://arxiv.org/abs/2306.03889}{{\ttfamily 2306.03889}}.

\bibitem{Cao:2016uwt}
J.~Cao, X.~Guo, Y.~He, P.~Wu and Y.~Zhang, \emph{{Diphoton signal of the light
  Higgs boson in natural NMSSM}},
  \href{https://doi.org/10.1103/PhysRevD.95.116001}{\emph{Phys. Rev. D}
  {\bfseries 95} (2017) 116001},
  [\href{https://arxiv.org/abs/1612.08522}{{\ttfamily 1612.08522}}].

\bibitem{Cao:2019ofo}
J.~Cao, X.~Jia, Y.~Yue, H.~Zhou and P.~Zhu, \emph{{96 GeV diphoton excess in
  seesaw extensions of the natural NMSSM}},
  \href{https://doi.org/10.1103/PhysRevD.101.055008}{\emph{Phys. Rev. D}
  {\bfseries 101} (2020) 055008},
  [\href{https://arxiv.org/abs/1908.07206}{{\ttfamily 1908.07206}}].

\bibitem{Choi:2019yrv}
K.~Choi, S.~H. Im, K.~S. Jeong and C.~B. Park, \emph{{Light Higgs bosons in the
  general NMSSM}},
  \href{https://doi.org/10.1140/epjc/s10052-019-7473-1}{\emph{Eur. Phys. J. C}
  {\bfseries 79} (2019) 956},
  [\href{https://arxiv.org/abs/1906.03389}{{\ttfamily 1906.03389}}].

\bibitem{Biekotter:2021qbc}
T.~Biek\"otter, A.~Grohsjean, S.~Heinemeyer, C.~Schwanenberger and G.~Weiglein,
  \emph{{Possible indications for new Higgs bosons in the reach of the LHC:
  N2HDM and NMSSM interpretations}},
  \href{https://doi.org/10.1140/epjc/s10052-022-10099-1}{\emph{Eur. Phys. J. C}
  {\bfseries 82} (2022) 178},
  [\href{https://arxiv.org/abs/2109.01128}{{\ttfamily 2109.01128}}].

\bibitem{Li:2022etb}
W.~Li, J.~Zhu, K.~Wang, S.~Ma, P.~Tian and H.~Qiao, \emph{{A light Higgs boson
  in the NMSSM confronted with the CMS di-photon and di-tau excesses}},
  \href{https://arxiv.org/abs/2212.11739}{{\ttfamily 2212.11739}}.

\bibitem{PhysRevD.100.035023}
J.~M. Cline and T.~Toma, \emph{Pseudo-goldstone dark matter confronts cosmic
  ray and collider anomalies},
  \href{https://doi.org/10.1103/PhysRevD.100.035023}{\emph{Phys. Rev. D}
  {\bfseries 100} (Aug, 2019) 035023}.

\bibitem{Biekotter:2017xmf}
T.~Biek\"otter, S.~Heinemeyer and C.~Mu\~noz, \emph{{Precise prediction for the
  Higgs-boson masses in the $\mu \nu $ SSM}},
  \href{https://doi.org/10.1140/epjc/s10052-018-5978-7}{\emph{Eur. Phys. J. C}
  {\bfseries 78} (2018) 504},
  [\href{https://arxiv.org/abs/1712.07475}{{\ttfamily 1712.07475}}].

\bibitem{Liu:2018ryo}
L.~Liu, H.~Qiao, K.~Wang and J.~Zhu, \emph{{A Light Scalar in the Minimal
  Dilaton Model in Light of LHC Constraints}},
  \href{https://doi.org/10.1088/1674-1137/43/2/023104}{\emph{Chin. Phys. C}
  {\bfseries 43} (2019) 023104},
  [\href{https://arxiv.org/abs/1812.00107}{{\ttfamily 1812.00107}}].

\bibitem{Liu:2018xsw}
D.~Liu, J.~Liu, C.~E.~M. Wagner and X.-P. Wang, \emph{{A Light Higgs at the LHC
  and the B-Anomalies}},
  \href{https://doi.org/10.1007/JHEP06(2018)150}{\emph{JHEP} {\bfseries 06}
  (2018) 150}, [\href{https://arxiv.org/abs/1805.01476}{{\ttfamily
  1805.01476}}].

\bibitem{Biekotter:2019gtq}
T.~Biek\"otter, S.~Heinemeyer and C.~Mu\~noz, \emph{{Precise prediction for the
  Higgs-Boson masses in the ${\mu}{\nu}$SSM with three right-handed neutrino
  superfields}},
  \href{https://doi.org/10.1140/epjc/s10052-019-7175-8}{\emph{Eur. Phys. J. C}
  {\bfseries 79} (2019) 667},
  [\href{https://arxiv.org/abs/1906.06173}{{\ttfamily 1906.06173}}].

\bibitem{Richard:2017kot}
F.~Richard, \emph{{Search for a light radion at HL-LHC and ILC250}},
  \href{https://arxiv.org/abs/1712.06410}{{\ttfamily 1712.06410}}.

\bibitem{Aguilar-Saavedra:2020wrj}
J.~A. Aguilar-Saavedra and F.~R. Joaquim, \emph{{Multiphoton signals of a (96
  GeV?) stealth boson}},
  \href{https://doi.org/10.1140/epjc/s10052-020-7952-4}{\emph{Eur. Phys. J. C}
  {\bfseries 80} (2020) 403},
  [\href{https://arxiv.org/abs/2002.07697}{{\ttfamily 2002.07697}}].

\bibitem{Biekotter:2020cjs}
T.~Biek\"otter, M.~Chakraborti and S.~Heinemeyer, \emph{{The
  \textquotedblleft{}96 GeV excess\textquotedblright{} at the LHC}},
  \href{https://doi.org/10.1142/S0217751X21420185}{\emph{Int. J. Mod. Phys. A}
  {\bfseries 36} (2021) 2142018},
  [\href{https://arxiv.org/abs/2003.05422}{{\ttfamily 2003.05422}}].

\bibitem{Coloretti:2023wng}
G.~Coloretti, A.~Crivellin, S.~Bhattacharya and B.~Mellado, \emph{{Searching
  for Low-Mass Resonances Decaying into $W$ Bosons}},
  \href{https://arxiv.org/abs/2302.07276}{{\ttfamily 2302.07276}}.

\bibitem{Bhattacharya:2023lmu}
S.~Bhattacharya, G.~Coloretti, A.~Crivellin, S.-E. Dahbi, Y.~Fang, M.~Kumar
  et~al., \emph{{Growing Excesses of New Scalars at the Electroweak Scale}},
  \href{https://arxiv.org/abs/2306.17209}{{\ttfamily 2306.17209}}.

\bibitem{Ashanujjaman:2023etj}
S.~Ashanujjaman, S.~Banik, G.~Coloretti, A.~Crivellin, B.~Mellado and A.-T.
  Mulaudzi, \emph{{$SU(2)_L$ triplet scalar as the origin of the 95 GeV
  excess?}},  \href{https://arxiv.org/abs/2306.15722}{{\ttfamily 2306.15722}}.

\bibitem{Escribano:2023hxj}
P.~Escribano, V.~M. Lozano and A.~Vicente, \emph{{A Scotogenic explanation for
  the 95 GeV excesses}},  \href{https://arxiv.org/abs/2306.03735}{{\ttfamily
  2306.03735}}.

\bibitem{Borah:2023hqw}
D.~Borah, S.~Mahapatra, P.~K. Paul and N.~Sahu, \emph{{Scotogenic
  U(1)L\ensuremath{\mu}-L\ensuremath{\tau} origin of (g-2)\ensuremath{\mu},
  W-mass anomaly and 95~GeV excess}},
  \href{https://doi.org/10.1103/PhysRevD.109.055021}{\emph{Phys. Rev. D}
  {\bfseries 109} (2024) 055021},
  [\href{https://arxiv.org/abs/2310.11953}{{\ttfamily 2310.11953}}].

\bibitem{Pich:2009sp}
A.~Pich and P.~Tuzon, \emph{{Yukawa Alignment in the Two-Higgs-Doublet Model}},
  \href{https://doi.org/10.1103/PhysRevD.80.091702}{\emph{Phys. Rev. D}
  {\bfseries 80} (2009) 091702},
  [\href{https://arxiv.org/abs/0908.1554}{{\ttfamily 0908.1554}}].

\bibitem{Staub:2013tta}
F.~Staub, \emph{{SARAH 4 : A tool for (not only SUSY) model builders}},
  \href{https://doi.org/10.1016/j.cpc.2014.02.018}{\emph{Comput. Phys. Commun.}
  {\bfseries 185} (2014) 1773--1790},
  [\href{https://arxiv.org/abs/1309.7223}{{\ttfamily 1309.7223}}].

\bibitem{Porod:2003um}
W.~Porod, \emph{{SPheno, a program for calculating supersymmetric spectra, SUSY
  particle decays and SUSY particle production at e+ e- colliders}},
  \href{https://doi.org/10.1016/S0010-4655(03)00222-4}{\emph{Comput. Phys.
  Commun.} {\bfseries 153} (2003) 275--315},
  [\href{https://arxiv.org/abs/hep-ph/0301101}{{\ttfamily hep-ph/0301101}}].

\bibitem{Bahl:2022igd}
H.~Bahl, T.~Biek\"otter, S.~Heinemeyer, C.~Li, S.~Paasch, G.~Weiglein et~al.,
  \emph{{HiggsTools: BSM scalar phenomenology with new versions of HiggsBounds
  and HiggsSignals}},
  \href{https://doi.org/10.1016/j.cpc.2023.108803}{\emph{Comput. Phys. Commun.}
  {\bfseries 291} (2023) 108803},
  [\href{https://arxiv.org/abs/2210.09332}{{\ttfamily 2210.09332}}].

\bibitem{Bechtle:2013xfa}
P.~Bechtle, S.~Heinemeyer, O.~St$\dot{a}$l, T.~Stefaniak and G.~Weiglein,
  \emph{{$HiggsSignals$: Confronting arbitrary Higgs sectors with measurements
  at the Tevatron and the LHC}},
  \href{https://doi.org/10.1140/epjc/s10052-013-2711-4}{\emph{Eur. Phys. J. C}
  {\bfseries 74} (2014) 2711},
  [\href{https://arxiv.org/abs/1305.1933}{{\ttfamily 1305.1933}}].

\bibitem{Bechtle:2013wla}
P.~Bechtle, O.~Brein, S.~Heinemeyer, O.~St\r{a}l, T.~Stefaniak, G.~Weiglein
  et~al., \emph{{$\mathsf{HiggsBounds}-4$: Improved Tests of Extended Higgs
  Sectors against Exclusion Bounds from LEP, the Tevatron and the LHC}},
  \href{https://doi.org/10.1140/epjc/s10052-013-2693-2}{\emph{Eur. Phys. J. C}
  {\bfseries 74} (2014) 2693},
  [\href{https://arxiv.org/abs/1311.0055}{{\ttfamily 1311.0055}}].

\bibitem{Bechtle:2020pkv}
P.~Bechtle, D.~Dercks, S.~Heinemeyer, T.~Klingl, T.~Stefaniak, G.~Weiglein
  et~al., \emph{{HiggsBounds-5: Testing Higgs Sectors in the LHC 13 TeV Era}},
  \href{https://doi.org/10.1140/epjc/s10052-020-08557-9}{\emph{Eur. Phys. J. C}
  {\bfseries 80} (2020) 1211},
  [\href{https://arxiv.org/abs/2006.06007}{{\ttfamily 2006.06007}}].

\bibitem{Bechtle:2020uwn}
P.~Bechtle, S.~Heinemeyer, T.~Klingl, T.~Stefaniak, G.~Weiglein and
  J.~Wittbrodt, \emph{{HiggsSignals-2: Probing new physics with precision Higgs
  measurements in the LHC 13 TeV era}},
  \href{https://doi.org/10.1140/epjc/s10052-021-08942-y}{\emph{Eur. Phys. J. C}
  {\bfseries 81} (2021) 145},
  [\href{https://arxiv.org/abs/2012.09197}{{\ttfamily 2012.09197}}].

\bibitem{Panda:2019dnf}
C.~D. Panda, C.~Meisenhelder, M.~Verma, D.~G. Ang, J.~Chow, Z.~Lasner et~al.,
  \emph{{Attaining the shot-noise-limit in the ACME measurement of the electron
  electric dipole moment}},
  \href{https://doi.org/10.1088/1361-6455/ab4a61}{\emph{J. Phys. B} {\bfseries
  52} (2019) 235003}, [\href{https://arxiv.org/abs/1907.00554}{{\ttfamily
  1907.00554}}].

\bibitem{Roussy:2022cmp}
T.~S. Roussy et~al., \emph{{An improved bound on the electron{\textquoteright}s
  electric dipole moment}},
  \href{https://doi.org/10.1126/science.adg4084}{\emph{Science} {\bfseries 381}
  (2023) adg4084}, [\href{https://arxiv.org/abs/2212.11841}{{\ttfamily
  2212.11841}}].

\bibitem{Abel:2020pzs}
C.~Abel et~al., \emph{{Measurement of the Permanent Electric Dipole Moment of
  the Neutron}},
  \href{https://doi.org/10.1103/PhysRevLett.124.081803}{\emph{Phys. Rev. Lett.}
  {\bfseries 124} (2020) 081803},
  [\href{https://arxiv.org/abs/2001.11966}{{\ttfamily 2001.11966}}].

\bibitem{Abe:2013qla}
T.~Abe, J.~Hisano, T.~Kitahara and K.~Tobioka, \emph{{Gauge invariant Barr-Zee
  type contributions to fermionic EDMs in the two-Higgs doublet models}},
  \href{https://doi.org/10.1007/JHEP01(2014)106}{\emph{JHEP} {\bfseries 01}
  (2014) 106}, [\href{https://arxiv.org/abs/1311.4704}{{\ttfamily 1311.4704}}].

\bibitem{Berge:2008dr}
S.~Berge and W.~Bernreuther, \emph{{Determining the CP parity of Higgs bosons
  at the LHC in the tau to 1-prong decay channels}},
  \href{https://doi.org/10.1016/j.physletb.2008.12.065}{\emph{Phys. Lett. B}
  {\bfseries 671} (2009) 470--476},
  [\href{https://arxiv.org/abs/0812.1910}{{\ttfamily 0812.1910}}].

\bibitem{Berge:2014sra}
S.~Berge, W.~Bernreuther and S.~Kirchner, \emph{{Determination of the Higgs
  CP-mixing angle in the tau decay channels at the LHC including the
  Drell{\textendash}Yan background}},
  \href{https://doi.org/10.1140/epjc/s10052-014-3164-0}{\emph{Eur. Phys. J. C}
  {\bfseries 74} (2014) 3164},
  [\href{https://arxiv.org/abs/1408.0798}{{\ttfamily 1408.0798}}].

\bibitem{CMS:2021sdq}
{\scshape CMS} collaboration, A.~Tumasyan et~al., \emph{{Analysis of the $CP$
  structure of the Yukawa coupling between the Higgs boson and $\tau$ leptons
  in proton-proton collisions at $ \sqrt{s} $ = 13 TeV}},
  \href{https://doi.org/10.1007/JHEP06(2022)012}{\emph{JHEP} {\bfseries 06}
  (2022) 012}, [\href{https://arxiv.org/abs/2110.04836}{{\ttfamily
  2110.04836}}].

\bibitem{ATLAS:2022akr}
{\scshape ATLAS} collaboration, G.~Aad et~al., \emph{{Measurement of the CP
  properties of Higgs boson interactions with $\tau $-leptons with the ATLAS
  detector}}, \href{https://doi.org/10.1140/epjc/s10052-023-11583-y}{\emph{Eur.
  Phys. J. C} {\bfseries 83} (2023) 563},
  [\href{https://arxiv.org/abs/2212.05833}{{\ttfamily 2212.05833}}].

\bibitem{Berge:2013jra}
S.~Berge, W.~Bernreuther and H.~Spiesberger, \emph{{Higgs CP properties using
  the $\tau$ decay modes at the ILC}},
  \href{https://doi.org/10.1016/j.physletb.2013.11.006}{\emph{Phys. Lett. B}
  {\bfseries 727} (2013) 488--495},
  [\href{https://arxiv.org/abs/1308.2674}{{\ttfamily 1308.2674}}].

\bibitem{Desch:2003rw}
K.~Desch, A.~Imhof, Z.~Was and M.~Worek, \emph{{Probing the CP nature of the
  Higgs boson at linear colliders with tau spin correlations: The Case of mixed
  scalar - pseudoscalar couplings}},
  \href{https://doi.org/10.1016/j.physletb.2003.10.074}{\emph{Phys. Lett. B}
  {\bfseries 579} (2004) 157--164},
  [\href{https://arxiv.org/abs/hep-ph/0307331}{{\ttfamily hep-ph/0307331}}].

\bibitem{Rouge:2005iy}
A.~Rouge, \emph{{CP violation in a light Higgs boson decay from tau-spin
  correlations at a linear collider}},
  \href{https://doi.org/10.1016/j.physletb.2005.05.076}{\emph{Phys. Lett. B}
  {\bfseries 619} (2005) 43--49},
  [\href{https://arxiv.org/abs/hep-ex/0505014}{{\ttfamily hep-ex/0505014}}].

\bibitem{Gritsan:2022php}
A.~V. Gritsan et~al., \emph{{Snowmass White Paper: Prospects of CP-violation
  measurements with the Higgs boson at future experiments}},
  \href{https://arxiv.org/abs/2205.07715}{{\ttfamily 2205.07715}}.

\bibitem{ilctalk}
{Ivana Vidakovic}, ``Cp violation in higgs to tautau decays at the cepc.''
  \url{https://indico.ihep.ac.cn/event/25300/contributions/203608/attachments/96851/128083/CP%20violation%20in%20Higgs%20decay%20to%20tautau%20at%20the%20CEPC.pdf},
  2025.

\bibitem{Li:2025ouv}
C.~Li and G.~Moortgat-Pick, \emph{{Determination of
  $\mathcal{C}\mathcal{P}$-violating HZZ interaction with polarised beams at
  the ILC}}, \href{https://doi.org/10.1140/epjc/s10052-025-13828-4}{\emph{Eur.
  Phys. J. C} {\bfseries 85} (2025) 99}.

\bibitem{ATLAS:2025jdz}
{\scshape ATLAS} collaboration, \emph{{ATLAS projected sensitivities for for
  the cross-section measurement and new physics searches in four-top-quark
  production at the HL-LHC}}, .

\bibitem{AharonyShapira:2021ize}
S.~Aharony~Shapira, \emph{{Current bounds on baryogenesis from complex Yukawa
  couplings of light fermions}},
  \href{https://doi.org/10.1103/PhysRevD.105.095037}{\emph{Phys. Rev. D}
  {\bfseries 105} (2022) 095037},
  [\href{https://arxiv.org/abs/2106.05338}{{\ttfamily 2106.05338}}].

\bibitem{Kannike:2012pe}
K.~Kannike, \emph{{Vacuum Stability Conditions From Copositivity Criteria}},
  \href{https://doi.org/10.1140/epjc/s10052-012-2093-z}{\emph{Eur. Phys. J. C}
  {\bfseries 72} (2012) 2093},
  [\href{https://arxiv.org/abs/1205.3781}{{\ttfamily 1205.3781}}].

\bibitem{Ivanov:2018jmz}
I.~P. Ivanov, M.~K{\"o}pke and M.~M{\"u}hlleitner, \emph{{Algorithmic
  Boundedness-From-Below Conditions for Generic Scalar Potentials}},
  \href{https://doi.org/10.1140/epjc/s10052-018-5893-y}{\emph{Eur. Phys. J. C}
  {\bfseries 78} (2018) 413},
  [\href{https://arxiv.org/abs/1802.07976}{{\ttfamily 1802.07976}}].

\bibitem{Passarino:1978jh}
G.~Passarino and M.~J.~G. Veltman, \emph{{One Loop Corrections for e+ e-
  Annihilation Into mu+ mu- in the Weinberg Model}},
  \href{https://doi.org/10.1016/0550-3213(79)90234-7}{\emph{Nucl. Phys. B}
  {\bfseries 160} (1979) 151--207}.

\bibitem{Leigh:1990kf}
R.~G. Leigh, S.~Paban and R.~M. Xu, \emph{{Electric dipole moment of
  electron}}, \href{https://doi.org/10.1016/0550-3213(91)90128-K}{\emph{Nucl.
  Phys. B} {\bfseries 352} (1991) 45--58}.

\bibitem{Bowser-Chao:1997kjp}
D.~Bowser-Chao, D.~Chang and W.-Y. Keung, \emph{{Electron electric dipole
  moment from CP violation in the charged Higgs sector}},
  \href{https://doi.org/10.1103/PhysRevLett.79.1988}{\emph{Phys. Rev. Lett.}
  {\bfseries 79} (1997) 1988--1991},
  [\href{https://arxiv.org/abs/hep-ph/9703435}{{\ttfamily hep-ph/9703435}}].

\bibitem{Jung:2013hka}
M.~Jung and A.~Pich, \emph{{Electric Dipole Moments in Two-Higgs-Doublet
  Models}}, \href{https://doi.org/10.1007/JHEP04(2014)076}{\emph{JHEP}
  {\bfseries 04} (2014) 076},
  [\href{https://arxiv.org/abs/1308.6283}{{\ttfamily 1308.6283}}].

\bibitem{Cheung:2014oaa}
K.~Cheung, J.~S. Lee, E.~Senaha and P.-Y. Tseng, \emph{{Confronting Higgcision
  with Electric Dipole Moments}},
  \href{https://doi.org/10.1007/JHEP06(2014)149}{\emph{JHEP} {\bfseries 06}
  (2014) 149}, [\href{https://arxiv.org/abs/1403.4775}{{\ttfamily 1403.4775}}].

\bibitem{Cheung:2020ugr}
K.~Cheung, A.~Jueid, Y.-N. Mao and S.~Moretti, \emph{{Two-Higgs-doublet model
  with soft $CP$ violation confronting electric dipole moments and colliders}},
  \href{https://doi.org/10.1103/PhysRevD.102.075029}{\emph{Phys. Rev. D}
  {\bfseries 102} (2020) 075029},
  [\href{https://arxiv.org/abs/2003.04178}{{\ttfamily 2003.04178}}].

\end{thebibliography}\endgroup

\end{document}